\documentclass[aps,prd,showpacs,onecolumn,preprintnumbers,notitlepage,
nofootinbib,amsfont,amssymb,10pt,singlespacing] {revtex4-1}
\usepackage{amsmath,bm}
\usepackage{times}
\usepackage{braket}
\usepackage{color,graphicx}
\usepackage{slashed}
\usepackage{hyperref}

\newcommand{\beq}{\begin{eqnarray}}
\newcommand{\eeq}{\end{eqnarray}}
\newcommand{\non}{\nonumber\\ }
\newcommand{\ov}{\overline}

\newcommand{\acp}{{\cal A}_{CP}}
\newcommand{\pb}{\phi_{B}}
\newcommand{\psl}{ P \hspace{-2.8truemm}/ }

\newcommand{\epsl}{\epsilon \hspace{-1.8truemm}/\,  }

\def\lsim{ {\ \lower-1.2pt\vbox{\hbox{\rlap{$<$}\lower6pt\vbox{\hbox{$\sim$}
}}}\ } }
\def\gsim{ {\ \lower-1.2pt\vbox{\hbox{\rlap{$>$}\lower6pt\vbox{\hbox{$\sim$}
}}}\ } }

\def \jhep{ J. High Energy Phys.  }

\definecolor{Red}{rgb}{1.,0.,0.}

\definecolor{Blue}{rgb}{0.,0.,1.}

\definecolor{nicered}{rgb}{0.7,0.1,0.2}
\definecolor{nicegreen}{rgb}{0.1,0.4,0.2}

\hypersetup{colorlinks,citecolor=nicegreen,linkcolor=nicered}

\begin{document}

\title{Nonleptonic decays of $B \to ( f_1(1285),f_1(1420) ) V$ in the perturbative QCD approach}
\author{Xin~Liu}
\email[Electronic address: ]{liuxin@jsnu.edu.cn}
\affiliation{School of Physics and Electronic Engineering, Jiangsu Normal University, Xuzhou, Jiangsu 221116,
People's Republic of China}

\author{Zhen-Jun~Xiao}
\email[Electronic address: ]{xiaozhenjun@njnu.edu.cn}
\affiliation{Department of Physics and Institute of Theoretical
Physics,\\ Nanjing Normal University, Nanjing, Jiangsu 210023,
People's Republic of China}

\author{Zhi-Tian~Zou}
\email[Electronic address: ]{zouzt@ytu.edu.cn}
\affiliation{Department of Physics, Yantai University, Yantai, Shandong 264005, People's Republic of China}


\date{\today}

\begin{abstract}

We investigate the branching ratios, the polarization fractions,
the direct {\it CP}-violating asymmetries, and the relative phases in
20 nonleptonic decay modes of $B \to f_1 V$ within the framework of
the perturbative QCD
approach at leading order with $f_1$
including two $^3\!P_1$-axial-vector states $f_1(1285)$ and $f_1(1420)$.
Here, $B$ denotes $B^+$, $B^0$, and $B_s^0$ mesons and $V$
stands for the lightest vector mesons $\rho$, $K^*$, $\omega$, and $\phi$ , respectively.
The $B_s^0 \to  f_1 V$ decays are studied theoretically for the first time in the literature.
Together with the angle $\phi_{f_1} \approx (24^{+3.2}_{-2.7})^\circ$ extracted from the measurement through
$B_{d/s} \to J/\psi f_1(1285)$ modes for the $f_1(1285)-f_1(1420)$ mixing system,
it is of great interest to find phenomenologically
some modes such as the tree-dominated $B^+ \to f_1 \rho^+$ and
the penguin-dominated $B^{+,0} \to f_1 K^{*+,0}, B_s^0 \to f_1 \phi$ with large branching ratios
around ${\cal O}(10^{-6})$ or even ${\cal O}(10^{-5})$, which are expected to be measurable at the LHCb and/or the
Belle-II experiments in the near future. The good agreement (sharp contrast)
of branching ratios and decay pattern for $B^+ \to f_1 \rho^+,
B^{+,0} \to f_1(1285) K^{*+,0} [B^{+,0} \to f_1(1420) K^{*+,0}]$ decays
between QCD factorization and perturbative QCD factorization predictions
can help us to distinguish these two rather different factorization approaches via precision measurements,
which would also be helpful for us in exploring the annihilation decay mechanism through
its important roles for the considered $B \to f_1 V$ decays.
\end{abstract}


\preprint{\footnotesize JSNU-PHY-TH-2016}
\maketitle

%
%

\section{Introduction}

The studies on nonleptonic $B$ meson
weak decays are generally expected to provide not only
good opportunities for testing the standard model(SM), but
also powerful means for probing both weak and strong dynamics, even different new physics(NP) scenarios
beyond the SM. It has been discussed
that the naive expectations of
polarization fractions, i.e., the longitudinal one $f_L \sim 1$
and the transverse two $f_\parallel \approx  f_\perp \sim
{\cal O}(m_V^2/m_B^2)$~\cite{Colangelo:2004rd,Kagan:2004uw} with $m_V\, (m_B)$ being the mass of the light vector ($B$) meson, are violated mainly in the penguin-dominated vector-vector $B$ meson decays~\cite{Beneke:2006hg,Cheng:2008gxa,Cheng:2009xz,Cheng:2009cn,Zou:2015iwa}, e.g., $f_L \sim f_T(=f_\parallel+f_\perp)$ in the famous $B \to \phi K^*$ process~\cite{Aubert:2003mm,Chen:2003jfa,Agashe:2014kda},
which has resulted in many investigations from various ways based
on different mechanisms, such as large penguin-induced
annihilation contributions~\cite{Kagan:2004uw}, form-factor
tuning~\cite{Li:2004mp}, final-state interactions~\cite{Cheng:2004ru,Colangelo:2004rd}, and even possible NP~\cite{Yang:2004pm}, to interpret anomalous polarizations
in those considered $B \to VV$ modes. Analogous to $B \to VV$ decays with rich physics involved in three polarization states, it is therefore of particular interest to explore the $B \to VA, AV$
($A$ is an axial-vector state) modes to shed light on the underlying helicity structure of the decay mechanism~\cite{Cheng:2008gxa}
through polarization studies.
Furthermore, stringent comparisons between theoretical predictions and experimental data for the physical observables may also help us to  further understand the hadronic structure of the involved
axial-vector bound states~\cite{Cheng:2007mx,Feldmann:2014iha,Liu:2014dxa,Liu:2014doa,Liu:2014jsa}.

Recently, the $B_{d/s} \to J/\psi f_1(1285)$ modes measured by the Large Hadron Collider beauty(LHCb) Collaboration
for the first time in the heavy $b$ flavor sector~\cite{Aaij:2013rja} motivated us to study the production of $^3\!P_1$-axial-vector $f_1(1285)$ and $f_1(1420)$ states  in the hadronic $B$ meson decays, such as $B_s^0 \to J/\psi f_1$~\cite{Liu:2014doa} and $B \to f_1 P$~\cite{Liu:2014jsa}   within the framework of perturbative QCD(pQCD) approach~\cite{Keum:2000ph} at leading order [Hereafter, for the sake of simplicity,
we will use $f_1$ to denote both $f_1(1285)$ and $f_1(1420)$ unless otherwise stated.]. Now, we will extend
this pQCD formalism to nonleptonic $B \to f_1 V$
decays, with $B$~\footnote{It is
noted that the pure annihilation-type $B_c \to f_1 V$ decays have
been studied by two of us(X.L. and Z.J.X.) in the pQCD approach
focusing on the predictions of the decay rates and
the polarization fractions~\cite{Liu:2010nz}.}($V$) being the $B^+$, $B^0$, and $B_s^0$( the lightest vector $\rho$, $K^*$, $\omega$, and $\phi$) states, in which the $B_s^0 \to  f_1 V$ decays are studied theoretically for the first time in the literature, although no data on these $B \to VA, AV$ type modes has been released so far. Though many efforts have been made to
develop the next-to-leading order pQCD formalism~\cite{Li:2010nn,Liu:2015sra}, because of a well-known fact
that leading order contributions dominate in the perturbation theory,
here we will still work at leading order to clarify the
physics for convenience. We will calculate the {\it CP}-averaged
branching ratios, the polarization fractions, the {\it CP}-violating
asymmetries, and the relative phases of 20 nonleptonic weak
decays of $B \to f_1 V$ by employing the low energy effective Hamiltonian~\cite{Buchalla:1995vs} and the pQCD approach based
on the $k_T$ factorization theorem. Assisted by the techniques
of $k_T$ resummation and threshold resummation, we can include
all possible contributions by explicitly evaluating the factorizable emission,
the nonfactorizable emission,
the factorizable annihilation,
and the nonfactorizable annihilation
Feynman diagrams in the pQCD approach with no end-point
singularities. The overall consistency between pQCD predictions
and experimental data for the $B \to PP$, $PV$, and $VV$ decays
is very good and indicates the advantage and reliability of the pQCD approach in estimating the hadronic matrix elements of $B$ meson decays.

In the quark model, the two $f_1$ states, i.e., $f_{1}(1285)$ and
its partner $f_1(1420)$, are classified specifically as
the light $p$-wave axial-vector flavorless mesons
carrying quantum
number $J^{PC} = 1^{++}$~\cite{Agashe:2014kda}.
In analogy to the pseudoscalar $\eta-\eta^{\prime}$
mixing~\cite{Agashe:2014kda}, these two axial-vector
$f_1$ states are also considered as a mixture induced by
nonstrange state $f_{1q} \equiv (u\bar u+ d\bar d)/\sqrt{2}$ and
strange one $f_{1s}\equiv s\bar s$ in the quark-flavor basis
and can be described as a $2\times 2$ rotation matrix
with mixing angle $\phi_{f_1}$ as follows
~\cite{Aaij:2013rja}:
 \beq
\left(
\begin{array}{c} f_1(1285)\\ f_1(1420) \\ \end{array} \right ) &=&
  \left( \begin{array}{cc}
 \cos{\phi_{f_1}} & -\sin{\phi_{f_1}} \\
 \sin{\phi_{f_1}} & \cos{\phi_{f_1}} \end{array} \right )
 \left( \begin{array}{c}  f_{1q}\\ f_{1s} \\ \end{array} \right )\;.
 \label{eq:mix-f1q-f1s}
 \eeq
In fact,
there also exists another mixing scheme called the singlet-octet basis with
flavor singlet state $f_1= (u \bar u + d \bar d + s \bar s)
/\sqrt{3}$ and flavor octet one $f_8= (u \bar u + d \bar d
- 2 s \bar s)/ \sqrt{6}$. The corresponding mixing angle
$\theta_{f_1}$ is related with $\phi_{f_1}$ via the equation
$\phi_{f_1}= \theta_i - \theta_{f_1}$, with $\theta_i$ being the
"ideal" mixing angle, specifically, $\theta_i = 35.3^\circ$.
It is therefore expected that $\phi_{f_1}$ can measure the
deviation from the ideal mixing.
Determination of the magnitude for the mixing angle $\phi_{f_1}$
is one of the key issues to understand the physical properties
of the $f_1$ states. Furthermore, it is essential to note that
$\phi_{f_1}$ also has an important role in constraining the mixing
angle $\theta_{K_1}$, which arises from the mixing between
two distinct types of
axial-vector $K_{1A}(^3\!P_1)$ and $K_{1B}(^1\!P_1)$ states, through
the Gell-Mann$-$Okubo mass formula~\cite{Cheng:2011pb,Agashe:2014kda}.
It is therefore definitely interesting to investigate the mixing
angle $\phi_{f_1}$ in different ways.
However, the value of $\phi_{f_1}$ is still a controversy
presently~\cite{Liu:2014doa,Liu:2014jsa}, though
there are several explorations that have been
performed at both theoretical and experimental aspects.
Of course, it is expected that this status will be greatly
improved with the successful
upgrade of LHC RUN-II and the scheduled running of Belle-II
experiments ever since the $f_1(1285)$ state, as well as the
value of $\phi_{f_1}$, has been measured preliminarily
in the $B$ decay system~\cite{Aaij:2013rja}.

Up to now, to our best knowledge, the nonleptonic $B^{+,0} \to f_1 V$ decays have been theoretically investigated by
G.~Calder${\rm \acute{o}}$n {\it et al.}~\cite{Calderon:2007nw}
in the naive factorization approach and by Cheng and Yang~\cite{Cheng:2008gxa}
within QCD factorization(QCDF), respectively. However,
the conclusion that $Br(B \to f_1 V)[{\cal O}(10^{-8}-10^{-6})]
< Br(B \to f_1 P)[{\cal O}(10^{-5})]$ predicted in Ref.~\cite{Calderon:2007nw}, seems to contradict our naive
expectation. As pointed out in Ref.~\cite{Cheng:2008gxa},
the authors believed
that, because of the existence of three polarization states for
the vector meson, the $B \to f_1 V$ decays may generally have larger
decay rates than the $B \to f_1 P$ ones correspondingly. Furthermore,
due to the similar QCD behavior between
vector and $^3\!P_1$-axial-vector states~\cite{Yang:2007zt},
the analogy between
$B \to f_1 V$ and $B \to (\omega, \phi) V$ decays can be naively anticipated.
For example, if $f_1(1285)$ is highly dominated by the $f_{1q}$
flavor state, then $Br(B^+ \to f_1(1285) \rho^+)$ can be comparable
with $Br(B^+ \to \omega \rho^+)$. Actually, because $f_1(1285)$
mixes with the $s\bar s$ component around 20\%
($\sim \sin^2\phi_{f_1})$ based on
Eq.~(\ref{eq:mix-f1q-f1s}) and the preliminary value $\phi_{f_1} \sim 24^\circ$
given by the LHCb Collaboration~\cite{Aaij:2013rja},
it is therefore estimated that
the decay rate of $B^+ \to f_1(1285) \rho^+$ may be somewhat smaller
than that of $B^+ \to \omega \rho^+$. As a matter of fact, the branching ratios of $B^+ \to f_1(1285) \rho^+$ predicted
within the QCDF and pQCD formalisms,
as far as the central values are concerned, are $(9-10) \times 10^{-6}$~\cite{Cheng:2008gxa} and $11.1 \times 10^{-6}$ in this work, respectively, which are indeed comparative and slightly smaller
than that of $B^+ \to \omega \rho^+$ with updated values
$16.9 \times 10^{-6}$~\cite{Cheng:2009cn} and
$12.1 \times 10^{-6}$~\cite{Zou:2015iwa} correspondingly.
Moreover, the polarization fractions for the $B^{+,0} \to f_1 V$
channels were also given within the framework of QCDF~\cite{Cheng:2008gxa}. But, frankly speaking, lack of
experimental constraints on the parametrized
hard-spectator scattering
and weak annihilation contributions in QCDF greatly weakens
the reliability of predictions for $B^{+,0} \to f_1 V$ decays,
which will limit the hints to relevant experiments, even to understand the physics hidden in relevant modes. It is therefore definitely interesting to investigate these aforementioned $B \to f_1 V$ decays
in other frameworks, e.g., the pQCD approach in the present work, to clarify the discrepancies and further distinguish the factorization approaches through experimental examinations with good precision.

The paper is organized as follows. In Sec.~\ref{sec:form}, we present
the formalism, hadron wave functions and analytic pQCD calculations
of 20 nonleptonic $B \to f_1 V$ decays.
The numerical results and phenomenological
analyses are addressed in Sec.~\ref{sec:randd} explicitly. Finally,
Sec.~\ref{sec:summary} contains the main conclusions and a short summary.


\section{ Formalism and perturbative calculations}\label{sec:form}

In this section, we first make a brief introduction to the pQCD formalism at leading order. For more details, the readers can
refer to the review article in Ref.~\cite{Keum:2000ph}.
Nowadays, the pQCD approach has been known as one of the important
factorization methods based on QCD dynamics to perturbatively
evaluate hadronic matrix elements in the decays of heavy
$b$ flavor mesons. The unique point of this pQCD approach is that
it picks up the transverse momentum $k_T$ of the valence quarks
in all the initial and final states, as a result of which
the calculations of hadronic matrix elements free of end-point
singularities always occur in the collinear factorization theorem
employed in the QCDF approach~\cite{Beneke:1999br} and soft-collinear effective theory(SCET)~\cite{Bauer:2000yr}.
Hence, all topologies of Feynman diagrams in the hadronic $B$ meson
decays are effectively calculable in the pQCD
framework, where three energy scales $m_W$(mass of $W$ boson), $m_b$(mass of $b$ quark) and
$t\approx \sqrt{m_b \Lambda_{\rm QCD}} $(factorization
hard-collinear scale with $\Lambda_{\rm QCD}$, the hadronic scale) are
involved~\cite{Keum:2000ph,Li:1994cka}.
Note that, unlike the QCDF approach~\cite{Beneke:2001ev},
the annihilation contributions in the pQCD
formalism can be calculated without introducing any parameters.
When $t$ is no less than the factorization scale, i.e.,
$\geq \sqrt{m_b \Lambda_{\rm QCD}}$, the running of Wilson
coefficients $C_i(t)$ will be perturbatively controlled
through the renormalization group equation.
The soft dynamics below $\sqrt{m_b \Lambda_{\rm QCD}}$
will be described by hadron wave
functions$\Phi$, which are nonperturbative but universal
for all channels and usually determined by
employing nonperturbative QCD techniques such as QCD sum rules and/or
lattice QCD or extracted experimentally
from other well-measured processes.
It is worth emphasizing that the physics between $m_b$
and $\sqrt{m_b \Lambda_{\rm QCD}}$ will be
absorbed into the so-called "hard kernel"$H$ and perturbatively
evaluated in the pQCD approach. The decay amplitude
for $B \to f_1 V$ decays in the pQCD approach can therefore be conceptually written as follows:
\beq
A(B \to f_1 V) &\sim &\int\!\! d x_1 d
x_2 d x_3 b_1 d b_1 b_2 d b_2 b_3 d b_3
\non && \cdot {\mathrm{Tr}}
\left [ C(t) \Phi_{B}(x_1, b_1) \Phi_{V}(x_2, b_2)
\Phi_{f_1}(x_3, b_3) H(x_i, b_i, t) S_t(x_i)\, e^{-S(t)} \right ]\;,
\label{eq:a2}
\eeq
where $x_i(i=1,2,3)$ is the momentum fraction of the valence quark
in the involved mesons; $b_i$ is the conjugate space coordinate
of $k_{iT}$; $t$ is the largest running energy
scale in hard kernel $H(x_i,b_i,t)$; Tr denotes the trace
over Dirac and SU(3) color indices;
$C(t)$ stands for the Wilson coefficients including
the large logarithms $\ln (m_W/t)$~\cite{Keum:2000ph}; and
$\Phi$ is
the wave function describing the hadronization of quarks and anti-quarks to the meson. The jet function $S_t(x_i)$ comes from
threshold resummation, which
exhibits a strong suppression effect in the small $x$
region~\cite{Li:2001ay,Li:2002mi}, while
the Sudakov factor $e^{-S(t)}$ arises from $k_T$ resummation, which
provides a strong suppression in the small $k_T$(or large $b$) region~\cite{Botts:1989kf,Li:1992nu}.
These resummation effects therefore guarantee the removal of the end-point singularities.
The detailed expressions for $S_t(x_i)$ and $e^{-S(t)}$ can be
easily found in the original Refs.~\cite{Li:2001ay,Li:2002mi,Botts:1989kf,Li:1992nu}.
Thus, with Eq.~(\ref{eq:a2}),
we can give the convoluted amplitudes of the $B \to f_1 V$ decays explicitly,
which will be presented in the next section, through
the evaluations of the hard kernel $H(x_i,b_i,t)$ at leading
order in the $\alpha_s$ expansion with
the pQCD approach.

\subsection{Hadron wave functions}\label{ssec:wf}

The heavy $B$ meson is usually treated as a heavy-light system and its light-cone wave function
can generally be defined as~\cite{Keum:2000ph,Lu:2002ny}
\beq
\Phi_{B}
&=& \frac{i }{\sqrt{2N_c}}
\left\{(\psl +m_{B})\gamma_5
 \phi_{B}(x, k_T) \right\}_{\alpha\beta}\;,
\label{eq:def-bq}
\eeq
where $\alpha,\beta$ are the color indices;
$P$ is the momentum of $B$ meson; $N_c$ is the color factor; and
$k_T$ is the intrinsic transverse momentum of the light quark in $B$ meson.

In Eq.~(\ref{eq:def-bq}), $\phi_{B}(x,k_T)$ is the $B$ meson distribution amplitude
and obeys the following normalization condition,
\beq
\int_0^1 dx \phi_{B}(x, b=0) &=& \frac{f_{B}}{2 \sqrt{2N_c}}\;,\label{eq:norm}
\eeq
where $b$ is the conjugate space coordinate of transverse momentum $k_T$ and $f_B$ is the decay constant of the $B$ meson.

The light-cone wave functions of light vector meson $V$
and axial-vector state $f_{1}$ have been given
in the QCD sum rule method up to twist-3 as~\cite{Ball:1998sk,Ball:2007rt}
 \beq
\Phi^{L}_{V}
 &=&  \frac{1}{\sqrt{2 N_c}} 
   \biggl\{ m_{V}\, {\epsl}_L \,\phi_{V}(x)  +
 {\epsl}_L \, \psl\,\phi_{V}^t(x)  + m_{V}\, \phi_{V}^s(x) \biggr\}_{\alpha\beta}\;,
 \eeq
 \beq
\Phi^{T}_{V}
 &=&  \frac{1}{\sqrt{2 N_c}} 
  \biggl\{ m_{V}\, {\epsl}_T\, \phi_{V}^v(x) +
{\epsl}_T\, \psl\, \phi_{V}^T(x)+m_{V}
i\epsilon_{\mu\nu\rho\sigma}\gamma_5\gamma^\mu {\epsl}_T^{\nu}
n^\rho v^\sigma \phi_{V}^a(x) \biggr\}_{\alpha\beta}\;,
 \eeq
 and~\cite{Yang:2007zt,Li:2009tx}
 \beq
\Phi^{L}_{f_{1}}
 &=&  \frac{ 1}{\sqrt{2 N_c}}\gamma_5 
  \biggl\{ m_{f_{1}}\, {\epsl}_L \,\phi_{f_{1}}(x)  +
 {\epsl}_L \, \psl\,\phi_{f_{1}}^t(x)  + m_{f_{1}}\, \phi_{f_{1}}^s(x) \biggr\}_{\alpha\beta}\;,
 \eeq
 \beq
\Phi^{T}_{f_{1}}
 &=&  \frac{ 1}{\sqrt{2 N_c}} \gamma_5 
  \biggl\{ m_{f_{1}}\, {\epsl}_T\, \phi_{f_{1}}^v(x) +
{\epsl}_T\, \psl\, \phi_{f_{1}}^T(x)+m_{f_{1}}
i\epsilon_{\mu\nu\rho\sigma}\gamma_5\gamma^\mu {\epsl}_T^{\nu}
n^\rho v^\sigma \phi_{f_{1}}^a(x) \biggr\}_{\alpha\beta}\;,
\eeq
for longitudinal and transverse polarizations, respectively,
with the polarization vectors $\epsilon_L$ and $\epsilon_T$ of $V$
or $f_{1}$,
satisfying $P \cdot \epsilon=0$, where
$x$ denotes the momentum
fraction carried by quarks in the meson; and $n=(1,0,{\bf 0}_T)$
and $v=(0,1,{\bf 0}_T)$ are dimensionless lightlike unit vectors;
and $m_{f_1}$ stands for the mass of light
axial-vector $f_1$ states.
We adopt the convention $\epsilon^{0123}=1$ for the
Levi-Civit$\grave{a}$ tensor $\epsilon^{\mu\nu\alpha\beta}$.
Note that the explicit expressions for all the above-mentioned
distribution amplitudes $\phi(x)$ with different twists can be found later in the Appendix.

\subsection{Perturbative calculations in the pQCD approach} \label{ssec:pcalc}

For the considered 20 $B \to f_1 V$ decays induced by the
$\bar b \to \bar q(q=d\ {\rm or}\ s) $ transition
at the quark level,
the related weak effective
Hamiltonian $H_{{\rm eff}}$ can be writen as~\cite{Buchalla:1995vs}
\beq
H_{\rm eff}\, &=&\, {G_F\over\sqrt{2}}
\biggl\{ V^*_{ub}V_{uq} [ C_1(\mu)O_1^{u}(\mu)
+C_2(\mu)O_2^{u}(\mu) ]
 - V^*_{tb}V_{tq} [ \sum_{i=3}^{10}C_i(\mu)O_i(\mu) ] \biggr\}+ {\rm H.c.}\;,
\label{eq:heff}
\eeq
with the Fermi constant $G_F=1.16639\times 10^{-5}{\rm
GeV}^{-2}$, Cabibbo-Kobayashi-Maskawa(CKM) matrix elements $V$,
and Wilson coefficients $C_i(\mu)$ at the renormalization scale
$\mu$. The local four-quark
operators $O_i(i=1,\cdots,10)$ are written as
\begin{enumerate}
\item[]{(1) Current-current(tree) operators}
\begin{eqnarray}
{\renewcommand\arraystretch{1.5}
\begin{array}{ll}
\displaystyle
O_1^{u}\, =\,
(\bar{q}_\alpha u_\beta)_{V-A}(\bar{u}_\beta b_\alpha)_{V-A}\;,
& \displaystyle
O_2^{u}\, =\, (\bar{q}_\alpha u_\alpha)_{V-A}(\bar{u}_\beta b_\beta)_{V-A}\;;
\end{array}}
\label{eq:operators-1}
\end{eqnarray}

\item[]{(2) QCD penguin operators}
\begin{eqnarray}
{\renewcommand\arraystretch{1.5}
\begin{array}{ll}
\displaystyle
O_3\, =\, (\bar{q}_\alpha b_\alpha)_{V-A}\sum_{q'}(\bar{q}'_\beta q'_\beta)_{V-A}\;,
& \displaystyle
O_4\, =\, (\bar{q}_\alpha b_\beta)_{V-A}\sum_{q'}(\bar{q}'_\beta q'_\alpha)_{V-A}\;,
\\
\displaystyle
O_5\, =\, (\bar{q}_\alpha b_\alpha)_{V-A}\sum_{q'}(\bar{q}'_\beta q'_\beta)_{V+A}\;,
& \displaystyle
O_6\, =\, (\bar{q}_\alpha b_\beta)_{V-A}\sum_{q'}(\bar{q}'_\beta q'_\alpha)_{V+A}\;;
\end{array}}
\label{eq:operators-2}
\end{eqnarray}

\item[]{(3) Electroweak penguin operators}
\begin{eqnarray}
{\renewcommand\arraystretch{1.5}
\begin{array}{ll}
\displaystyle
O_7\, =\,
\frac{3}{2}(\bar{q}_\alpha b_\alpha)_{V-A}\sum_{q'}e_{q'}(\bar{q}'_\beta q'_\beta)_{V+A}\;,
& \displaystyle
O_8\, =\,
\frac{3}{2}(\bar{q}_\alpha b_\beta)_{V-A}\sum_{q'}e_{q'}(\bar{q}'_\beta q'_\alpha)_{V+A}\;,
\\
\displaystyle
O_9\, =\,
\frac{3}{2}(\bar{q}_\alpha b_\alpha)_{V-A}\sum_{q'}e_{q'}(\bar{q}'_\beta q'_\beta)_{V-A}\;,
& \displaystyle
O_{10}\, =\,
\frac{3}{2}(\bar{q}_\alpha b_\beta)_{V-A}\sum_{q'}e_{q'}(\bar{q}'_\beta q'_\alpha)_{V-A}\;,
\end{array}}
\label{eq:operators-3}
\end{eqnarray}
\end{enumerate}
with the color indices $\alpha, \ \beta$ and the notations
$(\bar{q}'q')_{V\pm A} = \bar q' \gamma_\mu (1\pm \gamma_5)q'$.
The index $q'$ in the summation of the above operators runs
through $u,\;d,\;s$, $c$, and $b$.

\begin{figure}[!!htb]
\centering
\begin{tabular}{l}
\includegraphics[width=0.8\textwidth]{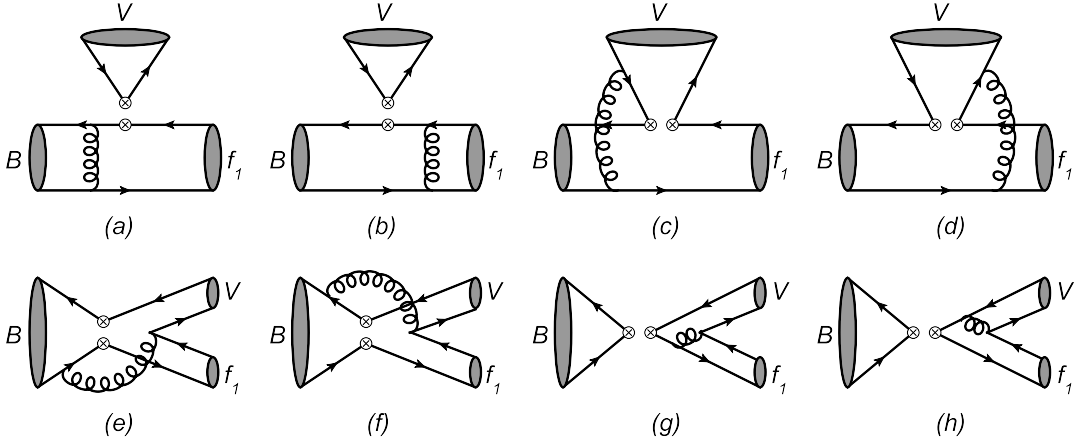}
\end{tabular}
\caption{Typical Feynman diagrams contributing to
 $B \to f_1 V$ decays in the pQCD approach
at leading order. The other diagrams contributing to
those considered decays can be easily obtained by
exchanging the positions of $f_1$ and $V$.}
  \label{fig:fig1}
\end{figure}

From the effective Hamiltonian
(\ref{eq:heff}), there are eight types
of diagrams contributing to $B
\to f_1 V$ decays in the pQCD approach at
leading order as illustrated in Fig.~\ref{fig:fig1}. The possible
contributions to the considered decays can be easily obtained
by exchanging the positions of $f_1$ and $V$.
We calculate the contributions arising from
various operators as shown in
Eqs.~(\ref{eq:operators-1})-(\ref{eq:operators-3}).
As presented in Ref.~\cite{Liu:2014dxa}[see Eqs.~(33)-(57) for details],
we have given the analytic $B \to VA$ decay amplitudes
only with a $B \to A$ transition. This part will be repeated in this
work, in order to present the analytically complete expressions
for $B \to VA$ and $AV$ decays.
It should be mentioned that, hereafter,
for the sake of simplicity, we will use
$F$ and $M$ to describe the factorizable
and nonfactorizable amplitudes induced by the $(V-A)(V-A)$ operators,
$F^{P_1}$ and $M^{P_1}$ to describe the factorizable
and nonfactorizable amplitudes arising from the $(V-A)(V+A)$ operators,
and $F^{P_2}$ and $M^{P_2}$ to describe the factorizable
and nonfactorizable amplitudes coming from the $(S-P)(S+P)$ operators
that are obtained by making a Fierz transformation from the $(V-A)(V+A)$ ones,
respectively. Furthermore, before starting the
perturbative calculations,
a comment should be given: in light of the successful clarification
of most branching ratios and polarization fractions in the $B \to
VV$ decays by keeping the terms proportional
to $r^2_V=m_V^2/m_B^2$
in the denominator of propagators for virtual
quarks and gluons with the pQCD approach~\cite{Zou:2015iwa}, we will
follow this treatment in the present work for 20 nonleptonic
$B \to f_1 V$ modes, i.e., retaining the similar terms with
$r_V^2$ and $r_{f_1}^2=m_{f_1}^2/m_B^2$, which could be
examined by future measurements to further clarify
its universality.

For the factorizable emission($fe$)
diagrams in Figs.\ref{fig:fig1}(a) and \ref{fig:fig1}(b),
the corresponding Feynman amplitudes with one longitudinal polarization($L$) and
two transverse polarizations($N$ and $T$) can be written as follows:
\beq
F_{fe}^{L}&=&- 8\pi C_F m_{B}^2 \int_0^1
d x_{1} dx_{3}\, \int_{0}^{\infty} b_1 db_1 b_3 db_3\, \pb(x_1,b_1)
\left\{ \left [(1+x_{3})\phi_{A}(x_{3}) +
r_{A}(1-2x_{3}) \right.\right. \non && \left. \left.\times
(\phi^{t}_{A}(x_3)+\phi^{s}_{A}(x_{3}))\right]E_{fe}(t_{a})
h_{fe}(x_{1},x_{3},b_{1},b_{3})
+2r_{A}\phi^{s}_{A}(x_3)E_{fe}(t_{b})
h_{fe}(x_{3},x_{1},b_{3},b_{1}) \right\}\;,
\label{eq:abl-va}
\eeq
\beq
F_{fe}^{N}&=& -8\pi C_F m_{B}^2 \int_0^1 d x_{1}
dx_{3}\, \int_{0}^{\infty} b_1 db_1 b_3 db_3\, \phi_{B}(x_1,b_1)
r_{V} \left\{ [\phi^{T}_{A}(x_3)
+2r_{A}\phi^{v}_{A}(x_3)+r_{A}x_{3}
\right. \non && \left. \times
(\phi^{v}_{A}(x_3)-\phi^{a}_{A}(x_3))]E_{fe}(t_{a})
h_{fe}(x_{1},x_{3},b_{1},b_{3})+r_{A}[\phi^{v}_{A}(x_3)+\phi^{a}_{A}(x_3)]
E_{fe}(t_{b}) h_{fe}(x_{3},x_{1},b_{3},b_{1})\right\}\;,
\label{eq:abn-va}
\eeq
\beq
 F_{fe}^{T}&=& -16\pi C_F m_{B}^2 \int_0^1 d x_{1} dx_{3}\,
\int_{0}^{\infty} b_1 db_1 b_3 db_3\, \phi_{B}(x_1,b_1)
r_{V}\left\{[\phi^{T}_{A}(x_3) +2r_{A}\phi^{a}_{A}(x_3)-r_{A}x_{3}
\right. \non &&
\left.\times
(\phi^{v}_{A}(x_3)-\phi^{a}_{A}(x_3))]E_{fe}(t_{a})  h_{fe}(x_{1},x_{3},b_{1},b_{3})
+r_{A}[\phi^{v}_{A}(x_3)+\phi^{a}_{A}(x_3)] E_{fe}(t_{b})h_{fe}(
x_{3},x_{1},b_{3},b_{1})\right\}\;, \label{eq:abt-va}
\eeq
where, in this work, $A$ will specifically denote
the axial-vector states $f_1(1285)$ and $f_1(1420)$
 and
$C_F=4/3$ is a color factor. For the hard functions $h$, the running hard scales $t$,
and the convolution functions $E(t)$, refer to the Appendix in Ref.~\cite{Zou:2015iwa}.

Since only the vector part of the
$(V+A)$ current contributes to the vector meson production, $
\langle A |V-A|B\rangle \langle V |V+A | 0 \rangle = \langle A
|V-A |B \rangle \langle V |V-A|0 \rangle,$ we have
 \beq
 F_{fe}^{P_1}= F_{fe} \; \label{eq:abp1-va} .
 \eeq

Because a vector meson cannot be produced via scalar and/or pseudoscalar currents, then the contribution arising from the $(S \pm P)$
operators is
 \beq
 F_{fe}^{P_2}= 0 \; \label{eq:abp2-va} .
 \eeq

For the nonfactorizable emission($nfe$) diagrams in Figs. \ref{fig:fig1}(c) and \ref{fig:fig1}(d),
the corresponding Feynman amplitudes are
\beq
 M_{nfe}^{L} &=& -\frac{16 \sqrt{6}}{3}\pi C_F m_{B}^2
\int_{0}^{1}d x_{1}d x_{2}\,d x_{3}\,\int_{0}^{\infty} b_1d b_1 b_2d
b_2\, \phi_{B}(x_1,b_1)\phi _{V}(x_{2}) \left\{\left[
(1-x_{2}) \phi_{A} ( x_{3} )
\right. \right. \non && \left.  +r_{A}x_3
(\phi^{t}_{A}(x_3)-\phi^{s}_{A}(x_3)) \right]E_{nfe}(t_c)
h_{nfe}^{c}(x_1,x_2,x_3,b_1,b_2) -\left[
(x_{2}+x_{3}) \phi_{A} ( x_{3} )
\right.  \non && \left.\left.  -r_{A} x_3(\phi^{t}_{A}(x_3)+\phi^{s}_{A}(x_3))
\right] E_{nfe}(t_d)h_{nfe}^{d}(x_1,x_2,x_3,b_1,b_2) \right\}\;,
\label{eq:cdl-va}
\eeq
\beq
M_{nfe}^{N} &=& -\frac{16 \sqrt{6}}{3}\pi C_F  m_{B}^2
\int_{0}^{1}d x_{1}d x_{2}\,d x_{3}\,\int_{0}^{\infty} b_1d b_1 b_2d
b_2\, \phi_{B}(x_1,b_1) r_{V}\left\{(1-x_2)
(\phi^{v}_{V}(x_2)+\phi^{a}_{V}(x_2) )
\right. \non && \left. \times
\phi^{T}_{A}(x_3) h_{nfe}^{c}(x_1,x_2,x_3,b_1,b_2) E_{nfe}(t_c)
+ \left[x_2(\phi^{v}_{V}(x_2) +\phi^a_{V}(x_2))\phi^{T}_{A}(x_3)
\right.\right. \non && \left. \left. -2r_{A}(x_2+x_3) (\phi^{v}_{V}(x_2) \phi^{v}_{A}(x_3)+
\phi^{a}_{V}(x_2) \phi^{a}_{A}(x_3) ) \right] E_{nfe}(t_d)
h_{nfe}^{d}(x_1,x_2,x_3,b_1,b_2) \right\}\;,
 \label{eq:cdn-va}
 \eeq
 \beq
 M_{nfe}^{T}&=& -\frac{32 \sqrt{6}}{3}\pi C_F m_{B}^2
\int_{0}^{1}d x_{1}d x_{2}\,d x_{3}\,\int_{0}^{\infty} b_1d b_1 b_2d
b_2\, \phi_{B}(x_1,b_1)   r_{V}\left\{
(1-x_2)(\phi^{v}_{V}(x_2)+\phi^{a}_{V}(x_2) )
\right. \non && \left. \times \phi^{T}_{A}(x_3)
h_{nfe}^{c}(x_1,x_2,x_3,b_1,b_2)
E_{nfe}(t_c)+   \left[
x_2(\phi^{v}_{V}(x_2)+\phi^a_{V}(x_2))\phi^{T}_{A}(x_3)
\right.\right. \non && \left.\left. -2r_{A}(x_2+x_3)
(\phi^{v}_{V}(x_2) \phi^{a}_{A}(x_3)+ \phi^{a}_{V}(x_2)
\phi^{v}_{A}(x_3) ) \right]E_{nfe}(t_d)
h_{nfe}^{d}(x_1,x_2,x_3,b_1,b_2) \right\}\; \label{eq:cdt-va},
\eeq
 \beq
 M_{nfe}^{L,P_1} &=& -\frac{16 \sqrt{6}}{3}\pi C_F  m_{B}^2
\int_{0}^{1}d x_{1}d x_{2}\,d x_{3}\,\int_{0}^{\infty} b_1d b_1 b_2d
b_2\, \phi_{B}(x_1,b_1)  r_{V}
\left\{\left[(1-x_2)(\phi^{t}_{V}(x_2)+\phi^{s}_{V}(x_2))
\right. \right.\non && \times
\left. \phi_{A}(x_3)  -r_{A}(1-x_2) (\phi^{t}_{V}(x_2)
+\phi^{s}_{V}(x_2))(\phi^{t}_{A}(x_3)-\phi^{s}_{A}(x_3))-r_{A}
x_3(\phi^{t}_{V}(x_2) -\phi^{s}_{V}(x_2)) \right. \non && \left. \times
(\phi^{t}_{A}(x_3)+\phi^{s}_{A}(x_3)) \right]
 E_{nfe}(t_{c})h_{nfe}^{c}(x_{1},x_{2},x_{3},b_{1},b_{2})
+\left[ x_2 ( \phi^{t}_{V}(x_2)-\phi^{s}_{V}(x_2)) \phi_{A}(x_3)
\right.  \non &&  \left. \left.
-r_{A}x_2(\phi^{t}_{V}(x_2) -\phi^{s}_{V}(x_2)) (\phi^{t}_{A}(x_3)-\phi^{s}_{A}(x_3))
-r_{A}x_{3}(\phi^{t}_{V}(x_2)+\phi^{s}_{V}(x_2))
(\phi^{t}_{A}(x_3)+\phi^{s}_{A}(x_3)) \right] \right. \non &&  \left. \times
E_{nfe}(t_d)h_{nfe}^{d}(x_{1},x_{2},x_{3},b_{1},b_{2})
\right\}\;,\label{eq:cdlp1-va}
\eeq
\beq
M_{nfe}^{N,P_1} &=&-\frac{16 \sqrt{6}}{3}\pi C_F m_{B}^2
\int_{0}^{1}d x_{1}d x_{2}\,d x_{3}\,\int_{0}^{\infty} b_1d b_1 b_2d
b_2\, \phi_{B}(x_1,b_1)
r_{A}x_{3} \phi^{T}_{V}(x_2) (\phi^{v}_{A}(x_3)-\phi^{a}_{A}(x_3))
 \non && \times
\left \{ E_{nfe}(t_{c})
h_{nfe}^{c}(x_{1},x_{2},x_{3},b_{1},b_{2})
+ E_{nfe}(t_{d}) h_{nfe}^{d}(x_{1},x_{2},x_{3},b_{1},b_{2}) \right\}\;,
\label{eq:cdnp1-va}
\eeq
\beq
M_{nfe}^{T,P_1}&=& 2M_{nfe}^{N,P_1}\;,
\label{eq:cdtp1-va}
\eeq
\beq
M_{nfe}^{L,P_2} &=& -\frac{16 \sqrt{6}}{3}\pi C_F  m_{B}^2
\int_{0}^{1}d x_{1}d x_{2}\,d x_{3}\,\int_{0}^{\infty} b_1d b_1 b_2d
b_2\, \phi_{B}(x_1,b_1)  \phi_{V}(x_2)\left\{
\left[(1-x_2+x_3)\phi_{A}(x_3)
\right.\right. \non &&  \left. \left.  - r_{A}x_3
(\phi^{t}_{A}(x_3)+\phi^{s}_{A}(x_3))\right]
E_{e}(t_{c}) h_{nfe}^{c}(x_{1},x_{2},x_{3},b_{1},b_{2})-
h_{nfe}^{d}(x_{1},x_{2},x_{3},b_{1},b_{2})E_{nfe}(t_{d})\right. \non && \times
\left[
x_2\phi_{A}(x_3)+r_{A}x_{3}(\phi^{t}_{A}(x_3)-\phi^{s}_{A}(x_3))
\right]\left.\right\}\;,
\label{eq:cdlp2-va}
\eeq
\beq
M_{nfe}^{N,P_2} &=& \frac{16 \sqrt{6}}{3}\pi C_F m_{B}^2
\int_{0}^{1}d x_{1}d x_{2}\,d x_{3}\,\int_{0}^{\infty} b_1d b_1 b_2d
b_2\, \phi_{B}(x_1,b_1)  r_{V}
\left\{\left[(1-x_2)(\phi^{v}_{V}(x_2)-\phi^{a}_{V}(x_2))
\right. \right.\non && \left. \left. \times
\phi^{T}_{A}(x_3)-2r_{A}(1-x_2+x_3) (\phi^{v}_{V}(x_2)
\phi^{v}_{A}(x_3)-\phi^{a}_{V}(x_2)\phi^{a}_{A}(x_3))
\right]h_{nfe}^{c}(x_{1},x_{2},x_{3},b_{1},b_{2})
\right. \non &&\left. \times
E_{nfe}(t_{c})
 + x_2(\phi^{v}_{V}(x_2)-\phi^{a}_{V}(x_2))
\phi^{T}_{A}(x_3)  E_{nfe}(t_{d})
h_{nfe}^{d}(x_{1},x_{2},x_{3},b_{1},b_{2})\right\}\;,
\label{eq:cdnp2-va}
\eeq
\beq
M_{nfe}^{T,P_2} &=& \frac{32 \sqrt{6}}{3}\pi C_F m_{B}^2
\int_{0}^{1}d x_{1}d x_{2}\,d x_{3}\,\int_{0}^{\infty} b_1d b_1 b_2d
b_2\, \phi_{B}(x_1,b_1) r_{V}
\left\{\left[(1-x_2)(\phi^{v}_{V}(x_2)-\phi^{a}_{V}(x_2))
\right. \right. \non && \left. \left.  \times
\phi^{T}_{A}(x_3)-2r_{A}(1-x_2+x_3) (\phi^{v}_{V}(x_2)
\phi^{a}_{A}(x_3)-\phi^{a}_{V}(x_2)\phi^{v}_{A}(x_3))
\right] h_{nfe}^{c}(x_{1},x_{2},x_{3},b_{1},b_{2})
\right. \non &&\left. \times
E_{nfe}(t_{c})+ x_2(\phi^{v}_{V}(x_2)-\phi^{a}_{V}(x_2))
\phi^{T}_{A}(x_3) E_{nfe}(t_{d})
h_{nfe}^{d}(x_{1},x_{2},x_{3},b_{1},b_{2})\right\}\;,
\label{eq:cdtp2-va}
 \eeq

For the nonfactorizable annihilation($nfa$) diagrams in Figs.
\ref{fig:fig1}(e) and \ref{fig:fig1}(f), we have
\beq
M_{nfa}^{L} &=& -\frac{16\sqrt{6}}{3}\pi C_F m_{B}^2 \int_{0}^{1}d x_{1}d x_{2}\,d
x_{3}\,\int_{0}^{\infty} b_1d b_1 b_2d b_2\, \phi_{B}(x_1,b_1)
\left\{\left[ (1-x_3) \phi_{V}(x_2)\phi_{A}(x_3) \right.\right.\non
&& \left. \left. + r_{V}r_{A}
\left((1+x_2-x_3)(\phi^{s}_{V}(x_2)\phi^{s}_{A}(x_3)-\phi^{t}_{V}(x_2)
\phi^{t}_{A}(x_3))
-(1-x_2-x_3)(\phi^{s}_{V}(x_2)\phi^{t}_{A}(x_3)
\right.\right.\right.\non && \left.\left. \left.
-\phi^{t}_{V}(x_2)\phi^{s}_{A}(x_3) ) \right) \right]E_{nfa}(t_e)
 h_{nfa}^{e}(x_1,x_2,x_3,b_1,b_2)
-\left[ x_2\phi_{V}(x_2)\phi_{A}(x_3) + 2 r_{V}
r_{A}(\phi^{t}_{V}(x_2) \right.\right. \non &&
\left.\left. \times \phi^{t}_{A}(x_3)+\phi^{s}_{V}(x_2)\phi^{s}_{A}(x_3))-
r_{V}r_{A}(1+x_2-x_3)(\phi^{t}_{V}(x_2)\phi^{t}_{A}(x_3)
-\phi^{s}_{V}(x_2)\phi^{s}_{A}(x_3)) + r_{V} r_{A} \right.\right.\non
&&\left.\left. \times
(1-x_2-x_3)(\phi^{s}_{V}(x_2)\phi^{t}_{A}(x_3)
-\phi^{t}_{V}(x_2)\phi^{s}_{A}(x_3))\right] E_{nfa}(t_f)
 h_{nfa}^{f}(x_1,x_2,x_3,b_1,b_2) \right\}\;,
\label{eq:efl-va}
\eeq
\beq
M_{nfa}^{N} &=& \frac{32 \sqrt{6}}{3}\pi C_F m_{B}^2
\int_{0}^{1}d x_{1}d x_{2}\,d x_{3}\,\int_{0}^{\infty} b_1d b_1 b_2d
b_2\, \phi_{B}(x_1,b_1) r_{V} r_{A} \non &&\times
\left[\phi^{v}_{V}(x_2)\phi^{v}_{A}(x_3)
+\phi^{a}_{V}(x_2)\phi^{a}_{A}(x_3)\right] E_{nfa}(t_f)
h_{nfa}^{f}(x_1,x_2,x_3,b_1,b_2) \;,
 \label{eq:efn-va}
 \eeq
 \beq
M_{nfa}^{T}&=& \frac{64 \sqrt{6}}{3}\pi C_F m_{B}^2
\int_{0}^{1}d x_{1}d x_{2}\,d x_{3}\,\int_{0}^{\infty} b_1d b_1 b_2d
b_2\, \phi_{B}(x_1,b_1) r_{V} r_{A} \non &&\times
\left[\phi^{v}_{V}(x_2)\phi^{a}_{A}(x_3)
+\phi^{a}_{V}(x_2)\phi^{v}_{A}(x_3)\right] E_{nfa}(t_f)
h_{nfa}^{f}(x_1,x_2,x_3,b_1,b_2) \;
\label{eq:eft-va},
\eeq
\beq
M_{nfa}^{L,P_1} &=& -\frac{16 \sqrt{6}}{3}\pi C_F  m_{B}^2
\int_{0}^{1}d x_{1}d x_{2}\,d x_{3}\,\int_{0}^{\infty} b_1d b_1 b_2d
b_2\, \phi_{B}(x_1,b_1) \left\{ \left[ r_{A}(1-x_3)
(\phi^{s}_{A}(x_3)- \phi^{t}_{A}(x_3)) \right. \right.
\non && \left. \left. \times
\phi_{V}(x_2)+r_{V}x_{2}(\phi^{t}_{V}(x_2)
+\phi^{s}_{V}(x_2))\phi_{A}(x_3)\right]
 E_{nfa}(t_{e})h_{nfa}^{e}(x_{1},x_{2},x_{3},b_{1},b_{2})-\left[ r_{V}(2-x_2)
\phi_{A}(x_3)
 \right.\right.\non && \left. \left. \times
 (\phi^{t}_{V}(x_2)+\phi^{s}_{V}(x_2))- r_{A}(1+x_3)\phi_{V}(x_2)
(\phi^{s}_{A}(x_3)-\phi^{t}_{A}(x_3))\right]
E_{nfa}(t_f)h_{nfa}^{f}(x_{1},x_{2},x_{3},b_{1},b_{2})
\right\}\;,\label{eq:eflp1-va}
\eeq
\beq
M_{nfa}^{N,P_1} &=&-\frac{16 \sqrt{6}}{3}\pi C_F  m_{B}^2
\int_{0}^{1}d x_{1}d x_{2}\,d x_{3}\,\int_{0}^{\infty} b_1d b_1 b_2d
b_2\, \phi_{B}(x_1,b_1)
\left\{\left[r_{V}x_2(\phi^{v}_{V}(x_2) +\phi^{a}_{V}(x_2))
\phi^T_{A}(x_3)
\right.\right. \non && \left. \left.
- r_{A} (1-x_3) \phi^T_{V}(x_2)
(\phi^{a}_{A}(x_3)-\phi^{v}_{A}(x_3))\right]
 E_{nfa}(t_{e})h_{nfa}^{e}(x_{1},x_{2},x_{3},b_{1},b_{2})
 +\left[r_{V}(2-x_2)
 \phi^{T}_{A}(x_3)
 \right.\right.\non && \left. \left.\times
 (\phi^{v}_{V}(x_2) +\phi^{a}_{V}(x_2))  -r_{A}(1+x_3)\phi^{T}_{V}(x_2)
(\phi^{a}_{A}(x_3)-\phi^{v}_{A}(x_3)) \right]
E_{nfa}(t_f)h_{nfa}^{f}(x_{1},x_{2},x_{3},b_{1},b_{2})
\right\}\;,\label{eq:efnp1-va}
\eeq
\beq
M_{nfa}^{T,P_1} &=& 2 M_{nfa}^{N,P_1}
\;,\label{eq:eftp1-va}
 \eeq
\beq
M_{nfa}^{L,P_2} &=& \frac{16\sqrt{6}}{3}\pi C_F m_{B}^2 \int_{0}^{1}d x_{1}d x_{2}\,d
x_{3}\,\int_{0}^{\infty} b_1d b_1 b_2d b_2\, \phi_{B}(x_1,b_1)
\left\{\left[ x_2 \phi_{V}(x_2)\phi_{A}(x_3) \right.\right.\non
&& \left. \left. + r_{V}r_{A}
\left((1+x_2-x_3)(\phi^{s}_{V}(x_2)\phi^{s}_{A}(x_3)-\phi^{t}_{V}(x_2)
\phi^{t}_{A}(x_3))
+(1-x_2-x_3)(\phi^{s}_{V}(x_2)\phi^{t}_{A}(x_3)
\right.\right.\right.\non && \left.\left. \left.
-\phi^{t}_{V}(x_2)\phi^{s}_{A}(x_3) ) \right) \right]E_{nfa}(t_e)
 h_{nfa}^{e}(x_1,x_2,x_3,b_1,b_2)
-\left[(1- x_3)\phi_{V}(x_2)\phi_{A}(x_3) + 2 r_{V}
r_{A}(\phi^{t}_{V}(x_2) \right.\right. \non &&
\left.\left. \times \phi^{t}_{A}(x_3)+\phi^{s}_{V}(x_2)\phi^{s}_{A}(x_3))-
r_{V}r_{A}(1+x_2-x_3)(\phi^{t}_{V}(x_2)\phi^{t}_{A}(x_3)
-\phi^{s}_{V}(x_2)\phi^{s}_{A}(x_3)) - r_{V} r_{A} \right.\right.\non
&&\left.\left. \times
(1-x_2-x_3)(\phi^{s}_{V}(x_2)\phi^{t}_{A}(x_3)
-\phi^{t}_{V}(x_2)\phi^{s}_{A}(x_3))\right] E_{nfa}(t_f)
 h_{nfa}^{f}(x_1,x_2,x_3,b_1,b_2) \right\}\;,
\label{eq:eflp2-va}
\eeq
\beq
M_{nfa}^{N,P_2} &=& - M_{nfa}^{N} \;,
 \label{eq:efnp2-va}
 \eeq
 \beq
M_{nfa}^{T,P_2}&=& M_{nfa}^{T} \;
\label{eq:eftp2-va}.
\eeq

For the factorizable annihilation($fa$) diagrams in Figs.
\ref{fig:fig1}(g) and \ref{fig:fig1}(h),
the contributions are
\beq
F_{fa}^{L}&=&- 8\pi C_F m_{B}^2 \int_0^1 d x_{2} dx_{3}\,
 \int_{0}^{\infty} b_2
db_2 b_3 db_3\, \left\{ \left [ x_{2}
\phi_{V}(x_{2})\phi_{A}(x_{3}) +2r_{V}r_{A}\phi^{s}_{A}(x_{3})
((1+x_{2}) \phi^{s}_{V}(x_2)
 \right. \right.\non && \left.\left. - (1-x_{2}) \phi^{t}_{V}(x_2))\right] E_{fa}(t_{g})
h_{fa}(x_{2},1-x_{3},b_{2},b_{3}) -
\left[(1-x_{3})\phi_{V}(x_2)\phi_{A}(x_3) + 2 r_{V} r_{A}
\phi^{s}_{V}(x_2) \right.\right. \non && \left. \left.  \times
(x_{3}\phi^{t}_{A}(x_3)+(2-x_{3})\phi^{s}_{A}(x_3))\right]E_{fa}(t_{h})
h_{fa}(1-x_{3},x_{2},b_{3},b_{2}) \right\}\;,
\label{eq:ghl-va}
\eeq
\beq
F_{fa}^{N}&=& -8\pi C_F m_{B}^2 \int_0^1 d x_{2}
dx_{3}\, \int_{0}^{\infty} b_2 db_2 b_3 db_3\,
r_{V}r_{A} \left\{ E_{fa}(t_{g})\left[(1+x_{2})(\phi^{v}_{V}(x_2)\phi^{v}_{A}(x_3)
+\phi^{a}_{V}(x_2)\phi^{a}_{A}(x_3))
\right. \right.  \non && \left.\left.
-(1-x_{2})(\phi^{v}_{V}(x_2)\phi^{a}_{A}(x_3)
+\phi^{a}_{V}(x_2)\phi^{v}_{A}(x_3))\right]
h_{fa}(x_{2},1-x_{3},b_{2},b_{3})-
\left[(2-x_{3})(\phi^{v}_{V}(x_2)\phi^{v}_{A}(x_3)\right.\right.\non
&& \left.\left.
+\phi^{a}_{V}(x_2)\phi^{a}_{A}(x_3))+x_3(\phi^{v}_{V}(x_2)\phi^{a}_{A}(x_3)
+\phi^{a}_{V}(x_2)\phi^{v}_{A}(x_3))\right] E_{fa}(t_{h})
h_{fa}( 1-x_{3},x_{2},b_{3},b_{2})\right\}\;,
\label{eq:ghn-va}
\eeq
\beq
F_{fa}^{T}&=& -16\pi C_F m_{B}^2 \int_0^1 d x_{2}
dx_{3}\, \int_{0}^{\infty} b_2 db_2 b_3 db_3\,
r_{V}r_{A}\left\{E_{fa}(t_{g})\left[(1+x_{2})
(\phi^{v}_{V}(x_2)\phi^{a}_{A}(x_3)
+\phi^{a}_{V}(x_2)\phi^{v}_{A}(x_3))
\right.\right. \non &&
\left. \left.
-(1-x_{2})(\phi^{v}_{V}(x_2)\phi^{v}_{A}(x_3)
+\phi^{a}_{V}(x_2)\phi^{a}_{A}(x_3))\right]
h_{fa}(x_{2},1-x_{3},b_{2},b_{3})
+\left[x_{3}(\phi^{v}_{V}(x_2)\phi^{v}_{A}(x_3)\right.\right. \non
&& \left. \left.
+\phi^{a}_{V}(x_2)\phi^{a}_{A}(x_3))+(2-x_3)
(\phi^{v}_{V}(x_2)\phi^{a}_{A}(x_3)
+\phi^{a}_{V}(x_2)\phi^{v}_{A}(x_3))\right] E_{fa}(t_{h})
 h_{fa}( 1-x_{3},x_{2},b_{3},b_{2})\right\}\;;
\label{eq:ght-va}
\eeq
\beq
F_{fa}^{L,P_1}&=&- F_{fa}^{L}\;;
\label{eq:ghlp1-va}
\eeq
\beq
F_{fa}^{N,P_1}&=&- F_{fa}^{N}\;;
\label{eq:ghnp1-va}
\eeq
\beq
F_{fa}^{T,P_1}&=& F_{fa}^{T}\;;
\label{eq:ghtp1-va}
\eeq
\beq
F_{fa}^{L,P_2}&=&- 16\pi C_F m_{B}^2
\int_0^1 d x_{2} dx_{3}\, \int_{0}^{\infty}  b_2
db_2 b_3 db_3\, \left\{ \left[2 r_{A}
\phi_{V}(x_2)\phi^{s}_{A}(x_3)- r_{V} {x_2} (\phi^{t}_{V}(x_2) -
\phi^{s}_{V}(x_2) )
 \right.\right. \non && \left.\left.\times
\phi_{A}(x_3) \right]h_{fa}(x_{2},1-x_{3},b_{2},b_{3}) E_{fa}(t_{g})
+ \left[2 r_{V} \phi^{s}_{V}(x_2) \phi_{A}(x_3)+ r_{A} (1-x_3)
\phi_{V}(x_2)
\right.\right.\non && \left.\left. \times
(\phi^{t}_{A}(x_3)+\phi^{s}_{A}(x_3))\right]
E_{fa}(t_{h}) h_{fa}(1-x_{3},x_{2},b_{3},b_{2})
\right\}\;, \label{eq:ghlp2-va}
\eeq
\beq
F_{fa}^{N,P_2}&=& -16\pi C_F
m_{B}^2 \int_0^1 d x_{2} dx_{3}\, \int_{0}^{\infty} b_2 db_2 b_3
db_3\,
\left\{r_{A}\phi^{T}_{V}(x_2)(\phi^{a}_{A}(x_3) -\phi^v_{A}(x_{3}))
 h_{fa}(x_{2},1-x_{3},b_{2},b_{3})
\right. \non && \left. \times
E_{fa}(t_{g})
+ r_{V}(\phi^{v}_{V}(x_2)
+\phi^a_{V}(x_{2}))\phi^{T}_{A}(x_3) E_{fa}(t_{h}) h_{fa}(
1-x_{3},x_{2},b_{3},b_{2})\right\}\;,
\label{eq:ghnp2-va}
\eeq
\beq
F_{fa}^{T,P_2}&=& 2 F_{fa}^{N,P_2}.
\label{eq:ghtp2-va}
\eeq

When we exchange the positions of vector and axial-vector states
in Fig.~\ref{fig:fig1}, the amplitudes $F'$, $M'$, ${F'}^{P_1}$, ${M'}^{P_1}$, ${F'}^{P_2}$, and ${M'}^{P_2}$ arising from new Feynman
diagrams can be easily and correspondingly obtained as follows:
\beq
{F'}_{fe}^{L} &=&- 8\pi C_F m_{B}^2 \int_0^1
d x_{1} dx_{3}\, \int_{0}^{\infty} b_1 db_1 b_3 db_3\, \pb(x_1,b_1)
\left\{ \left [(1+x_{3})\phi_{V}(x_{3}) +
r_{V}(1-2x_{3}) \right.\right. \non && \left. \left.\times
(\phi^{t}_{V}(x_3)+\phi^{s}_{V}(x_{3}))\right]E_{fe}(t'_{a})
h_{fe}(x_{1},x_{3},b_{1},b_{3})
+2r_{V}\phi^{s}_{V}(x_3)E_{fe}(t'_{b})
h_{fe}(x_{3},x_{1},b_{3},b_{1}) \right\}\;,
\label{eq:abl-av}
\eeq
\beq
{F'}_{fe}^{N}&=& -8\pi C_F m_{B}^2 \int_0^1 d x_{1}
dx_{3}\, \int_{0}^{\infty} b_1 db_1 b_3 db_3\, \phi_{B}(x_1,b_1)
r_{A} \left\{ [\phi^{T}_{V}(x_3)
+2r_{V}\phi^{v}_{V}(x_3)+r_{V}x_{3}
\right. \non && \left. \times
(\phi^{v}_{V}(x_3)-\phi^{a}_{V}(x_3))]E_{fe}(t'_{a})
h_{fe}(x_{1},x_{3},b_{1},b_{3})+r_{V}[\phi^{v}_{V}(x_3)+\phi^{a}_{V}(x_3)]
E_{fe}(t'_{b}) h_{fe}(x_{3},x_{1},b_{3},b_{1})\right\}\;,
\label{eq:abn-av}
\eeq
\beq
 {F'}_{fe}^{T}&=& -16\pi C_F m_{B}^2 \int_0^1 d x_{1} dx_{3}\,
\int_{0}^{\infty} b_1 db_1 b_3 db_3\, \phi_{B}(x_1,b_1)
r_{A}\left\{[\phi^{T}_{V}(x_3) +2r_{V}\phi^{a}_{V}(x_3)-r_{V}x_{3}
\right. \non &&
\left.\times
(\phi^{v}_{V}(x_3)-\phi^{a}_{V}(x_3))]E_{fe}(t'_{a})  h_{fe}(x_{1},x_{3},b_{1},b_{3})
+r_{V}[\phi^{v}_{V}(x_3)+\phi^{a}_{V}(x_3)] E_{fe}(t'_{b})h_{fe}(
x_{3},x_{1},b_{3},b_{1})\right\}\;. \label{eq:abt-av}
\eeq
For the hard functions $h_{i}$, the running hard scales $t'_i$,
and the convolution functions $E_{i}(t')$, refer to Ref.~\cite{Zou:2015iwa}.

Since only the aixal-vector part of the
$(V+A)$ current contributes to the production of axial-vector states,
then $\langle V |V-A|B\rangle \langle A |V+A | 0 \rangle = - \langle V
|V-A |B \rangle \langle A |V-A|0 \rangle,$ which means
 \beq
 {F'}_{fe}^{P_1}= - {F'}_{fe} \; \label{eq:abp1-av} .
 \eeq

Analogously, because an axial-vector state also cannot be produced via scalar and/or pseudoscalar currents, then the contribution from the $(S \pm P)$ operators is
 \beq
 {F'}_{fe}^{P_2}= 0 \; \label{eq:abp2-av} .
 \eeq

The rest Feynman amplitudes can be presented explicitly as follows:
\beq
 {M'}_{nfe}^{L} &=& -\frac{16 \sqrt{6}}{3}\pi C_F m_{B}^2
\int_{0}^{1}d x_{1}d x_{2}\,d x_{3}\,\int_{0}^{\infty} b_1d b_1 b_2d
b_2\, \phi_{B}(x_1,b_1)\phi_{A}(x_{2}) \left\{\left[
(1-x_{2}) \phi_{V} (x_{3})
\right. \right. \non && \left.  +r_{V} x_3
(\phi^{t}_{V}(x_3)-\phi^{s}_{V}(x_3)) \right]E_{nfe}(t'_c)
h_{nfe}^{c}(x_1,x_2,x_3,b_1,b_2) -\left[
(x_{2}+x_{3}) \phi_{V} (x_{3})
\right.  \non && \left.\left.  -r_{V} x_3(\phi^{t}_{V}(x_3)+\phi^{s}_{V}(x_3))
\right] E_{nfe}(t'_d)h_{nfe}^{d}(x_1,x_2,x_3,b_1,b_2) \right\}\;,
\label{eq:cdl-av}
\eeq
\beq
{M'}_{nfe}^{N} &=& -\frac{16 \sqrt{6}}{3}\pi C_F  m_{B}^2
\int_{0}^{1}d x_{1}d x_{2}\,d x_{3}\,\int_{0}^{\infty} b_1d b_1 b_2d
b_2\, \phi_{B}(x_1,b_1) r_{A}\left\{(1-x_2)
(\phi^{v}_{A}(x_2)+\phi^{a}_{A}(x_2) )
\right. \non && \left. \times
\phi^{T}_{V}(x_3) h_{nfe}^{c}(x_1,x_2,x_3,b_1,b_2) E_{nfe}(t'_c)
+ \left[x_2(\phi^{v}_{A}(x_2) +\phi^a_{A}(x_2))\phi^{T}_{V}(x_3)
\right.\right. \non && \left. \left. -2r_{V}(x_2+x_3) (\phi^{v}_{A}(x_2) \phi^{v}_{V}(x_3)+
\phi^{a}_{A}(x_2) \phi^{a}_{V}(x_3) ) \right] E_{nfe}(t'_d)
h_{nfe}^{d}(x_1,x_2,x_3,b_1,b_2) \right\}\;,
 \label{eq:cdn-av}
 \eeq
 \beq
 {M'}_{nfe}^{T}&=& -\frac{32 \sqrt{6}}{3}\pi C_F m_{B}^2
\int_{0}^{1}d x_{1}d x_{2}\,d x_{3}\,\int_{0}^{\infty} b_1d b_1 b_2d
b_2\, \phi_{B}(x_1,b_1)   r_{A}\left\{
(1-x_2)(\phi^{v}_{A}(x_2)+\phi^{a}_{A}(x_2) )
\right. \non && \left. \times \phi^{T}_{V}(x_3)
h_{nfe}^{c}(x_1,x_2,x_3,b_1,b_2)
E_{nfe}(t'_c)+   \left[
x_2(\phi^{v}_{A}(x_2)+\phi^a_{A}(x_2))\phi^{T}_{V}(x_3)
\right.\right. \non && \left.\left. -2r_{V}(x_2+x_3)
(\phi^{v}_{A}(x_2) \phi^{a}_{V}(x_3)+ \phi^{a}_{A}(x_2)
\phi^{v}_{V}(x_3) ) \right]E_{nfe}(t'_d)
h_{nfe}^{d}(x_1,x_2,x_3,b_1,b_2) \right\}\; \label{eq:cdt-av}.
\eeq
 \beq
 {M'}_{nfe}^{L,P_1} &=& \frac{16 \sqrt{6}}{3}\pi C_F  m_{B}^2
\int_{0}^{1}d x_{1}d x_{2}\,d x_{3}\,\int_{0}^{\infty} b_1d b_1 b_2d
b_2\, \phi_{B}(x_1,b_1)  r_{A}
\left\{\left[(1-x_2)(\phi^{t}_{A}(x_2)+\phi^{s}_{A}(x_2))
\right. \right.\non && \times
\left. \phi_{V}(x_3)  -r_{V}(1-x_2) (\phi^{t}_{A}(x_2)
+\phi^{s}_{A}(x_2))(\phi^{t}_{V}(x_3)-\phi^{s}_{V}(x_3))-r_{V}
x_3(\phi^{t}_{A}(x_2) -\phi^{s}_{A}(x_2)) \right. \non && \left. \times
(\phi^{t}_{V}(x_3)+\phi^{s}_{V}(x_3)) \right]
 E_{nfe}(t'_{c})h_{nfe}^{c}(x_{1},x_{2},x_{3},b_{1},b_{2})
+\left[ x_2 ( \phi^{t}_{A}(x_2)-\phi^{s}_{A}(x_2)) \phi_{V}(x_3)
\right.  \non &&  \left. \left.
-r_{V}x_2(\phi^{t}_{A}(x_2) -\phi^{s}_{A}(x_2)) (\phi^{t}_{V}(x_3)-\phi^{s}_{V}(x_3))
-r_{V}x_{3}(\phi^{t}_{A}(x_2)+\phi^{s}_{A}(x_2))
(\phi^{t}_{V}(x_3)+\phi^{s}_{V}(x_3)) \right] \right. \non &&  \left. \times
E_{nfe}(t'_d)h_{nfe}^{d}(x_{1},x_{2},x_{3},b_{1},b_{2})
\right\}\;,\label{eq:cdlp1-av}
\eeq
\beq
{M'}_{nfe}^{N,P_1} &=& \frac{16 \sqrt{6}}{3}\pi C_F m_{B}^2
\int_{0}^{1}d x_{1}d x_{2}\,d x_{3}\,\int_{0}^{\infty} b_1d b_1 b_2d
b_2\, \phi_{B}(x_1,b_1)
r_{V}x_{3} \phi^{T}_{A}(x_2) (\phi^{v}_{V}(x_3)-\phi^{a}_{V}(x_3))
 \non && \times
\left \{ E_{nfe}(t'_{c})
h_{nfe}^{c}(x_{1},x_{2},x_{3},b_{1},b_{2})
+ E_{nfe}(t'_{d}) h_{nfe}^{d}(x_{1},x_{2},x_{3},b_{1},b_{2}) \right\}\;,
\label{eq:cdnp1-av}
\eeq
\beq
{M'}_{nfe}^{T,P_1}&=& 2 {M'}_{nfe}^{N,P_1}\;,
\label{eq:cdtp1-av}
\eeq
\beq
{M'}_{nfe}^{L,P_2} &=& \frac{16 \sqrt{6}}{3}\pi C_F  m_{B}^2
\int_{0}^{1}d x_{1}d x_{2}\,d x_{3}\,\int_{0}^{\infty} b_1d b_1 b_2d
b_2\, \phi_{B}(x_1,b_1)  \phi_{A}(x_2)\left\{
\left[(1-x_2+x_3)\phi_{V}(x_3)
\right.\right. \non &&  \left. \left.  - r_{V}x_3
(\phi^{t}_{V}(x_3)+\phi^{s}_{V}(x_3))\right]
E_{e}(t'_{c}) h_{nfe}^{c}(x_{1},x_{2},x_{3},b_{1},b_{2})-
h_{nfe}^{d}(x_{1},x_{2},x_{3},b_{1},b_{2})E_{nfe}(t'_{d})\right. \non && \times
\left[
x_2\phi_{V}(x_3)+r_{V}x_{3}(\phi^{t}_{V}(x_3)-\phi^{s}_{V}(x_3))
\right]\left.\right\}\;,
\label{eq:cdlp2-av}
\eeq
\beq
{M'}_{nfe}^{N,P_2} &=& -\frac{16 \sqrt{6}}{3}\pi C_F m_{B}^2
\int_{0}^{1}d x_{1}d x_{2}\,d x_{3}\,\int_{0}^{\infty} b_1d b_1 b_2d
b_2\, \phi_{B}(x_1,b_1)  r_{A}
\left\{\left[(1-x_2)(\phi^{v}_{A}(x_2)-\phi^{a}_{A}(x_2))
\right. \right.\non && \left. \left. \times
\phi^{T}_{V}(x_3)-2r_{V}(1-x_2+x_3) (\phi^{v}_{A}(x_2)
\phi^{v}_{V}(x_3)-\phi^{a}_{A}(x_2)\phi^{a}_{V}(x_3))
\right]h_{nfe}^{c}(x_{1},x_{2},x_{3},b_{1},b_{2})
\right. \non &&\left. \times
E_{nfe}(t'_{c})
 + x_2(\phi^{v}_{A}(x_2)-\phi^{a}_{A}(x_2))
\phi^{T}_{V}(x_3)  E_{nfe}(t'_{d})
h_{nfe}^{d}(x_{1},x_{2},x_{3},b_{1},b_{2})\right\}\;,
\label{eq:cdnp2-av}
\eeq
\beq
{M'}_{nfe}^{T,P_2} &=& -\frac{32 \sqrt{6}}{3}\pi C_F m_{B}^2
\int_{0}^{1}d x_{1}d x_{2}\,d x_{3}\,\int_{0}^{\infty} b_1d b_1 b_2d
b_2\, \phi_{B}(x_1,b_1) r_{A}
\left\{\left[(1-x_2)(\phi^{v}_{A}(x_2)-\phi^{a}_{A}(x_2))
\right. \right. \non && \left. \left.  \times
\phi^{T}_{V}(x_3)-2r_{V}(1-x_2+x_3) (\phi^{v}_{A}(x_2)
\phi^{a}_{V}(x_3)-\phi^{a}_{A}(x_2)\phi^{v}_{V}(x_3))
\right] h_{nfe}^{c}(x_{1},x_{2},x_{3},b_{1},b_{2})
\right. \non &&\left. \times
E_{nfe}(t'_{c})+ x_2(\phi^{v}_{A}(x_2)-\phi^{a}_{A}(x_2))
\phi^{T}_{V}(x_3) E_{nfe}(t'_{d})
h_{nfe}^{d}(x_{1},x_{2},x_{3},b_{1},b_{2})\right\}\;,
\label{eq:cdtp2-av}
 \eeq
\beq
{M'}_{nfa}^{L} &=& -\frac{16\sqrt{6}}{3}\pi C_F m_{B}^2 \int_{0}^{1}d x_{1}d x_{2}\,d
x_{3}\,\int_{0}^{\infty} b_1d b_1 b_2d b_2\, \phi_{B}(x_1,b_1)
\left\{\left[ (1-x_3) \phi_{A}(x_2)\phi_{V}(x_3) \right.\right.\non
&& \left. \left. - r_{A}r_{V}
\left((1+x_2-x_3)(\phi^{s}_{A}(x_2)\phi^{s}_{V}(x_3)-\phi^{t}_{A}(x_2)
\phi^{t}_{V}(x_3))
-(1-x_2-x_3)(\phi^{s}_{A}(x_2)\phi^{t}_{V}(x_3)
\right.\right.\right.\non && \left.\left. \left.
-\phi^{t}_{A}(x_2)\phi^{s}_{V}(x_3) ) \right) \right]E_{nfa}(t'_e)
 h_{nfa}^{e}(x_1,x_2,x_3,b_1,b_2)
-\left[ x_2\phi_{A}(x_2)\phi_{V}(x_3) - 2 r_{A}
r_{V}(\phi^{t}_{A}(x_2) \right.\right. \non &&
\left.\left. \times \phi^{t}_{V}(x_3)+\phi^{s}_{A}(x_2)\phi^{s}_{V}(x_3))+
r_{A}r_{V}(1+x_2-x_3)(\phi^{t}_{A}(x_2)\phi^{t}_{V}(x_3)
-\phi^{s}_{A}(x_2)\phi^{s}_{V}(x_3)) - r_{A} r_{V} \right.\right.\non
&&\left.\left. \times
(1-x_2-x_3)(\phi^{s}_{A}(x_2)\phi^{t}_{V}(x_3)
-\phi^{t}_{A}(x_2)\phi^{s}_{V}(x_3))\right] E_{nfa}(t'_f)
 h_{nfa}^{f}(x_1,x_2,x_3,b_1,b_2) \right\}\;,
\label{eq:efl-av}
\eeq
\beq
{M'}_{nfa}^{N} &=& \frac{32 \sqrt{6}}{3}\pi C_F m_{B}^2
\int_{0}^{1}d x_{1}d x_{2}\,d x_{3}\,\int_{0}^{\infty} b_1d b_1 b_2d
b_2\, \phi_{B}(x_1,b_1) r_{A} r_{V} \non &&\times
\left[\phi^{v}_{A}(x_2)\phi^{v}_{V}(x_3)
+\phi^{a}_{A}(x_2)\phi^{a}_{V}(x_3)\right] E_{nfa}(t'_f)
h_{nfa}^{f}(x_1,x_2,x_3,b_1,b_2) \;,
 \label{eq:efn-av}
 \eeq
 \beq
{M'}_{nfa}^{T}&=& \frac{64 \sqrt{6}}{3}\pi C_F m_{B}^2
\int_{0}^{1}d x_{1}d x_{2}\,d x_{3}\,\int_{0}^{\infty} b_1d b_1 b_2d
b_2\, \phi_{B}(x_1,b_1) r_{A} r_{V} \non &&\times
\left[\phi^{v}_{A}(x_2)\phi^{a}_{V}(x_3)
+\phi^{a}_{A}(x_2)\phi^{v}_{V}(x_3)\right] E_{nfa}(t'_f)
h_{nfa}^{f}(x_1,x_2,x_3,b_1,b_2) \;
\label{eq:eft-av}.
\eeq
\beq
{M'}_{nfa}^{L,P_1} &=& \frac{16 \sqrt{6}}{3}\pi C_F  m_{B}^2
\int_{0}^{1}d x_{1}d x_{2}\,d x_{3}\,\int_{0}^{\infty} b_1d b_1 b_2d
b_2\, \phi_{B}(x_1,b_1) \left\{ \left[ r_{V}(1-x_3)
(\phi^{s}_{V}(x_3)- \phi^{t}_{V}(x_3)) \right. \right.
\non && \left. \left. \times
\phi_{A}(x_2)+r_{A}x_{2}(\phi^{t}_{A}(x_2)
+\phi^{s}_{A}(x_2))\phi_{V}(x_3)\right]
 E_{nfa}(t'_{e})h_{nfa}^{e}(x_{1},x_{2},x_{3},b_{1},b_{2})+\left[ r_{A}(2-x_2)
\phi_{V}(x_3)
 \right.\right.\non && \left. \left. \times
 (\phi^{t}_{A}(x_2)+\phi^{s}_{A}(x_2))+ r_{V}(1+x_3)\phi_{A}(x_2)
(\phi^{s}_{V}(x_3)-\phi^{t}_{V}(x_3))\right]
E_{nfa}(t'_f)h_{nfa}^{f}(x_{1},x_{2},x_{3},b_{1},b_{2})
\right\}\;,\label{eq:eflp1-av}
\eeq
\beq
{M'}_{nfa}^{N,P_1} &=&-\frac{16 \sqrt{6}}{3}\pi C_F  m_{B}^2
\int_{0}^{1}d x_{1}d x_{2}\,d x_{3}\,\int_{0}^{\infty} b_1d b_1 b_2d
b_2\, \phi_{B}(x_1,b_1)
\left\{\left[r_{A}x_2(\phi^{v}_{A}(x_2) +\phi^{a}_{A}(x_2))
\phi^T_{V}(x_3)
\right.\right. \non && \left. \left.
+ r_{A} (1-x_3) \phi^T_{A}(x_2)
(\phi^{a}_{V}(x_3)-\phi^{v}_{V}(x_3))\right]
 E_{nfa}(t'_{e})h_{nfa}^{e}(x_{1},x_{2},x_{3},b_{1},b_{2})
 +\left[r_{A}(2-x_2)
 \phi^{T}_{V}(x_3)
 \right.\right.\non && \left. \left.\times
 (\phi^{v}_{A}(x_2) +\phi^{a}_{A}(x_2))  +r_{V}(1+x_3)\phi^{T}_{A}(x_2)
(\phi^{a}_{V}(x_3)-\phi^{v}_{V}(x_3)) \right]
E_{nfa}(t'_f)h_{nfa}^{f}(x_{1},x_{2},x_{3},b_{1},b_{2})
\right\}\;,\label{eq:efnp1-av}
\eeq
\beq
{M'}_{nfa}^{T,P_1} &=& 2 {M'}_{nfa}^{N,P_1}
\;,\label{eq:eftp1-av}
 \eeq
\beq
{M'}_{nfa}^{L,P_2} &=& \frac{16\sqrt{6}}{3}\pi C_F m_{B}^2 \int_{0}^{1}d x_{1}d x_{2}\,d
x_{3}\,\int_{0}^{\infty} b_1d b_1 b_2d b_2\, \phi_{B}(x_1,b_1)
\left\{\left[ x_2 \phi_{A}(x_2)\phi_{V}(x_3) \right.\right.\non
&& \left. \left. - r_{A}r_{V}
\left((1+x_2-x_3)(\phi^{s}_{A}(x_2)\phi^{s}_{V}(x_3)-\phi^{t}_{A}(x_2)
\phi^{t}_{V}(x_3))
+(1-x_2-x_3)(\phi^{s}_{A}(x_2)\phi^{t}_{V}(x_3)
\right.\right.\right.\non && \left.\left. \left.
-\phi^{t}_{A}(x_2)\phi^{s}_{V}(x_3) ) \right) \right]E_{nfa}(t'_e)
 h_{nfa}^{e}(x_1,x_2,x_3,b_1,b_2)
-\left[(1- x_3)\phi_{A}(x_2)\phi_{V}(x_3) - 2 r_{A}
r_{V}(\phi^{t}_{A}(x_2) \right.\right. \non &&
\left.\left. \times \phi^{t}_{V}(x_3)+\phi^{s}_{A}(x_2)\phi^{s}_{V}(x_3))+
r_{A}r_{V}(1+x_2-x_3)(\phi^{t}_{A}(x_2)\phi^{t}_{V}(x_3)
-\phi^{s}_{A}(x_2)\phi^{s}_{V}(x_3)) + r_{A} r_{V} \right.\right.\non
&&\left.\left. \times
(1-x_2-x_3)(\phi^{s}_{A}(x_2)\phi^{t}_{V}(x_3)
-\phi^{t}_{A}(x_2)\phi^{s}_{V}(x_3))\right] E_{nfa}(t'_f)
 h_{nfa}^{f}(x_1,x_2,x_3,b_1,b_2) \right\}\;,
\label{eq:eflp2-av}
\eeq
\beq
{M'}_{nfa}^{N,P_2} &=& - {M'}_{nfa}^{N} \;,
 \label{eq:efnp2-av}
 \eeq
 \beq
{M'}_{nfa}^{T,P_2}&=& {M'}_{nfa}^{T} \;
\label{eq:eftp2-av}.
\eeq
\beq
{F'}_{fa}^{L}&=&- 8\pi C_F m_{B}^2 \int_0^1 d x_{2} dx_{3}\,
 \int_{0}^{\infty} b_2
db_2 b_3 db_3\, \left\{ \left [ x_{2}
\phi_{A}(x_{2})\phi_{V}(x_{3}) -2r_{A}r_{V}\phi^{s}_{V}(x_{3})
((1+x_{2}) \phi^{s}_{A}(x_2)
 \right. \right.\non && \left.\left. - (1-x_{2}) \phi^{t}_{A}(x_2))\right] E_{fa}(t'_{g})
h_{fa}(x_{2},1-x_{3},b_{2},b_{3}) -
\left[(1-x_{3})\phi_{A}(x_2)\phi_{V}(x_3) - 2 r_{A} r_{V}
\phi^{s}_{A}(x_2) \right.\right. \non && \left. \left.  \times
(x_{3}\phi^{t}_{V}(x_3)+(2-x_{3})\phi^{s}_{V}(x_3))\right]E_{fa}(t'_{h})
h_{fa}(1-x_{3},x_{2},b_{3},b_{2}) \right\}\;,
\label{eq:ghl-av}
\eeq
\beq
{F'}_{fa}^{N}&=& -8\pi C_F m_{B}^2 \int_0^1 d x_{2}
dx_{3}\, \int_{0}^{\infty} b_2 db_2 b_3 db_3\,
r_{A}r_{V} \left\{ E_{fa}(t'_{g})\left[(1+x_{2})(\phi^{v}_{A}(x_2)\phi^{v}_{V}(x_3)
+\phi^{a}_{A}(x_2)\phi^{a}_{V}(x_3))
\right. \right.  \non && \left.\left.
-(1-x_{2})(\phi^{v}_{A}(x_2)\phi^{a}_{V}(x_3)
+\phi^{a}_{A}(x_2)\phi^{v}_{V}(x_3))\right]
h_{fa}(x_{2},1-x_{3},b_{2},b_{3})-
\left[(2-x_{3})(\phi^{v}_{A}(x_2)\phi^{v}_{V}(x_3)\right.\right.\non
&& \left.\left.
+\phi^{a}_{A}(x_2)\phi^{a}_{V}(x_3))+x_3(\phi^{v}_{A}(x_2)\phi^{a}_{V}(x_3)
+\phi^{a}_{A}(x_2)\phi^{v}_{V}(x_3))\right] E_{fa}(t'_{h})
h_{fa}( 1-x_{3},x_{2},b_{3},b_{2})\right\}\;,
\label{eq:ghn-av}
\eeq
\beq
{F'}_{fa}^{T}&=& -16\pi C_F m_{B}^2 \int_0^1 d x_{2}
dx_{3}\, \int_{0}^{\infty} b_2 db_2 b_3 db_3\,
r_{A}r_{V}\left\{E_{fa}(t'_{g})\left[(1+x_{2})
(\phi^{v}_{A}(x_2)\phi^{a}_{V}(x_3)
+\phi^{a}_{A}(x_2)\phi^{v}_{V}(x_3))
\right.\right. \non &&
\left. \left.
-(1-x_{2})(\phi^{v}_{A}(x_2)\phi^{v}_{V}(x_3)
+\phi^{a}_{A}(x_2)\phi^{a}_{V}(x_3))\right]
h_{fa}(x_{2},1-x_{3},b_{2},b_{3})
-\left[x_{3}(\phi^{v}_{A}(x_2)\phi^{v}_{V}(x_3)\right.\right. \non
&& \left. \left.
+\phi^{a}_{A}(x_2)\phi^{a}_{V}(x_3))+(2-x_3)
(\phi^{v}_{A}(x_2)\phi^{a}_{V}(x_3)
+\phi^{a}_{A}(x_2)\phi^{v}_{V}(x_3))\right] E_{fa}(t'_{h})
 h_{fa}( 1-x_{3},x_{2},b_{3},b_{2})\right\}\;;
\label{eq:ght-av}
\eeq
\beq
{F'}_{fa}^{L,P_1}&=&- {F'}_{fa}^{L}\;;
\label{eq:ghlp1-av}
\eeq
\beq
{F'}_{fa}^{N,P_1}&=&- {F'}_{fa}^{N}\;;
\label{eq:ghnp1-av}
\eeq
\beq
{F'}_{fa}^{T,P_1}&=& {F'}_{fa}^{T}\;;
\label{eq:ghtp1-av}
\eeq
\beq
{F'}_{fa}^{L,P_2}&=&- 16\pi C_F m_{B}^2
\int_0^1 d x_{2} dx_{3}\, \int_{0}^{\infty}  b_2
db_2 b_3 db_3\, \left\{ \left[2 r_{V}
\phi_{A}(x_2)\phi^{s}_{V}(x_3)+ r_{A} {x_2} (\phi^{t}_{A}(x_2) -
\phi^{s}_{A}(x_2) )
 \right.\right. \non && \left.\left.\times
\phi_{V}(x_3) \right]h_{fa}(x_{2},1-x_{3},b_{2},b_{3}) E_{fa}(t'_{g})
- \left[2 r_{A} \phi^{s}_{A}(x_2) \phi_{V}(x_3)- r_{V} (1-x_3)
\phi_{A}(x_2)
\right.\right.\non && \left.\left. \times
(\phi^{t}_{V}(x_3)+\phi^{s}_{V}(x_3))\right]
E_{fa}(t'_{h}) h_{fa}(1-x_{3},x_{2},b_{3},b_{2})
\right\}\;, \label{eq:ghlp2-av}
\eeq
\beq
{F'}_{fa}^{N,P_2}&=& 16\pi C_F
m_{B}^2 \int_0^1 d x_{2} dx_{3}\, \int_{0}^{\infty} b_2 db_2 b_3
db_3\,
\left\{r_{V}\phi^{T}_{A}(x_2)(\phi^{a}_{V}(x_3) -\phi^v_{V}(x_{3}))
 h_{fa}(x_{2},1-x_{3},b_{2},b_{3})
\right. \non && \left. \times
E_{fa}(t'_{g})
- r_{A}(\phi^{v}_{A}(x_2)
+\phi^a_{A}(x_{2}))\phi^{T}_{V}(x_3) E_{fa}(t'_{h}) h_{fa}(
1-x_{3},x_{2},b_{3},b_{2})\right\}\;,
\label{eq:ghnp2-av}
\eeq
\beq
{F'}_{fa}^{T,P_2}&=& 2 {F'}_{fa}^{N,P_2};
\label{eq:ghtp2-av}
\eeq

Thus, by combining various contributions from different diagrams as
presented in Eqs.~(\ref{eq:abl-va})-(\ref{eq:ghtp2-av}) and the mixing pattern in Eq.~(\ref{eq:mix-f1q-f1s}),
the total decay amplitudes for 10 nonleptonic decays of $B \to f_1(1285) V$ can be written as follows (the superscript $h$ in
the following formulas describes the helicity amplitudes with longitudinal, normal, and transverse polarizations, respectively):
\begin{itemize}

\item[1.]{ $B^+ \to f_1(1285) (\rho^+,K^{*+})$ decays}
\beq
 A^h(B^+ \to f_1(1285) \rho^+) &=&
\biggl\{ [a_1] (f_\rho F^h_{fe}+ f_B F^h_{fa}+ f_B {F'}^{h}_{fa}) + [a_2] f_{f_{1q}} {F'}^{h}_{fe}
 + [C_1] (M^h_{nfe} + M^h_{nfa} + {M'}^{h}_{nfa})
 \non &&  +[C_2] {M'}^{h}_{nfe}
\biggr\} \lambda_u^d \zeta_{f_{1q}}
- \lambda_t^d  \zeta_{f_{1q}}\biggl\{ [a_4 + a_{10}] (f_\rho F^h_{fe}+
f_B F^h_{fa} + f_B {F'}^{h}_{fa}) +
[a_6 + a_8]\non && \times (f_B F_{fa}^{h,P_2}+ f_B F_{fa}^{\prime h,P_2}) + [C_3+ C_9] (M^h_{nfe}+ M^h_{nfa}
+ {M'}^{h}_{nfa})+ [C_5 + C_7] \non && \times (M_{nfe}^{h,P_1} + M_{nfa}^{h,P_1} +
{M'}_{nfa}^{h,P_1}) + [2a_3+ a_4 -2 a_5 -\frac{1}{2} (a_7 - a_9 + a_{10})] f_{f_{1q}} {F'}^{h}_{fe}\non &&  + [C_3+ 2 C_4 - \frac{1}{2} (C_9 -C_{10})] {M'}^{h}_{nfe} +[C_5 - \frac{1}{2} C_7] {M'}_{nfe}^{h,P_1}
+[2 C_6 +\frac{1}{2} C_8] {M'}_{nfe}^{h,P_2}
\biggr\} \non &&
-\lambda_t^d  \zeta_{f_{1s}}
\biggl\{ [a_3 -a_5 + \frac{1}{2} (a_7 - a_9)] f_{f_{1s}} {F'}^{h}_{fe} + [C_4 -\frac{1}{2} C_{10}] {M'}^{h}_{nfe} \non &&
+[C_6 - \frac{1}{2} C_8] {M'}_{nfe}^{h,P_2} \biggr\}
\label{eq:f1285rhop};
\eeq
\beq
 A^h(B^+ \to f_1(1285) K^{*+}) &=&
\lambda_u^s \biggl\{ [a_1] \biggl((f_{K^*} F^h_{fe}+ f_B F^h_{fa})
\zeta_{f_{1q}} + f_B {F'}^{h}_{fa} \zeta_{f_{1s}}
\biggr) + [a_2] f_{f_{1q}} {F'}^{h}_{fe}  \zeta_{f_{1q}}
 + [C_1]  \non &&  \times
  \biggl({M'}^{h}_{nfa} \zeta_{f_{1s}} +
 (M^h_{nfe} + M^h_{nfa})  \zeta_{f_{1q}}
 \biggr) +[C_2] {M'}^{h}_{nfe}  \zeta_{f_{1q}}
\biggr\}  - \lambda_t^s \biggl\{ [a_4 + a_{10}]
\non &&  \times
\biggl((f_{K^*} F^h_{fe}+
f_B F^h_{fa})  \zeta_{f_{1q}}
+ f_B {F'}^{h}_{fa} \zeta_{f_{1s}} \biggr) +
\biggl( f_B F_{fa}^{h,P_2}  \zeta_{f_{1q}}
+ f_B {F'}_{fa}^{h,P_2} \zeta_{f_{1s}}
\biggr)\non &&  \times [a_6 + a_8]
+ [C_3+ C_9]
 \biggl({M'}^{h}_{nfa} \zeta_{f_{1s}} +
(M^h_{nfe}+ M^h_{nfa})  \zeta_{f_{1q}}
\biggr)+[C_5
+ C_7] \non && \times  \biggl((M_{nfe}^{h,P_1} + M_{nfa}^{h,P_1})
 \zeta_{f_{1q}} +
{M'}_{nfa}^{h,P_1} \zeta_{f_{1s}} \biggr)
+ \biggl([2a_3 -2 a_5 -\frac{1}{2} (a_7 - a_9)] f_{f_{1q}}  {F'}^{h}_{fe}\non &&
+ [  2 C_4 + \frac{1}{2} C_{10}]{M'}^{h}_{nfe}
+[2 C_6 +\frac{1}{2} C_8] {M'}_{nfe}^{h,P_2}\biggr)
\zeta_{f_{1q}} +\biggl( [C_3+C_4 -\frac{1}{2}(C_9+ C_{10})]
\non && \times
{M'}^{h}_{nfe} + [a_3 +a_4 -a_5 + \frac{1}{2} (a_7 - a_9
 - a_{10})] f_{f_{1s}} {F'}^{h}_{fe}
 +[C_5 - \frac{1}{2} C_7] {M'}_{nfe}^{h,P_1}\non &&
+[C_6 - \frac{1}{2} C_8]  {M'}_{nfe}^{h,P_2}\biggr)
\zeta_{f_{1s}} \biggr\}
\label{eq:f1285kstp};
\eeq

\item[2.]{$B^0 \to f_1(1285) (\rho^0, K^{*0}, \omega, \phi)$ decays}
\beq
\sqrt{2} A^h(B^0 \to f_1(1285) \rho^0) &=&
\biggl\{ a_2 (f_\rho F^h_{fe}+ f_B F^h_{fa}+ f_B {F'}^{h}_{fa} -f_{f_{1q}} {F'}^{h}_{fe})
 + C_2 (M^h_{nfe} + M^h_{nfa} + {M'}^{h}_{nfa} -  {M'}^{h}_{nfe})
\biggr\} \non &&
\times
\lambda_u^d \zeta_{f_{1q}}
- \lambda_t^d \zeta_{f_{1q}}
\biggl\{ [-a_4 -\frac{1}{2}(3 a_7- 3a_{9}- a_{10})] f_\rho F^h_{fe}+[-a_4 +\frac{1}{2}(3 a_7+ 3a_{9}+ a_{10})]
\non && \times (f_B F^h_{fa} + f_B {F'}^{h}_{fa})- [2 a_3 +a_4- 2a_5 -\frac{1}{2}(a_7- a_{9}+ a_{10})]f_{f_{1q}} {F'}^{h}_{fe}
-[a_6 - \frac{1}{2} a_8] \non && \times
( f_B F_{fa}^{h,P_2}+ f_B {F'}_{fa}^{h,P_2}) + [-C_3+ \frac{1}{2}(C_9+3 C_{10})] (M^h_{nfe}+ M^h_{nfa}
+ {M'}^{h}_{nfa})
+[\frac{3}{2} C_8] \non &&\times
(M_{nfe}^{h,P_2}+ M_{nfa}^{h,P_2}
+ {M'}_{nfa}^{h,P_2})
-[C_5 -\frac{1}{2} C_7]  (M_{nfe}^{h,P_1} + M_{nfa}^{h,P_1} +
{M'}_{nfa}^{h,P_1}+{M'}_{nfe}^{h,P_1})\non &&  -[C_3 + 2 C_4 - \frac{1}{2} (C_9 -C_{10})] {M'}^h_{nfe} -[2 C_6 + \frac{1}{2} C_8] {M'}_{nfe}^{h,P_2}
\biggr\}
-\lambda_t^d \biggl\{ -[a_3 -a_5 + \frac{1}{2}
\non && \times (a_7 - a_9)] f_{f_{1s}} {F'}^{h}_{fe}
 - [C_4 -\frac{1}{2} C_{10}] {M'}^h_{nfe}
-[C_6 - \frac{1}{2} C_8] {M'}_{nfe}^{h,P_2} \biggr\}  \zeta_{f_{1s}}
\label{eq:f1285rho-d};
\eeq
\beq
 A^h(B^0 \to f_1(1285) K^{*0}) &=&
\lambda_u^s \biggl\{
[a_2] f_{f_{1q}} {F'}^h_{fe}
 +[C_2] {M'}^h_{nfe}
\biggr\} \zeta_{f_{1q}}
- \lambda_t^s \biggl\{ [a_4 -\frac{1}{2} a_{10}]
\biggl((f_{K^*} F^h_{fe}+
f_B F^h_{fa}) \zeta_{f_{1q}} \non &&
+\zeta_{f_{1s}} f_B {F'}^h_{fa} \biggr) +
[a_6 -\frac{1}{2} a_8] \biggl( f_B F_{fa}^{h,P_2} \zeta_{f_{1q}}
+ f_B {F'}_{fa}^{h,P_2}\zeta_{f_{1s}}
\biggr) + [C_3-\frac{1}{2} C_9]
\non && \times \biggl((M^h_{nfe}+ M^h_{nfa})
\zeta_{f_{1q}} + {M'}^h_{nfa}\zeta_{f_{1s}}
\biggr)
+ [C_5 -\frac{1}{2} C_7]  \biggl((M_{nfe}^{h,P_1}
 + M_{nfa}^{h,P_1}) \zeta_{f_{1q}}
\non &&
+ {M'}_{nfa}^{h,P_1}\zeta_{f_{1s}}
\biggr) + \biggl([2a_3 -2 a_5 -\frac{1}{2} (a_7 - a_9)] f_{f_{1q}} {F'}^h_{fe}
+ [  2 C_4 + \frac{1}{2} C_{10}] {M'}^h_{nfe}
\non &&
+[2 C_6 +\frac{1}{2} C_8] {M'}_{nfe}^{h,P_2}\biggr) \zeta_{f_{1q}}
+\biggl( [a_3 +a_4 -a_5 + \frac{1}{2} (a_7 - a_9 - a_{10})]
f_{f_{1s}} {F'}^h_{fe}
\non &&+ [C_3+C_4 -\frac{1}{2}(C_9+ C_{10})] {M'}^h_{nfe}
+[C_5 - \frac{1}{2} C_7] {M'}_{nfe}^{h,P_1}
+[C_6 - \frac{1}{2} C_8] \non &&
\times {M'}_{nfe}^{h,P_2}\biggr)
\zeta_{f_{1s}} \biggr\}
\label{eq:f1285kst0};
\eeq
\beq
\sqrt{2} A^h(B^0 \to f_1(1285) \omega) &=&
\lambda_u^d \biggl\{ a_2 (f_{\omega} F^h_{fe}+ f_B F^h_{fa}+ f_B {F'}^h_{fa} +f_{f_{1q}} {F'}^h_{fe})
 + C_2 (M^h_{nfe} + M^h_{nfa} + {M'}^h_{nfa} \non && +  {M'}^h_{nfe})
\biggr\}\cdot\zeta_{f_{1q}}
- \lambda_t^d \biggl\{[2 a_3 +a_4- 2a_5 -\frac{1}{2}(a_7- a_{9}+ a_{10})](f_{\omega} F^h_{fe} +f_{f_{1q}} {F'}^h_{fe}) \non && + [2 a_3+a_4+ 2a_5 +\frac{1}{2}(a_7+ a_{9}- a_{10})] (f_B F^h_{fa} + f_B {F'}^h_{fa})
+ ( f_B F_{fa}^{h,P_2}+ f_B {F'}_{fa}^{h,P_2})
\non && \times [a_6 - \frac{1}{2} a_8]
 +[C_3 + 2 C_4 - \frac{1}{2} (C_9 -C_{10})] (M^h_{nfe}+ {M'}^h_{nfe} + M^h_{nfa}
+ {M'}^h_{nfa})\non &&
+[C_5 -\frac{1}{2} C_7] (M_{nfe}^{h,P_1} + M_{nfa}^{h,P_1} +
{M'}_{nfa}^{h,P_1}+{M'}_{nfe}^{h,P_1})+[2 C_6 + \frac{1}{2} C_8] (M_{nfe}^{h,P_2}+ {M'}_{nfe}^{h,P_2}\non &&
+ M_{nfa}^{h,P_2}
+ {M'}_{nfa}^{h,P_2})
\biggr\}\cdot  \zeta_{f_{1q}}
-\lambda_t^d \biggl\{ [a_3 -a_5 + \frac{1}{2}
(a_7 - a_9)] f_{f_{1s}} {F'}^h_{fe}
+ [C_4 -\frac{1}{2} C_{10}] \non && \times
{M'}^h_{nfe} +[C_6 - \frac{1}{2} C_8]
{M'}_{nfe}^{h,P_2} \biggr\}\cdot \zeta_{f_{1s}}
\label{eq:f1285ome-d};
\eeq
\beq
 A^h(B^0 \to f_1(1285) \phi) &=&
-\lambda_t^d \biggl\{ [a_3 -a_5 + \frac{1}{2} (a_7 - a_9)]
f_{\phi} F^h_{fe} \zeta_{f_{1q}}
+[a_3+a_5- \frac{1}{2} (a_7 + a_9)]
(f_B F^h_{fa}+ f_B {F'}^h_{fa}) \zeta_{f_{1s}}
  \non && +[C_4 -\frac{1}{2} C_{10}] \biggl(M^h_{nfe}
\zeta_{f_{1q}}
+(M^h_{nfa}+{M'}^h_{nfa}) \zeta_{f_{1s}} \biggr)
+[C_6 - \frac{1}{2} C_8] \biggl(
(M_{nfa}^{h,P_2} +{M'}_{nfa}^{h,P_2}) \zeta_{f_{1s}}
\non &&  +M_{nfe}^{h,P_2} \zeta_{f_{1q}} \biggr) \biggr\}
\label{eq:f1285phi-d};
\eeq

\item[3.]{$B_s^0 \to f_1(1285) (\rho^0, \bar{K}^{*0}, \omega, \phi)$ decays}
\beq
\sqrt{2} A^h(B_s^0 \to f_1(1285) \rho^0) &=&\lambda_u^s
\biggl\{ a_2 \biggr(f_\rho F^h_{fe}\zeta_{f_{1s}}
+ (f_B F^h_{fa}+ f_B {F'}^h_{fa})\zeta_{f_{1q}}
\biggl)
 + C_2 \biggr(M^h_{nfe} \zeta_{f_{1s}} + (M^h_{nfa} + {M'}^h_{nfa})
 \non &&\cdot \zeta_{f_{1q}} \biggl)
\biggr\}  - \lambda_t^s\cdot \frac{3}{2} \cdot
\biggl\{[a_{9}- a_{7}] f_\rho F^h_{fe}
\zeta_{f_{1s}} + [ a_7+ a_{9}](f_B F^h_{fa} + f_B {F'}^h_{fa})
\zeta_{f_{1q}} + C_{10}
\non &&  \times
\biggr(M^h_{nfe}\zeta_{f_{1s}}
+ (M^h_{nfa}
+ {M'}^h_{nfa})\zeta_{f_{1q}}
\biggl) +
C_8 \biggr(M_{nfe}^{h,P_2}\zeta_{f_{1s}} + (M_{nfa}^{h,P_2}
+ {M'}_{nfa}^{h,P_2})\zeta_{f_{1q}} \biggl) \biggr\}
\label{eq:f1285rho-s};
\eeq
\beq
 A^h(B_s^0 \to f_1(1285) \bar K^{*0}) &=&
\lambda_u^d \biggl\{
a_2 f_{f_{1q}} {F'}^h_{fe}
 +C_2 {M'}^h_{nfe}
\biggr\}\cdot \zeta_{f_{1q}}
- \lambda_t^d \biggl\{ [a_4 -\frac{1}{2} a_{10}] \biggl( (f_{K^*}
F^h_{fe}+
f_B F^h_{fa})\cdot \zeta_{f_{1s}}\non &&
+ f_B {F'}^h_{fa}
\cdot \zeta_{f_{1q}} \biggr)+
[a_6 -\frac{1}{2} a_8] \biggl(f_B F_{fa}^{h,P_2}
\zeta_{f_{1s}}
+ f_B {F'}_{fa}^{h,P_2}\zeta_{f_{1q}}
\biggr) + [C_3-\frac{1}{2} C_9] \non &&
\times \biggl(
 {M'}^h_{nfa}\zeta_{f_{1q}}
+(M^h_{nfe}+ M^h_{nfa})
\zeta_{f_{1s}}  \biggr)
+ [C_5 -\frac{1}{2} C_7]  \biggl((M_{nfe}^{h,P_1} + M_{nfa}^{h,P_1}) \zeta_{f_{1s}}  +\zeta_{f_{1q}}\non &&
\times {M'}_{nfa}^{h,P_1}  \biggr)
+ \biggl(  [2a_3 + a_4
-2 a_5 -\frac{1}{2} (a_7 - a_9+ a_{10})] f_{f_{1q}} {F'}^h_{fe}
+ [ C_3 + 2 C_4- \frac{1}{2} \non &&
\times (C_9 - C_{10})] {M'}^h_{nfe}
+[C_5 - \frac{1}{2} C_7] {M'}_{nfe}^{h,P_1}
+[2 C_6 +\frac{1}{2} C_8] {M'}_{nfe}^{h,P_2}\biggr)
\zeta_{f_{1q}}
+ \biggl( [a_3  \non &&  -a_5
+ \frac{1}{2} (a_7 - a_9 )]f_{f_{1s}} {F'}^h_{fe}
+ [C_4  -\frac{1}{2} C_{10}] {M'}^h_{nfe}
+[C_6 - \frac{1}{2} C_8]  {M'}_{nfe}^{h,P_2}\biggr)
\zeta_{f_{1s}} \biggr\}
\label{eq:f1285kst0b};
\eeq
\beq
\sqrt{2} A^h(B_s^0 \to f_1(1285) \omega) &=&
\biggl\{
\zeta_{f_{1s}}
\cdot (a_2 f_{\omega} F^h_{fe}+C_2 M^h_{nfe})
+ \zeta_{f_{1q}} \cdot \biggl( a_2 (f_B F^h_{fa}+ f_B {F'}^h_{fa})
 + C_2 ( M^h_{nfa} + {M'}^h_{nfa} )\biggr)
\biggr\}\non &&
\times \lambda_u^s - \lambda_t^s \biggl\{
\zeta_{f_{1q}}  \cdot \biggl( (2C_4 + \frac{1}{2} C_{10})
(M^h_{nfa} + {M'}^h_{nfa})
+ (2C_6 + \frac{1}{2} C_{8}) (M_{nfa}^{h,P_2} + {M'}_{nfa}^{h,P_2})
\non &&
+ (2a_3+2a_5+\frac{1}{2}(a_7+a_9)) (f_B F^h_{fa}
+ f_B {F'}^h_{fa}) \biggr)
+ \zeta_{f_{1s}}
\cdot  \biggl( (2a_3 -2a_5 - \frac{1}{2}(a_7 - a_9))
\non && \times f_{\omega} F^h_{fe}
+(2C_4 + \frac{1}{2} C_{10}) M^h_{nfe}
+ (2C_6 + \frac{1}{2} C_8) M_{nfe}^{h,P_2} \biggr) \biggr\}
\label{eq:f1285ome-s};
\eeq
\beq
A^h(B_s^0 \to f_1(1285) \phi) &=&
\lambda_u^s \biggl\{ \zeta_{f_{1q}}
\cdot (a_2 f_{f_{1q}} {F'}^h_{fe}+ C_2
{M'}^h_{nfe}) \biggr\}
- \lambda_t^s \biggl\{
\zeta_{f_{1s}}
\cdot\biggl( (a_3+a_4-a_5+\frac{1}{2}
(a_7-a_9-a_{10}))
\non && \times
(f_{\phi} F^h_{fe} + f_{f_{1s}} {F'}^h_{fe} )
+(a_6-\frac{1}{2}a_8) (f_B F_{fa}^{h,P_2} + f_B {F'}_{fa}^{h,P_2} ) +(C_3 +C_4-\frac{1}{2}(C_9+C_{10}))
\non && \times
(M^h_{nfe} + {M'}^h_{nfe}+M^h_{nfa} + {M'}^h_{nfa})
+(C_5
- \frac{1}{2} C_7) (M_{nfe}^{h,P_1} + {M'}_{nfe}^{h,P_1}
+M_{nfa}^{h,P_1} + {M'}_{nfa}^{h,P_1})\non &&
+(C_6 - \frac{1}{2} C_8) (M_{nfe}^{h,P_2} + {M'}_{nfe}^{h,P_2}
+M_{nfa}^{h,P_2} + {M'}_{nfa}^{h,P_2})
+(a_3+a_4+a_5-\frac{1}{2}(a_7+a_9+a_{10}))
\non &&\times
(f_B F^h_{fa} +f_B {F'}^h_{fa}) \biggr)
+ \zeta_{f_{1q}}
\cdot \biggl( (2a_3-2a_5 - \frac{1}{2}(a_7 - a_9)) f_{1s} {F'}^h_{fe}
+(2C_4
+ \frac{1}{2} C_{10}) {M'}^h_{nfe} \non &&  + (2C_6 + \frac{1}{2} C_8) {M'}_{nfe}^{h,P_2} \biggr) \biggr\}
\label{eq:f1285phi-s};
\eeq

\end{itemize}
where $\lambda_u^{d(s)}=V^*_{ub} V_{ud(s)}$ and  $\lambda_t^{d(s)}=
V^*_{tb} V_{td(s)}$, and $\zeta_{f_{1q}} = \cos\phi_{f_1}/\sqrt{2}$ and
$\zeta_{f_{1s}}= -\sin\phi_{f_1}$.
Also, $a_i$ is the standard combination
of the Wilson coefficients $C_i$ defined as follows:
\beq
a_1 &=& C_2 + \frac{C_1}{3};\quad
a_2 = C_1 + \frac{C_2}{3};
\quad a_i= C_i+C_{i\pm 1}/3,\quad  i=3-10.
\eeq
where $C_2 \sim 1$ is
the largest one among all the Wilson coefficients and the upper (lower)
sign applies, when $i$ is odd (even).
When we make the replacements with  $\zeta_{f_{1q}} \to \zeta'_{f_{1q}} = \sin\phi_{f_1}
/\sqrt{2}$ and $\zeta_{f_{1s}} \to \zeta'_{f_{1s}} = \cos\phi_{f_1}$ in the above equations, i.e., Eqs.~(\ref{eq:f1285rhop})-(\ref{eq:f1285phi-s}), the decay amplitudes of other 10 $B \to f_1(1420) V$ modes
will be straightforwardly obtained.

\section{Numerical Results and Discussions} \label{sec:randd}

In this section, we will present numerically the pQCD predictions
of the {\it CP}-averaged branching ratios, the polarization fractions, the {\it CP}-violating asymmetries, and the relative
phases for those considered 20 nonleptonic
$B \to f_1 V$ decays. Some
comments are essentially given on the input quantities
for axial-vector $f_1$ states:
\begin{itemize}
\item[]{(a) $f_{1q(s)}$ state distribution amplitude}

In light of the similar
behavior between vector and $^3\!P_1$-axial-vector mesons~\cite{Yang:2007zt} and the same form
for $\rho$ and $\omega$ distribution amplitudes in the
vector meson sector but with different decay constants $f_{\rho}$ and $f_{\omega}$, we argue that the $f_{1q}$ distribution amplitude can be taken with the
same one as that of the $a_1(1260)$ state with decay constant
$f_{f_{1q}} = 0.193$~GeV
~\cite{Verma:2011yw}. While, for simplicity, we adopt the same distribution amplitude as the flavor singlet $f_1$ state [not to be confused with
the abbreviation $f_1$ of $f_1(1285)$ and $f_1(1420)$ mesons]~\cite{Liu:2014doa} for the $f_{1s}$ state
with decay constant $f_{f_{1s}}= 0.230$~GeV~\cite{Verma:2011yw}.

\item[]{(b) $f_{1q(s)}$ state mass and mixing angle}

As mentioned in the Introduction, the value of
the mixing angle $\phi_{f_1}=(24.0^{+3.2}_{-2.7})^\circ$
has been
measured preliminarily by the LHCb Collaboration in 2013
in the heavy $b$ flavor sector~\cite{Aaij:2013rja}. Because of the good agreement between this measurement
and the latest update $(27 \pm 2)^\circ$ in lattice QCD
calculations~\cite{Dudek:2013yja}, we will adopt experimental
data $\phi_{f_1}= 24.0^\circ$ to predict the quantities numerically in this work.
On the other hand, as exhibited in Ref.~\cite{Liu:2014jsa}, the predictions of $Br(B^{+,0} \to A P)_{\rm pQCD}$ with the measured angle are generally consistent with those $Br(B^{+,0} \to A P)_{\rm QCDF}$ based on the same mixing
matrix for the $f_1(1285)-f_1(1420)$ system with $\alpha_{^3\!P_1} \sim 18^\circ$, i.e., the second entry $\theta_{^3\!P_1} \sim 53^\circ$
in the flavor singlet-octet basis~\cite{Cheng:2007mx}. Moreover, for the masses of two $f_{1q}$ and $f_{1s}$ states, we adopt $m_{f_{1q}} \sim m_{f_1(1285)}$
and $m_{f_{1s}} \sim m_{f_1(1420)}$ for convenience.
\end{itemize}

In numerical calculations, central values of the input parameters will be
used implicitly unless otherwise stated. The relevant QCD scale~({\rm GeV}), masses~({\rm GeV}),
and $B$ meson lifetime({\rm ps}) are the following
~\cite{Aaij:2013rja,Keum:2000ph,Yang:2007zt,Verma:2011yw,Beringer:1900zz}
\beq
 \Lambda_{\overline{\mathrm{MS}}}^{(f=4)} &=& 0.250\; , \quad m_W = 80.41\;,
 \quad  m_{B}= 5.28\;, \quad  m_{B_s}= 5.37\;, \quad  m_b = 4.8 \;; \non
  f_{f_{1q}}&=& 0.193^{+0.043}_{-0.038}\;,
  \quad f_{f_{1s}} = 0.230 \pm 0.009\;,
\quad m_{f_{1q}}= 1.28\;,
\quad m_{f_{1s}}= 1.42\;;
\non
  \tau_{B^+} &=& 1.641\;,  \quad \tau_{B^0}= 1.519\;,
   \quad  \tau_{B_s^0}= 1.497\;, \quad \phi_{f_1} = (24.0^{+3.2}_{-2.7})^\circ\;.
\label{eq:mass}
\eeq

For the CKM matrix elements, we adopt the Wolfenstein
parametrization at leading order~\cite{Wolfenstein:1983yz}
and the updated parameters
$A=0.814$, $\lambda=0.22537$, $\bar{\rho}=0.117 \pm 0.021$,
and $\bar{\eta}=0.353\pm 0.013$~\cite{Agashe:2014kda}.

\subsection{{\it CP}-averaged branching ratios}

For the considered $B \to f_1 V$ decays,
the decay rate can be written as
\beq
\Gamma =\frac{G_{F}^{2}|\bf{P_c}|}{16 \pi m^{2}_{B} }
\sum_{\sigma=L,N,T} A^{(\sigma)\dagger }  A^{(\sigma)}\;
\label{dr1}
\eeq
where $|\bf{P_c}|\equiv |\bf{P_{2z}}|=|\bf{P_{3z}}|$ is the momentum of either the
outgoing axial-vector meson or vector meson and $A^{(\sigma)}$ can be found, for example, in Eqs.~(\ref{eq:f1285rhop})-(\ref{eq:f1285phi-s}).
Using the decay amplitudes obtained in last section, it is straightforward to calculate the {\it CP}-averaged
branching ratios with uncertainties for the considered decays in the pQCD approach.

\begin{table}[hbt]
\caption{ Theoretical predictions of physical quantities of $B^+ \to f_1 \rho^+$ decays obtained in the pQCD approach with mixing angle
$\phi_{f_1} = 24^\circ$ in the quark-flavor($f_{1q}-f_{1s}$) basis. For comparison, we also quote the estimations in the framework of QCDF approach with mixing angle $\theta_{^3\!P_1} \sim 53^\circ$
in the flavor
singlet-octet($f_1-f_8$) basis.}
\label{tab:f1rhop}
 \begin{center}\vspace{-0.3cm}{
\begin{tabular}[t]{c|c||c|c||c|c}
\hline  \hline
   \multicolumn{2}{c||}{Decay Modes}   &  \multicolumn{2}{c||}{$B^+ \to f_1(1285)\rho^+$} &  \multicolumn{2}{c}{$B^+ \to f_1(1420) \rho^+$} \\
   \hline
 Parameter  & Definition & This work &   QCDF~\cite{Cheng:2008gxa}
 & This work &   QCDF~\cite{Cheng:2008gxa}  \\
\hline \hline
  BR($10^{-6}$)        & $\Gamma/ \Gamma_{\rm total}$
  &$11.1^{+3.2+5.4+6.0+0.4+0.2+0.8}_{-2.5-4.0-4.8-0.6-0.3-0.9}
  $
  &$8.9^{+5.1+0.4}_{-3.2-0.3}$
  &$\hspace{0.18cm}
 2.3^{+0.7+1.1+1.2+0.6+0.0+0.2}_{-0.5-0.8-0.9-0.4-0.0-0.2}
  $
  &$1.3^{+0.6+0.2}_{-0.3-0.0}$
 \\
 \hline \hline
 $f_L(\%)$      & $|{\cal A}_L|^2$
 &$96.3^{+0.2+0.2+0.4+0.0+0.1+0.0}_{-0.1-0.2-0.3-0.0-0.1-0.0}
  $
 &$90^{+4}_{-3}$
 &$90.5^{+0.0+1.7+1.8+1.2+1.2+0.7}_{-0.1-2.5-3.7-1.4-1.8-0.8}
  $
 &$93^{+4}_{-3}$
 \\
 $f_{||}(\%)$   & $|{\cal A}_{||}|^2$
 &$\hspace{0.18cm}
 2.3^{+0.0+0.1+0.2+0.0+0.1+0.0}_{-0.1-0.1-0.2-0.0-0.1-0.0}
  $
 &$-$
 &$\hspace{0.18cm}
 5.5^{+0.0+1.3+2.0+0.7+1.0+0.4}_{-0.1-0.9-1.1-0.7-0.7-0.4}
  $
 &$-$
  \\
 $f_{\perp}(\%)$& $|{\cal A}_\perp|^2$
 &$\hspace{0.18cm}
 1.4^{+0.1+0.1+0.1+0.0+0.1+0.0}_{-0.1-0.1-0.1-0.0-0.0-0.0}
$
 &$-$
 &$\hspace{0.18cm}
 4.1^{+0.0+1.1+1.6+0.6+0.8+0.3}_{-0.1-0.9-0.9-0.6-0.6-0.4}
  $
 &$-$
 \\
 \hline \hline
 $\phi_{||}$(rad)& $\arg\frac{{\cal A}_{||}}{{\cal A}_L}$
 &$\hspace{0.18cm}
 3.1^{+0.0+0.0+0.0+0.0+0.0+0.0}_{-0.0-0.1-0.1-0.0-0.0-0.0}
$
 &$-$
 &$\hspace{0.18cm}
 3.1^{+0.1+0.1+0.2+0.1+0.1+0.0}_{-0.0-0.0-0.0-0.0-0.0-0.0}
  $
 &$-$
 \\
 $\phi_{\perp}$(rad)& $\arg\frac{{\cal A}_{\perp}}{{\cal A}_L}$
 &$\hspace{0.18cm}
 3.1^{+0.0+0.0+0.1+0.0+0.0+0.0}_{-0.0-0.0-0.0-0.0-0.0-0.0}
  $
 &$-$
 &$\hspace{0.18cm}
 3.2^{+0.0+0.0+0.0+0.0+0.0+0.0}_{-0.0-0.0-0.1-0.0-0.0-0.0}
  $
 &$-$
  \\
  \hline \hline
 $\acp^{\rm dir}(\%)$& $\frac{\overline{\Gamma}-\Gamma}{\overline{\Gamma}+\Gamma}$
 &$-6.7^{+0.1+0.3+2.1+0.1+0.5+0.4}_{-0.0-0.2-2.9-0.0-0.5-0.3}
  $
 &$-$
 &$-3.7^{+0.4+0.7+1.8+0.3+0.6+0.1}_{-0.4-0.7-2.1-0.4-0.8-0.1}
  $
 &$-$
 \\
 $\acp^{\rm dir}(L)(\%)$& $\frac{\bar{f}_L-f_L}{\bar{f}_L+f_L}$
 &$-7.0^{+0.1+0.1+2.1+0.1+0.5+0.4}_{-0.0-0.1-2.8-0.0-0.6-0.3}
  $
 & $-$
 &$-5.4^{+0.7+0.4+1.8+0.2+1.0+0.2}_{-0.6-0.4-2.1-0.2-1.4-0.3}
  $
 & $-$
 \\
 $\acp^{\rm dir}(||)(\%)$& $\frac{\bar{f}_{||}-f_{||}}{\bar{f}_{||}+f_{||}}$
 &$\hspace{0.18cm}
 0.7^{+0.6+2.8+2.8+0.7+2.1+0.0}_{-0.4-3.5-3.8-0.8-1.2-0.0}
  $
 &$-$
 &$13.8^{+1.6+3.7+10.9+0.4+0.8+0.7}_{-1.8-3.7-11.0-0.6-0.6-0.8}
  $
 &$-$
 \\
 $\acp^{\rm dir}(\perp)(\%)$& $\frac{\bar{f}_\perp-f_\perp}{\bar{f}_\perp+f_\perp}$
 &$\hspace{0.18cm}
 1.3^{+0.7+3.0+3.0+0.7+2.4+0.1}_{-0.5-3.9-4.1-0.8-1.3-0.0}
$
 &$-$
 &$10.5^{+2.5+4.0+11.9+0.5+0.5+0.5}_{-3.2-3.9-12.2-0.6-0.3-0.6}
  $
 &$-$
 \\ \hline \hline
\end{tabular}}
\end{center}
\end{table}

The numerical results of the physical quantities are presented in Tables~\ref{tab:f1rhop}-\ref{tab:f1phi-s},
in which the six major errors are induced by the uncertainties
of the shape parameter
$\omega_b = 0.40 \pm 0.04\ (\omega_b = 0.50 \pm 0.05)$~GeV
in the $B^{+,0}\ (B_s^0)$ meson wave function; of the combined
decay constants $f_{M}$ from the $^3\!P_1$-axial-vector state
decay constants $f_{f_{1q}}=0.193^{+0.043}_{-0.038}$ and
$f_{f_{1s}}=0.230 \pm 0.009$~GeV and
vector meson decay constants
$f_V$ and $f_V^T$;
of the combined Gegenbauer moments $a_i^M$ from
$a_2^{\parallel}$ and $a_1^{\perp}$
for the axial-vector $f_{1q}$ and $f_{1s}$ states
and from $a_{(1)2V}^{\parallel,\perp}$ for the light
vector meson in both longitudinal and
transverse polarizations; of the mixing angle $\phi_{f_1}= (24.0^{+3.2}_{-2.7})^\circ$ for the $f_1(1285)-f_1(1420)$
mixing system; of the maximal
running hard scale $t_{\rm max}$;
and of the combined CKM matrix elements from
parameters $\bar \rho$ and $\bar \eta$, respectively.
It is worth mentioning that, though parts of next-to-leading order
corrections to two-body hadronic $B$ meson decays have been
proposed in the pQCD approach~\cite{Li:2010nn,Liu:2015sra}, the higher order QCD contributions
to $B \to VV$ modes beyond leading order are not yet available presently.
Therefore, as displayed in the above-mentioned tables,
the higher order contributions in this work are simply
investigated by exploring
the variation of hard scale $t_{\rm max}$, i.e., from $0.8t$ to $1.2t$
(not changing $1/b_i, i= 1,2,3$), in the hard kernel, which have
been counted into one of the sources of theoretical uncertainties.
It looks like the penguin-dominated decays such as
$B^{+,0} \to f_1 K^{*+,0}$, $B^0 \to f_1 \phi$, and
$B_s^0 \to f_1 (\bar K^{*0}, \omega, \phi)$
are more sensitive to the potential
higher order corrections, as can be clearly seen in
Tables~\ref{tab:f1kstp}, \ref{tab:f1kst0}, \ref{tab:f1phi-d}, \ref{tab:f1kst0b}, \ref{tab:f1omg-s}, and
\ref{tab:f1phi-s},
correspondingly.

\begin{table}[hbt]
\caption{ Same as Table~\ref{tab:f1rhop} but for $B^+ \to f_1 K^{*+}$ decays.}
\label{tab:f1kstp}
 \begin{center}\vspace{-0.3cm}{
\begin{tabular}[t]{c|c||c|c||c|c}
\hline  \hline
   \multicolumn{2}{c||}{Decay Modes}   &  \multicolumn{2}{c||}{$B^+ \to f_1(1285) K^{*+}$} &  \multicolumn{2}{c}{$B^+ \to f_1(1420) K^{*+}$}  \\
   \hline
 Parameter  & Definition & This work &   QCDF~\cite{Cheng:2008gxa}
 & This work &   QCDF~\cite{Cheng:2008gxa}  \\
\hline \hline
  BR($10^{-6}$)        & $\Gamma/ \Gamma_{\rm total}$
  &$6.4^{+0.5+2.4+1.6+0.3+2.1+0.1}_{-0.3-1.7-1.3-0.2-1.2-0.0}
  $
  &$5.7^{+3.8+21.4}_{-2.2-4.8}$
  &$\hspace{0.18cm}
 4.5^{+0.7+0.4+1.3+0.2+0.8+0.0}_{-0.6-0.4-1.2-0.3-0.5-0.1}
  $
  &$15.6^{+10.9+10.4}_{-5.2-4.7}$
 \\
 \hline \hline
 $f_L(\%)$      & $|{\cal A}_L|^2$
 &$23.5^{+0.8+2.3+4.8+1.3+1.8+0.5}_{-0.5-1.6-3.2-1.0-1.3-0.5}
  $
 &$47^{+49}_{-45}$
 &$69.3^{+1.0+0.9+10.2+0.5+4.8+0.4}_{-1.2-1.3-10.4-0.6-6.6-0.3}
  $
 &$64^{+37}_{-61}$
 \\
 $f_{||}(\%)$   & $|{\cal A}_{||}|^2$
 &$42.1^{+0.2+0.9+1.8+0.6+0.8+0.3}_{-0.4-1.2-2.4-0.7-1.0-0.2}
  $
 &$-$
 &$16.5^{+0.8+0.8+5.9+0.4+3.5+0.2}_{-0.6-0.7-5.7-0.4-2.6-0.2}
  $
 &$-$
  \\
 $f_{\perp}(\%)$& $|{\cal A}_\perp|^2$
 &$34.4^{+0.2+0.7+1.5+0.4+0.6+0.2}_{-0.4-1.1-2.4-0.6-0.8-0.2}
$
 &$-$
 &$14.2^{+0.5+0.5+3.8+0.2+3.0+0.2}_{-0.4-0.3-4.4-0.1-2.2-0.2}
  $
 &$-$
 \\
 \hline \hline
 $\phi_{||}$(rad)& $\arg\frac{{\cal A}_{||}}{{\cal A}_L}$
 &$\hspace{0.18cm}
 4.4^{+0.0+0.1+0.1+0.0+0.1+0.1}_{-1.3-0.2-1.8-0.0-0.2-0.1}
$
 &$-$
 &$\hspace{0.18cm}
 3.6^{+0.1+0.2+0.3+0.1+0.1+0.1}_{-0.0-0.1-0.1-0.0-0.1-0.0}
  $
 &$-$
 \\
 $\phi_{\perp}$(rad)& $\arg\frac{{\cal A}_{\perp}}{{\cal A}_L}$
 &$\hspace{0.18cm}
 4.4^{+0.0+0.1+0.1+0.0+0.1+0.1}_{-1.3-0.2-1.8-0.0-0.2-0.1}
  $
 &$-$
 &$\hspace{0.18cm}
 3.6^{+0.0+0.1+0.2+0.0+0.1+0.0}_{-0.1-0.1-0.3-0.1-0.1-0.0}
  $
 &$-$
  \\
  \hline \hline
 $\acp^{\rm dir}(\%)$& $\frac{\overline{\Gamma}-\Gamma}{\overline{\Gamma}+\Gamma}$
 &$-16.0^{+0.9+1.0+4.4+0.3+2.3+0.5}_{-0.9-0.9-4.2-0.3-2.2-0.5}
  $
 &$-$
 &$13.9^{+0.9+3.0+3.7+2.0+0.5+0.5}_{-0.8-2.8-4.0-1.7-0.8-0.4}
  $
 &$-$
 \\
 $\acp^{\rm dir}(L)(\%)$& $\frac{\bar{f}_L-f_L}{\bar{f}_L+f_L}$
 &$-94.5^{+3.3+7.3+20.7+4.1+8.0+1.4}_{-1.1-4.4-3.7-2.8-4.0-1.2}
  $
 & $-$
 &$25.4^{+1.1+4.9+2.3+3.4+1.5+1.0}_{-0.9-4.7-3.7-2.8-1.1-0.9}
  $
 & $-$
 \\
 $\acp^{\rm dir}(||)(\%)$& $\frac{\bar{f}_{||}-f_{||}}{\bar{f}_{||}+f_{||}}$
 &$\hspace{0.18cm}
 8.2^{+0.3+0.5+2.1+0.1+1.0+0.3}_{-0.3-0.5-2.1-0.1-1.0-0.3}
  $
 &$-$
 &$-14.1^{+1.1+3.0+4.9+1.8+2.2+0.5}_{-1.1-2.9-5.6-2.1-2.1-0.6}
  $
 &$-$
 \\
 $\acp^{\rm dir}(\perp)(\%)$& $\frac{\bar{f}_\perp-f_\perp}{\bar{f}_\perp+f_\perp}$
 &$\hspace{0.18cm}
 7.9^{+0.4+0.6+2.1+0.1+0.8+0.3}_{-0.3-0.4-2.0-0.1-0.9-0.2}
$
 &$-$
 &$-9.7^{+1.0+2.2+4.1+1.3+1.5+0.4}_{-0.9-2.0-4.0-1.4-1.4-0.3}
  $
 &$-$
 \\ \hline \hline
\end{tabular}}
\end{center}
\end{table}

\begin{enumerate}

\item[(1)]
According to the effective Hamiltonian shown in Eq.~(\ref{eq:heff}),
the considered 20 nonleptonic $B \to f_1 V$ decays contain two kinds of transitions, i.e., the $\bar b \to \bar d$ one with $\Delta S = 0$ and the 
$\bar b \to \bar s$ one with $\Delta S = 1$(here, the capital $S$ describes strange flavor number), in which
$B^{+,0} \to f_1 (\rho, \omega, \phi)$ and $B_s^0 \to f_1 \bar{K}^{*0}$
belong to the former class, while $B^{+,0} \to f_1 K^{*+,0}$ and $B_s^0 \to f_1 (\rho, \omega, \phi)$ are classified into the latter one. Also, in principle, if the decays with
these two kinds of transitions are dominated by the penguin amplitudes,
it can be roughly anticipated that
because $|\lambda_t^d|:|\lambda_t^s| \sim 0.22:1$
in magnitude, $Br(B \to f_1 V)_{\bar b \to \bar d}$ is basically
less than $Br(B \to f_1 V)_{\bar b \to \bar s}$.
Undoubtedly, the tree-dominated $B^+ \to f_1 \rho^+$ modes are exceptional.
A convincing example is directly observed from the ratios between $B^0 \to f_1 K^{*0}$ and $B_s^0 \to f_1 \bar K^{*0}$ decay rates. From the numerical branching ratios predicted in the pQCD
approach as given in Tables
~\ref{tab:f1kst0} and \ref{tab:f1kst0b},  the ratios $R^{d/s}_{f_{(1285)} K^*}$ and $R^{d/s}_{f_{(1420)} K^*}$ can be written as
\beq
R^{d/s}_{f_1{(1285)}K^*} &\equiv& \frac{Br(B^0 \to f_1(1285) K^{*0})_{\rm pQCD}}{Br(B_s^0 \to f_1(1285) \bar K^{*0})_{\rm pQCD}} \sim 9\;,
\quad
R^{d/s}_{f_1{(1420)}K^*} \equiv \frac{Br(B^0 \to f_1(1420) K^{*0})_{\rm pQCD}}{Br(B_s^0 \to f_1(1420) \bar K^{*0})_{\rm pQCD}} \sim 13\;,
\eeq
where, for the sake of simplicity, only central values are quoted
for clarification. The difference between these two ratios $R^{d/s}_{f_1(1285)K^*}$ and $R^{d/s}_{f_1(1420)K^*}$ is mainly induced
by the fact that $f_1(1285)[f_1(1420)]$ has a dominant $u\bar u+ d\bar d(s\bar s)$ component with $\cos\phi \sim 0.9$, which confirms
somewhat large tree contaminations in $B_{d/s} \to f_1(1285) K^{*0}$
decays. Numerically, in terms of central values, $Br(B^0 \to f_1(1285)
[f_1(1420)] K^{*0})$ varies from $4.96(4.37) \times 10^{-6}$ to
$5.08(4.34) \times 10^{-6}$, while $Br(B_s^0 \to f_1(1285)
[f_1(1420)] \bar K^{*0})$ changes from $5.47(3.40) \times 10^{-7}$ to
$1.99(2.84) \times 10^{-7}$ by neglecting the tree contributions.

\begin{table}[t]
\caption{ Same as Table~\ref{tab:f1rhop} but for $B^0 \to f_1 \rho^0$ decays.}
\label{tab:f1rho0-d}
 \begin{center}\vspace{-0.3cm}{
\begin{tabular}[t]{c|c||c|c||c|c}
\hline  \hline
   \multicolumn{2}{c||}{Decay Modes}   &  \multicolumn{2}{c||}{$B^0 \to f_1(1285) \rho^0$} &  \multicolumn{2}{c}{$B^0 \to f_1(1420) \rho^0$}  \\
   \hline
 Parameter  & Definition & This work &   QCDF~\cite{Cheng:2008gxa}
 & This work &   QCDF~\cite{Cheng:2008gxa}  \\
\hline \hline
  BR($10^{-7}$)        & $\Gamma/ \Gamma_{\rm total}$
  &$1.1^{+0.3+0.5+0.8+0.1+0.1+0.1}_{-0.2-0.3-0.2-0.0-0.0-0.0}
  $
  &$2.0^{+1.0+3.0}_{-1.0-0.0}$
  &$\hspace{0.18cm}
 0.7^{+0.2+0.1+0.1+0.0+0.2+0.0}_{-0.2-0.1-0.1-0.0-0.2-0.0}
  $
  &$0.4^{+1.2+0.8}_{-0.3-0.0}$
 \\
 \hline \hline
 $f_L(\%)$      & $|{\cal A}_L|^2$
 &$90.5^{+0.1+1.6+5.4+0.0+1.1+0.7}_{-0.0-2.0-12.8-0.3-1.1-0.8}
  $
 &$71^{+9}_{-36}$
 &$\hspace{0.18cm}
7.2^{+1.8+3.7+5.6+2.0+4.1+0.1}_{-1.1-2.4-1.7-1.7-2.6-0.0}
  $
 &$87^{+8}_{-40}$
 \\
 $f_{||}(\%)$   & $|{\cal A}_{||}|^2$
 &$\hspace{0.03cm}
4.5^{+0.0+1.1+6.8+0.2+0.5+0.4}_{-0.1-0.8-2.1-0.1-0.5-0.4}
  $
 &$-$
 &$49.3^{+0.4+1.0+0.9+0.7+1.3+0.0}_{-0.9-1.8-2.8-1.0-2.1-0.1}
  $
 &$-$
  \\
 $f_{\perp}(\%)$& $|{\cal A}_\perp|^2$
 &$\hspace{0.03cm}
5.0^{+0.1+1.0+6.1+0.1+0.6+0.4}_{-0.1-0.8-3.3-0.0-0.6-0.4}
$
 &$-$
 &$43.5^{+0.7+1.5+0.7+1.0+1.3+0.1}_{-0.9-1.9-2.8-1.0-2.0-0.0}
  $
 &$-$
 \\
 \hline \hline
 $\phi_{||}$(rad)& $\arg\frac{{\cal A}_{||}}{{\cal A}_L}$
 &$\hspace{0.03cm}
3.3^{+0.1+0.3+0.4+0.1+0.1+0.0}_{-0.0-0.1-0.1-0.0-0.0-0.0}
$
 &$-$
 &$\hspace{0.18cm}
3.5^{+0.0+0.4+0.2+0.2+0.1+0.0}_{-0.0-0.1-0.4-0.1-0.0-0.0}
  $
 &$-$
 \\
 $\phi_{\perp}$(rad)& $\arg\frac{{\cal A}_{\perp}}{{\cal A}_L}$
 &$\hspace{0.03cm}
3.3^{+0.1+0.2+0.4+0.1+0.1+0.1}_{-0.0-0.0-0.1-0.0-0.0-0.0}
  $
 &$-$
 &$\hspace{0.18cm}
 3.5^{+0.0+0.4+0.2+0.2+0.1+0.0}_{-0.0-0.1-0.3-0.1-0.0-0.0}
  $
 &$-$
  \\
  \hline \hline
 $\acp^{\rm dir}(\%)$& $\frac{\overline{\Gamma}-\Gamma}{\overline{\Gamma}+\Gamma}$
 &$18.0^{+12.9+3.9+40.6+2.3+1.6+0.6}_{-12.0-4.5-27.5-2.6-1.4-0.6}
  $
 &$-$
 &$24.1^{+0.5+7.5+17.2+4.5+5.1+1.1}_{-0.4-6.7-22.4-3.7-5.4-1.3}
  $
 &$-$
 \\
 $\acp^{\rm dir}(L)(\%)$& $\frac{\bar{f}_L-f_L}{\bar{f}_L+f_L}$
 &$24.7^{+13.7+1.3+39.2+0.5+3.0+1.1}_{-12.7-1.5-32.5-0.5-2.9-1.0}
  $
 & $-$
 &$-72.5^{+24.1+27.2+29.5+16.1+19.2+2.8}_{-20.8-26.1-14.7-18.2-18.6-2.7}
  $
 & $-$
 \\
 $\acp^{\rm dir}(||)(\%)$& $\frac{\bar{f}_{||}-f_{||}}{\bar{f}_{||}+f_{||}}$
 &$ -56.6^{+4.9+31.4+40.2+19.5+5.5+2.6}_{-5.2-26.4-11.4-17.8-2.3-2.7}
  $
 &$-$
 &$\hspace{0.18cm}
29.8^{+0.6+6.6+20.4+3.7+3.1+1.4}_{-0.6-6.2-23.3-3.2-3.4-1.5}
  $
 &$-$
 \\
 $\acp^{\rm dir}(\perp)(\%)$& $\frac{\bar{f}_\perp-f_\perp}{\bar{f}_\perp+f_\perp}$
 &$-36.9^{+6.2+30.0+27.3+19.4+7.0+1.9}_{-6.6-30.8-11.8-20.3-3.5-1.9}
$
 &$-$
 &$\hspace{0.18cm}
 33.6^{+0.7+7.0+19.5+4.1+3.8+1.7}_{-0.9-6.6-22.8-7.0-4.0-1.6}
  $
 &$-$
 \\ \hline \hline
\end{tabular}}
\end{center}
\end{table}

\item[(2)]
Based on the theoretical predictions given at leading order in the pQCD approach, as collected in Tables~\ref{tab:f1rhop}-\ref{tab:f1phi-s}, large {\it CP}-averaged branching ratios of the order of $10^{-6}-10^{-5}$ can be found in the channels such as $B^+ \to f_1 (\rho^+, K^{*+})$, $B^0 \to f_1 K^{*0}$, $B^0 \to f_1(1285) \omega$, and $B_s^0 \to f_1 \phi$, which can be
detected at the LHCb and Belle-II experiments in the near future. Of course, relative to the $B_s^0 \to \phi \phi$ decay, it is of particular interest to study the $B_s-\bar B_s$
mixing phase and even possible NP
through the
detectable $B_s^0 \to f_1 \phi$ decays with large decay rates complementarily, which is mainly because these two modes contain
the tiny and safely negligible tree pollution. More relevant 
discussions will be given below.

\item[(3)]
From Table~\ref{tab:f1rhop}, one can easily find that the {\it CP}-averaged branching ratios of color-allowed tree-dominated
$B^+ \to f_1 \rho^+$ decays are
\beq
Br(B^+ \to f_1(1285) \rho^+)_{\rm pQCD} &=& 11.1^{+8.7}_{-6.8} \times 10^{-6}\;, \qquad \;\;
Br(B^+ \to f_1(1420) \rho^+)_{\rm pQCD} = 2.3^{+1.9}_{-1.4} \times 10^{-6}\;;
\eeq
where various errors arising from the input parameters have been added in quadrature. It is known that the $B^+ \to f_1 \rho^+$ decays are induced by the interferences between $B^+ \to f_{1q} \rho^+$ and $f_{1s} \rho^+$ modes. The values of the branching ratios indicate a constructive(destructive) interference in the
$B^+ \to f_1(1285)[f_1(1420)] \rho^+$ decay. In fact, due to the
dominance of $f_{1q}(f_{1s})$ in the $f_1(1285)[f_1(1420)]$ state,
it is therefore naturally expected that $Br(B^+ \to f_1(1285)[f_1(1420)] \rho^+)_{\rm pQCD}$ is more like $Br(B^+ \to \omega[\phi] \rho^+)$. However, relative to $B^+ \to \phi \rho^+$ decay, the $B^+ \to f_1(1420) \rho^+$ mode receives an extra and significant interference from the dominant factorizable $B^+ \to f_{1q}$ transition with a factor $(\sin\phi_{f_1}) \sim 0.4$, which finally results in a larger $Br(B^+ \to f_1(1420) \rho^+)$ than $Br(B^+ \to \phi \rho^+)$ as it should be. Careful analysis of the decay amplitudes with three polarizations presented in Table~\ref{tab:DAs-f1rho-u} confirms
the above-mentioned arguments.

The $B^+ \to f_1 \rho^+$ decays have been investigated within the framework of the QCDF approach\cite{Cheng:2008gxa}.~\footnote{In light of the crude predictions given in Ref.~\cite{Calderon:2007nw} and the
consistent results presented in Refs.~\cite{Cheng:2007mx} and
~\cite{Liu:2014jsa} for the branching ratios of $B \to f_1 P$
decays, we will mainly focus on the theoretical predictions of
$B^{+,0} \to f_1 V$ modes obtained with QCDF and make
comprehensive analyses and comparisons in this work.}
The branching ratios were predicted as follows:
\beq
Br(B^+ \to f_1(1285) \rho^+)_{\rm QCDF} &=& 8.9^{+5.1}_{-3.2} \times 10^{-6}\;, \qquad \;\;
Br(B^+ \to f_1(1420) \rho^+)_{\rm QCDF} = 1.3^{+0.6}_{-0.3} \times 10^{-6}\;; \label{eq:f1rhop-qcdf}
\eeq
where the errors are also added in quadrature. Note that, as
discussed in Ref.~\cite{Liu:2014jsa}, the QCDF predictions only  with the mixing angle $\theta_{^3\!P_1} \sim 53.2^\circ$ are
basically consistent with the pQCD ones for $B^{+,0} \to f_1 P$
decay rates. Therefore, as listed in Eq.~(\ref{eq:f1rhop-qcdf}),
we still quote the theoretical predictions
for $B \to f_1 V$ decays
with $\theta_{^3\!P_1} \sim 53.2^\circ$ to make
concrete comparisons with those in the pQCD approach.
One can easily observe the good agreement
of the $B^+ \to f_1 \rho^+$ decay rates predicted in both the 
QCDF and pQCD approaches within uncertainties.

\begin{table}[t]
\caption{ Same as Table~\ref{tab:f1rhop} but for $B^0 \to f_1 K^{*0}$ decays.}
\label{tab:f1kst0}
 \begin{center}\vspace{-0.3cm}{
\begin{tabular}[t]{c|c||c|c||c|c}
\hline  \hline
   \multicolumn{2}{c||}{Decay Modes}   &  \multicolumn{2}{c||}{$B^0 \to f_1(1285) K^{*0}$}  &  \multicolumn{2}{c}{$B^0 \to f_1(1420) K^{*0}$}\\
   \hline
 Parameter  & Definition & This work &   QCDF~\cite{Cheng:2008gxa}
 & This work &   QCDF~\cite{Cheng:2008gxa}  \\
\hline \hline
  BR($10^{-6}$)        & $\Gamma/ \Gamma_{\rm total}$
  &$\hspace{0.18cm}
 5.0^{+0.1+1.6+1.3+0.2+1.7+0.0}_{-0.2-1.3-1.2-0.2-1.1-0.1}
  $
  &$5.1^{+3.6+20.0}_{-2.1-4.7}$
  &$\hspace{0.18cm}
 4.4^{+0.6+0.4+1.4+0.2+0.7+0.0}_{-0.6-0.4-1.2-0.3-0.5-0.0}
  $
  &$14.9^{+10.2+10.1}_{-5.0-4.6}$
 \\
 \hline \hline
 $f_L(\%)$      & $|{\cal A}_L|^2$
 &$15.8^{+0.9+2.8+5.8+1.6+0.7+0.1}_{-1.0-1.8-2.4-1.2-0.2-0.1}
  $
 &$45^{+55}_{-50}$
 &$71.0^{+1.3+1.7+10.9+1.2+4.4+0.1}_{-1.7-2.2-11.1-1.0-6.3-0.1}
  $
 &$64^{+38}_{-61}$
 \\
 $f_{||}(\%)$   & $|{\cal A}_{||}|^2$
 &$46.1^{+0.5+0.9+1.3+0.5+0.1+0.0}_{-0.5-1.4-3.3-0.8-0.4-0.1}
  $
 &$-$
 &$16.0^{+1.0+1.4+6.4+0.8+3.4+0.0}_{-0.9-1.2-6.3-0.8-2.4-0.1}
  $
 &$-$
  \\
 $f_{\perp}(\%)$& $|{\cal A}_\perp|^2$
 &$38.1^{+0.5+1.1+1.1+0.7+0.1+0.1}_{-0.4-1.4-2.6-0.8-0.3-0.0}
$
 &$-$
 &$13.0^{+0.6+0.9+4.7+0.5+2.9+0.1}_{-0.4-0.6-4.5-0.4-2.0-0.0}
  $
 &$-$
 \\
 \hline \hline
 $\phi_{||}$(rad)& $\arg\frac{{\cal A}_{||}}{{\cal A}_L}$
 &$\hspace{0.18cm}
 3.9^{+0.1+0.1+0.5+0.0+0.1+0.0}_{-0.1-0.2-0.4-0.1-0.1-0.0}
$
 &$-$
 &$\hspace{0.18cm}
 3.7^{+0.1+0.2+0.3+0.1+0.1+0.0}_{-0.0-0.1-0.1-0.0-0.1-0.0}
  $
 &$-$
 \\
 $\phi_{\perp}$(rad)& $\arg\frac{{\cal A}_{\perp}}{{\cal A}_L}$
 &$\hspace{0.18cm}
 3.9^{+0.1+0.1+0.5+0.0+0.1+0.0}_{-0.1-0.1-0.4-0.1-0.1-0.0}
  $
 &$-$
 &$\hspace{0.18cm}
 3.7^{+0.0+0.0+0.1+0.0+0.0+0.0}_{-0.1-0.3-0.4-0.1-0.2-0.0}
  $
 &$-$
  \\
  \hline \hline
 $\acp^{\rm dir}(\%)$& $\frac{\overline{\Gamma}-\Gamma}{\overline{\Gamma}+\Gamma}$
 &$-7.8^{+0.8+0.2+2.0+0.1+1.2+0.3}_{-0.9-0.0-1.8-0.0-1.0-0.3}
  $
 &$-$
 &$\hspace{0.18cm}
 4.7^{+0.0+0.9+0.2+0.6+0.8+0.2}_{-0.0-0.9-0.4-0.5-1.0-0.2}
  $
 &$-$
 \\
 $\acp^{\rm dir}(L)(\%)$& $\frac{\bar{f}_L-f_L}{\bar{f}_L+f_L}$
 &$\hspace{0.18cm}
 1.7^{+0.0+3.3+6.0+2.0+2.7+0.1}_{-0.2-2.6-10.6-1.7-2.4-0.0}
  $
 & $-$
 &$\hspace{0.18cm}
 3.4^{+0.0+0.9+0.3+0.5+1.0+0.1}_{-0.1-0.8-0.5-0.5-1.6-0.2}
  $
 & $-$
 \\
 $\acp^{\rm dir}(||)(\%)$& $\frac{\bar{f}_{||}-f_{||}}{\bar{f}_{||}+f_{||}}$
 &$-9.3^{+0.9+0.5+0.9+0.3+0.9+0.4}_{-0.9-0.4-0.9-0.2-0.8-0.3}
  $
 &$-$
 &$\hspace{0.18cm}
 7.9^{+0.3+1.6+2.0+1.1+0.7+0.3}_{-0.4-1.6-1.8-0.9-0.8-0.3}
  $
 &$-$
 \\
 $\acp^{\rm dir}(\perp)(\%)$& $\frac{\bar{f}_\perp-f_\perp}{\bar{f}_\perp+f_\perp}$
 &$-9.9^{+0.8+0.4+0.7+0.2+1.0+0.3}_{-1.0-0.5-0.9-0.2-1.0-0.4}
$
 &$-$
 &$\hspace{0.18cm}
 8.0^{+0.1+1.2+1.2+0.8+0.8+0.3}_{-0.2-1.4-1.5-0.8-0.8-0.3}
  $
 &$-$
 \\ \hline \hline
\end{tabular}}
\end{center}
\end{table}

\begin{table}[b]
\caption{ Same as Table~\ref{tab:f1rhop} but for $B^0 \to f_1 \omega$ decays.}
\label{tab:f1omg-d}
 \begin{center}\vspace{-0.3cm}{
\begin{tabular}[t]{c|c||c|c||c|c}
\hline  \hline
   \multicolumn{2}{c||}{Decay Modes}   &  \multicolumn{2}{c||}{$B^0 \to f_1(1285) \omega$} &  \multicolumn{2}{c}{$B^0 \to f_1(1420) \omega$}  \\
   \hline
 Parameter  & Definition & This work &   QCDF~\cite{Cheng:2008gxa}
 & This work &   QCDF~\cite{Cheng:2008gxa}  \\
\hline \hline
  BR($10^{-6}$)        & $\Gamma/ \Gamma_{\rm total}$
  &$1.0^{+0.2+0.5+0.3+0.0+0.1+0.1}_{-0.2-0.3-0.1-0.0-0.0-0.0}
  $
  &$0.9^{+1.0+2.2}_{-0.4-0.1}$
  &$\hspace{0.18cm}
 0.2^{+0.0+0.1+0.0+0.0+0.0+0.0}_{-0.0-0.1-0.0-0.0-0.0-0.0}
  $
  &$0.1^{+0.2+0.3}_{-0.1-0.0}$
 \\
 \hline \hline
 $f_L(\%)$      & $|{\cal A}_L|^2$
 &$60.1^{+2.3+1.2+8.1+0.0+2.4+0.5}_{-2.4-1.3-7.6-0.1-1.6-0.6}
  $
 &$86^{+7}_{-62}$
 &$
45.3^{+3.2+3.9+9.7+2.3+4.4+1.4}_{-3.4-4.7-9.3-2.4-3.0-1.4}
  $
 &$86^{+4}_{-76}$
 \\
 $f_{||}(\%)$   & $|{\cal A}_{||}|^2$
 &$
20.1^{+1.3+0.7+4.0+0.1+1.0+0.3}_{-1.2-0.6-4.2-0.0-1.3-0.2}
  $
 &$-$
 &$28.3^{+1.8+2.5+4.8+1.3+1.7+0.7}_{-1.8-2.3-5.1-1.3-2.5-0.9}
  $
 &$-$
  \\
 $f_{\perp}(\%)$& $|{\cal A}_\perp|^2$
 &$
19.8^{+1.1+0.6+3.5+0.1+0.6+0.3}_{-1.1-0.6-3.9-0.1-1.1-0.3}
$
 &$-$
 &$26.5^{+1.5+2.0+4.4+1.0+1.2+0.6}_{-1.5-1.8-4.7-1.0-2.0-0.7}
  $
 &$-$
 \\
 \hline \hline
 $\phi_{||}$(rad)& $\arg\frac{{\cal A}_{||}}{{\cal A}_L}$
 &$\hspace{0.18cm}
1.7^{+0.1+0.1+1.5+0.0+1.3+0.1}_{-0.0-0.0-0.1-0.0-0.0-0.0}
$
 &$-$
 &$\hspace{0.18cm}
3.2^{+0.0+0.0+0.1+0.0+0.2+0.0}_{-0.1-0.0-0.2-0.0-0.2-0.0}
  $
 &$-$
 \\
 $\phi_{\perp}$(rad)& $\arg\frac{{\cal A}_{\perp}}{{\cal A}_L}$
 &$\hspace{0.18cm}
1.7^{+0.1+0.1+0.3+0.0+2.9+0.1}_{-0.0-0.0-0.1-0.0-0.0-0.0}
  $
 &$-$
 &$\hspace{0.18cm}
 3.2^{+0.0+0.0+0.1+0.0+0.2+0.0}_{-0.1-0.0-0.2-0.0-0.2-0.0}
  $
 &$-$
  \\
  \hline \hline
 $\acp^{\rm dir}(\%)$& $\frac{\overline{\Gamma}-\Gamma}{\overline{\Gamma}+\Gamma}$
 &$-59.3^{+0.2+1.6+4.2+0.6+4.5+1.8}_{-0.0-1.7-1.8-0.6-1.0-1.5}
  $
 &$-$
 &$-6.0^{+2.8+12.2+18.7+6.5+9.2+0.2}_{-2.7-11.2-17.3-6.7-6.5-0.3}
  $
 &$-$
 \\
 $\acp^{\rm dir}(L)(\%)$& $\frac{\bar{f}_L-f_L}{\bar{f}_L+f_L}$
 &$-88.7^{+2.8+1.2+11.7+0.8+6.0+1.6}_{-2.7-1.6-6.3-0.9-0.0-1.6}
  $
 & $-$
 &$-7.3^{+5.6+25.1+17.9+13.5+24.2+0.3}_{-4.5-19.5-20.3-12.6-13.8-0.4}
  $
 & $-$
 \\
 $\acp^{\rm dir}(||)(\%)$& $\frac{\bar{f}_{||}-f_{||}}{\bar{f}_{||}+f_{||}}$
 &$-15.8^{+0.0+1.5+5.2+0.1+0.4+0.6}_{-0.1-1.7-3.8-0.1-0.3-0.7}
  $
 &$-$
 &$
-4.3^{+0.7+3.7+21.3+1.2+0.9+0.2}_{-0.9-4.1-17.6-1.3-1.9-0.3}
  $
 &$-$
 \\
 $\acp^{\rm dir}(\perp)(\%)$& $\frac{\bar{f}_\perp-f_\perp}{\bar{f}_\perp+f_\perp}$
 &$-14.3^{+0.1+1.4+5.3+0.1+0.5+0.6}_{-0.0-1.4-5.8-0.0-0.4-0.5}
$
 &$-$
 &$
 -5.6^{+0.7+3.5+19.7+1.0+0.8+0.3}_{-0.9-3.9-16.2-1.1-1.9-0.4}
  $
 &$-$
 \\ \hline \hline
\end{tabular}}
\end{center}
\end{table}

\item[(4)]
According to Table~\ref{tab:f1kstp}, the {\it CP}-averaged branching ratios
of $B^+ \to f_1 K^{*+}$ decays can be written as
\beq
Br(B^+ \to f_1(1285) K^{*+})_{\rm pQCD} &=& 6.4^{+3.6}_{-2.5} \times 10^{-6}\;,
\qquad
Br(B^+ \to f_1(1420) K^{*+})_{\rm pQCD} = 4.5^{+1.7}_{-1.5} \times 10^{-6}\;;
\eeq
Here, we have added all the errors in quadrature. For the former
$B^+ \to f_1(1285) K^{*+}$ decay, our predicted branching ratio
is in good consistency
with the value $5.7^{+21.7}_{-5.3} \times 10^{-6}$ derived in the QCDF
approach within theoretical errors. Generally speaking,
in light of the constructive or destructive interference
between $f_{1q} V$ and $f_{1s} V$ states, the latter
$Br(B^+ \to f_1(1420) K^{*+})$ is naturally expected to be
larger or smaller than $Br(B^+ \to f_1(1285) K^{*+})$
in principle. Although
$Br(B^+ \to f_1(1285) K^{*+})_{\rm pQCD}$ is, in terms of
the central values, somewhat larger than
$Br(B^+ \to f_1(1420) K^{*+})_{\rm pQCD}$,
the pQCD predictions of the $B^+ \to f_1 K^{*+}$ decay rates
within errors
are approximately equivalent to each other in this work, which
make a sharp contrast to the pattern obtained in the framework of QCDF.
The authors predicted the $B^+ \to f_1(1420) K^{*+}$ branching fraction
as $Br(B^+ \to f_1(1420) K^{*+})_{\rm QCDF}
= 15.6^{+15.1}_{-7.0} \times 10^{-6}$~\cite{Cheng:2008gxa}.
It
seems that the predicted branching ratio for $B^+ \to f_1(1420) K^{*+}$ indicates a
strongly constructive(moderately destructive) interference in QCDF(pQCD) between
$B^+ \to f_{1q} K^{*+}$ and
$B^+ \to f_{1s} K^{*+}$ channels. In order to understand
the branching ratios of $B^+ \to f_1 K^{*+}$ decays, different
from those QCDF predictions, the numerical values of decay
amplitudes are presented in Table~\ref{tab:DAs-f1kst-u}
explicitly involving three polarizations within
the pQCD framework. One can easily see the
dominated $B^+ \to f_{1q} K^{*+}(B^+ \to f_{1s} K^{*+})$
contributions induced by the dominance of $f_{1q}(f_{1s})$
in the $f_1(1285)[f_1(1420)]$ state[see Eq.~(\ref{eq:mix-f1q-f1s})
with $\phi_{f_1} \sim 24^\circ$]
and the moderately constructive(destructive)
interferences between
$B^+ \to f_{1q} K^{*+}$ and
$B^+ \to f_{1s} K^{*+}$
in the $B^+ \to f_1(1285)[f_1(1420)] K^{*+}$ decays in the pQCD approach.

\begin{table}[t]
\caption{ Same as Table~\ref{tab:f1rhop} but for $B^0 \to f_1 \phi$ decays.}
\label{tab:f1phi-d}
 \begin{center}\vspace{-0.3cm}{
\begin{tabular}[t]{c|c||c|c||c|c}
\hline  \hline
   \multicolumn{2}{c||}{Decay Modes}   &  \multicolumn{2}{c||}{$B^0 \to f_1(1285) \phi$} &  \multicolumn{2}{c}{$B^0 \to f_1(1420) \phi$}  \\
   \hline
 Parameter  & Definition & This work &   QCDF~\cite{Cheng:2008gxa}
 & This work &   QCDF~\cite{Cheng:2008gxa}  \\
\hline \hline
  BR($10^{-9}$)        & $\Gamma/ \Gamma_{\rm total}$
  &$8.9^{+1.8+3.3+3.4+0.3+2.2+0.4}_{-1.4-2.3-2.2-0.2-1.4-0.3}
  $
  &$2.0^{+2.0+9.0}_{-1.0-0.0}$
  &$
 3.7^{+0..2+0.3+2.6+0.2+0.9+0.1}_{-0.4-0.5-2.1-0.3-0.9-0.2}
  $
  &$0.8^{+0.9+0.9}_{-0.1-0.1}$
 \\
 \hline \hline
 $f_L(\%)$      & $|{\cal A}_L|^2$
 &$68.9^{+0.9+3.9+19.5+2.5+1.7+0.0}_{-0.9-3.3-17.7-2.1-2.4-0.0}
  $
 &$90^{+3}_{-71}$
 &$
85.9^{+1.6+5.7+11.4+3.6+0.0+0.0}_{-2.0-7.7-16.7-5.1-1.1-0.0}
  $
 &$98^{+2}_{-44}$
 \\
 $f_{||}(\%)$   & $|{\cal A}_{||}|^2$
 &$
17.3^{+0.5+1.9+9.9+1.2+1.3+0.0}_{-0.4-2.0-10.5-1.3-0.9-0.0}
  $
 &$-$
 &$\hspace{0.18cm}
7.4^{+1.1+4.3+9.0+2.8+0.6+0.0}_{-0.8-3.0-6.2-1.9-0.0-0.0}
  $
 &$-$
  \\
 $f_{\perp}(\%)$& $|{\cal A}_\perp|^2$
 &$
13.7^{+0.5+1.5+7.9+1.0+1.2+0.0}_{-0.4-1.7-8.5-1.1-0.8-0.0}
$
 &$-$
 &$\hspace{0.18cm}
6.7^{+0.9+3.5+7.7+2.3+0.5+0.0}_{-0.7-2.6-5.3-1.7-0.0-0.0}
  $
 &$-$
 \\
 \hline \hline
 $\phi_{||}$(rad)& $\arg\frac{{\cal A}_{||}}{{\cal A}_L}$
 &$\hspace{0.18cm}
3.7^{+0.0+0.0+0.0+0.0+0.0+0.0}_{-0.1-0.1-0.1-0.0-0.0-0.0}
$
 &$-$
 &$\hspace{0.18cm}
4.3^{+0.1+0.0+0.1+0.0+0.0+0.0}_{-0.1-0.1-0.2-0.0-0.0-0.0}
  $
 &$-$
 \\
 $\phi_{\perp}$(rad)& $\arg\frac{{\cal A}_{\perp}}{{\cal A}_L}$
 &$\hspace{0.18cm}
3.7^{+0.0+0.0+0.1+0.0+0.0+0.0}_{-0.0-0.0-0.1-0.0-0.0-0.0}
  $
 &$-$
 &$\hspace{0.18cm}
 4.4^{+0.1+0.0+0.1+0.0+0.0+0.0}_{-0.1-0.1-0.2-0.0-0.0-0.0}
  $
 &$-$
  \\
  \hline \hline
 $\acp^{\rm dir}(\%)$& $\frac{\overline{\Gamma}-\Gamma}{\overline{\Gamma}+\Gamma}$
 &$ \sim 0.0
  $
 &$-$
 &$ \sim 0.0
  $
 &$-$
 \\
 $\acp^{\rm dir}(L)(\%)$& $\frac{\bar{f}_L-f_L}{\bar{f}_L+f_L}$
 &$ \sim 0.0
  $
 & $-$
 &$ \sim 0.0
  $
 & $-$
 \\
 $\acp^{\rm dir}(||)(\%)$& $\frac{\bar{f}_{||}-f_{||}}{\bar{f}_{||}+f_{||}}$
 &$ \sim 0.0
  $
 &$-$
 &$ \sim 0.0
  $
 &$-$
 \\
 $\acp^{\rm dir}(\perp)(\%)$& $\frac{\bar{f}_\perp-f_\perp}{\bar{f}_\perp+f_\perp}$
 &$  \sim 0.0
$
 &$-$
 &$ \sim 0.0
  $
 &$-$
 \\ \hline \hline
\end{tabular}}
\end{center}
\end{table}

\begin{table}[b]
\caption{ Same as Table~\ref{tab:f1rhop} but for $B_s^0 \to f_1 \rho^0$ decays.}
\label{tab:f1rho0-s}
 \begin{center}\vspace{-0.3cm}{
\begin{tabular}[t]{c|c||c|c||c|c}
\hline  \hline
   \multicolumn{2}{c||}{Decay Modes}   &  \multicolumn{2}{c||}{$B_s^0 \to f_1(1285) \rho^0$} &  \multicolumn{2}{c}{$B_s^0 \to f_1(1420) \rho^0$}  \\
   \hline
 Parameter  & Definition & This work &   QCDF
 & This work &   QCDF   \\
\hline \hline
  BR($10^{-7}$)        & $\Gamma/ \Gamma_{\rm total}$
  &$0.5^{+0.2+0.1+0.3+0.1+0.1+0.0}_{-0.1-0.0-0.2-0.1-0.0-0.0}
  $
  &$-$
  &$
 2.5^{+0.8+0.2+1.4+0.1+0.2+0.0}_{-0.6-0.2-1.1-0.1-0.2-0.1}
  $
  &$-$
 \\
 \hline \hline
 $f_L(\%)$      & $|{\cal A}_L|^2$
 &$79.8^{+0.3+0.3+1.9+0.2+0.2+0.8}_{-0.3-0.0-3.6-0.1-0.1-0.8}
  $
 &$-$
 &$80.8^{+0.0+0.1+1.6+0.1+0.1+0.8}_{-0.0-0.1-2.7-0.0-0.1-0.8}
  $
 &$-$
 \\
 $f_{||}(\%)$   & $|{\cal A}_{||}|^2$
 &$
10.9^{+0.1+0.0+1.9+0.0+0.0+0.3}_{-0.2-0.2-1.0-0.1-0.1-0.4}
  $
 &$-$
 &$10.4^{+0.1+0.1+1.5+0.1+0.1+0.4}_{-0.0-0.0-0.8-0.0-0.0-0.4}
  $
 &$-$
  \\
 $f_{\perp}(\%)$& $|{\cal A}_\perp|^2$
 &$\hspace{0.18cm}
9.3^{+0.1+0.0+1.7+0.1+0.1+0.3}_{-0.1-0.1-0.8-0.1-0.1-0.3}
$
 &$-$
 &$\hspace{0.18cm}
8.7^{+0.1+0.2+1.4+0.1+0.1+0.4}_{-0.0-0.0-0.7-0.0-0.0-0.3}
  $
 &$-$
 \\
 \hline \hline
 $\phi_{||}$(rad)& $\arg\frac{{\cal A}_{||}}{{\cal A}_L}$
 &$\hspace{0.18cm}
3.1^{+0.0+0.1+0.1+0.0+0.0+0.0}_{-0.0-0.0-0.1-0.0-0.0-0.0}
$
 &$-$
 &$\hspace{0.18cm}
2.9^{+0.1+0.0+0.2+0.0+0.0+0.1}_{-0.0-0.0-0.0-0.0-0.0-0.0}
  $
 &$-$
 \\
 $\phi_{\perp}$(rad)& $\arg\frac{{\cal A}_{\perp}}{{\cal A}_L}$
 &$\hspace{0.18cm}
3.1^{+0.0+0.1+0.1+0.0+0.0+0.0}_{-0.0-0.0-0.1-0.0-0.0-0.0}
  $
 &$-$
 &$\hspace{0.18cm}
 3.0^{+0.0+0.0+0.1+0.0+0.0+0.0}_{-0.1-0.0-0.1-0.0-0.0-0.0}
  $
 &$-$
  \\
  \hline \hline
 $\acp^{\rm dir}(\%)$& $\frac{\overline{\Gamma}-\Gamma}{\overline{\Gamma}+\Gamma}$
 &$-26.4^{+3.3+2.2+3.8+5.2+1.5+1.0}_{-3.3-8.1-3.3-5.1-1.4-0.9}
  $
 &$-$
 &$23.7^{+2.0+1.9+15.6+1.3+1.8+0.8}_{-2.0-1.6-9.4-1.0-1.9-0.8}
  $
 &$-$
 \\
 $\acp^{\rm dir}(L)(\%)$& $\frac{\bar{f}_L-f_L}{\bar{f}_L+f_L}$
 &$-30.6^{+3.6+2.5+5.8+6.4+1.8+1.1}_{-3.7-10.1-8.1-6.5-1.8-1.2}
  $
 & $-$
 &$31.8^{+3.0+2.2+17.2+1.5+2.2+1.1}_{-3.0-2.0-10.4-1.3-1.5-1.1}
  $
 & $-$
 \\
 $\acp^{\rm dir}(||)(\%)$& $\frac{\bar{f}_{||}-f_{||}}{\bar{f}_{||}+f_{||}}$
 &$-13.8^{+2.5+0.1+12.7+0.4+0.5+0.7}_{-2.9-0.8-7.2-0.5-0.6-0.8}
  $
 &$-$
 &$
-9.6^{+2.2+0.1+14.8+0.1+0.3+0.5}_{-2.5-0.2-8.4-0.1-0.4-0.6}
  $
 &$-$
 \\
 $\acp^{\rm dir}(\perp)(\%)$& $\frac{\bar{f}_\perp-f_\perp}{\bar{f}_\perp+f_\perp}$
 &$-4.2^{+1.6+1.2+15.4+0.8+0.2+0.2}_{-1.6-0.3-8.9-0.7-0.4-0.2}
$
 &$-$
 &$
-10.8^{+2.4+0.3+13.8+0.2+0.1+0.6}_{-2.4-0.2-7.6-0.1-0.2-0.6}
  $
 &$-$
 \\ \hline \hline
\end{tabular}}
\end{center}
\end{table}

However, it should be pointed out
that when the very large errors are taken into account, $Br(B^+ \to f_1(1285) K^{*+})_{\rm QCDF} \sim Br(B^+ \to f_1(1420) K^{*+})_{\rm QCDF}$ can be observed. Moreover, objectively speaking, as discussed in Ref.~\cite{Cheng:2009cn}, different
predictions of $B \to VV$ decays have been theoretically
obtained by fitting
the parameters through different well-measured channels such as
$B \to \phi K^*$~\cite{Beneke:2006hg} and $B \to \rho K^*$~\cite{Cheng:2008gxa,Cheng:2009cn}, respectively, because of
inevitable end-point singularities
in the framework of QCDF.
This indefiniteness may render misunderstandings of the dynamics
involved in these kinds of decays with polarizations.
It will be very interesting and probably a challenge for the
theorists to further understand the QCD dynamics of
axial-vector $f_1$ mesons and the decay mechanism of $B \to f_1 K^*$ with
helicity in depth
once the experiments at LHCb and/or Belle-II confirm the aforementioned decay rates and decay pattern in the near future.

Similar phenomena also occur in the $B^0 \to f_1 K^{*0}$ modes(see Table~\ref{tab:f1kst0}), in which few contributions arising from the color-suppressed
tree amplitudes are involved. Specifically, the branching ratios will numerically decrease(increase) from $6.43(4.46) \times 10^{-6}$ to $5.65(4.61) \times 10^{-6}$ for $B^+ \to f_1(1285)[f_1(1420)] K^{*+}$ decay, and increase(decrease) from $4.96(4.37) \times 10^{-6}$ to $5.08(4.34) \times 10^{-6}$ for the $B^0 \to f_1(1285)[f_1(1420)] K^{*0}$ mode, when the contributions induced
by tree operators are turned off.
The stringent tests on the {\it CP}-averaged branching ratios for $B \to f_1 K^*$ decays predicted in the QCDF and pQCD approaches may provide an experimental check on these two competing frameworks.

\item[(5)]
As discussed in Refs.~\cite{Yang:2007zt,Cheng:2008gxa},
the behavior of axial-vector $^3\!P_1$ states is similar
to that of vector mesons, which will consequently result in the branching ratio
of $B \to f_1(1285)[f_1(1420)] K^*$ analogous to that of $B \to \omega[\phi] K^*$ decays in the pQCD approach as expected, if the $f_1(1285)[f_1(1420)]$ state is almost governed by the $f_{1q}(f_{1s})$ component.
However, from Tables~\ref{tab:f1kstp}, \ref{tab:f1kst0}, and \ref{tab:DAs-f1kst-u}, it can be clearly observed that
the predicted branching ratios of $B \to f_1(1285)[f_1(1420)] K^*$ decays in this work are larger(smaller) than those of $B \to \omega[\phi] K^*$ decays~\cite{Beneke:2006hg,Cheng:2009cn,Agashe:2014kda,Zou:2015iwa}.
The underlying reason is that, for the $B^0 \to f_1(1285)[f_1(1420)] K^{*0}$ mode for example, a constructive(destructive) interference arising from
$B^0 \to f_{1s}[f_{1q}] K^{*0}$(as can be seen in Table~\ref{tab:DAs-f1kst-u}) with
a factor $\sin\phi_{f_1} \sim 0.4$ will
enhance(reduce) the amplitude of $B^0 \to f_{1q}[f_{1s}] K^{*0}$, which finally leads to somewhat larger(smaller) branching ratio $5.0^{+2.7}_{-2.1}[4.4^{+1.7}_{-1.5}] \times 10^{-6}$ than that of $B^0 \to \omega[\phi] K^{*0}$, with $2.0^{+3.1}_{-1.4}[9.3^{+11.4}_{-6.5}] \times 10^{-6}$ in~\cite{Beneke:2006hg}, $2.5^{+2.5}_{-1.6}[9.5^{+12.0}_{-6.0}] \times 10^{-6}$ in~\cite{Cheng:2009cn}, $4.7^{+2.6}_{-2.0}[9.8^{+4.9}_{-3.8}] \times 10^{-6}$ in~\cite{Zou:2015iwa}, and $2.0 \pm 0.5[10.0 \pm 0.5] \times 10^{-6}$ in~\cite{Agashe:2014kda}, respectively.

\begin{table}[t]
\caption{ Same as Table~\ref{tab:f1rhop} but for $B_s^0 \to f_1 \bar K^{*0}$ decays.}
\label{tab:f1kst0b}
 \begin{center}\vspace{-0.3cm}{
\begin{tabular}[t]{c|c||c|c||c|c}
\hline  \hline
   \multicolumn{2}{c||}{Decay Modes}   &  \multicolumn{2}{c||}{$B_s^0 \to f_1(1285) \bar K^{*0}$}  &  \multicolumn{2}{c}{$B_s^0 \to f_1(1420) \bar K^{*0}$}\\
   \hline
 Parameter  & Definition & This work &   QCDF
 & This work &   QCDF   \\
\hline \hline
  BR($10^{-7}$)        & $\Gamma/ \Gamma_{\rm total}$
  &$5.5^{+1.0+2.2+1.0+0.0+1.1+0.3}_{-0.8-1.7-0.9-0.0-0.6-0.3}
  $
  &$-$
  &$3.4^{+0.6+0.4+1.8+0.0+0.6+0.0}_{-0.5-0.3-1.3-0.0-0.4-0.0}
  $
  &$-$
 \\
 \hline \hline
 $f_L(\%)$      & $|{\cal A}_L|^2$
 &$39.2^{+0.0+1.6+8.4+0.4+3.2+0.9}_{-0.3-1.6-8.2-0.4-1.5-0.8}
  $
 &$-$
 &$51.1^{+2.7+4.3+12.0+0.6+5.2+0.5}_{-2.8-4.5-15.5-0.7-6.5-0.6}
  $
 &$-$
 \\
 $f_{||}(\%)$   & $|{\cal A}_{||}|^2$
 &$31.8^{+0.2+0.9+4.3+0.3+1.1+0.5}_{-0.0-0.9-4.4-0.3-1.9-0.4}
  $
 &$-$
 &$25.8^{+1.5+2.6+8.2+0.5+3.6+0.3}_{-1.4-2.5-6.4-0.4-2.8-0.2}
  $
 &$-$
  \\
 $f_{\perp}(\%)$& $|{\cal A}_\perp|^2$
 &$29.0^{+0.2+0.6+3.8+0.1+0.4+0.5}_{-0.1-0.8-4.0-0.2-1.3-0.5}
$
 &$-$
 &$23.1^{+1.3+2.0+7.4+0.2+2.9+0.3}_{-1.2-2.0-5.7-0.2-2.4-0.3}
  $
 &$-$
 \\
 \hline \hline
 $\phi_{||}$(rad)& $\arg\frac{{\cal A}_{||}}{{\cal A}_L}$
 &$\hspace{0.18cm}
 3.0^{+1.0+1.1+2.1+1.3+1.3+1.3}_{-0.0-0.1-0.1-0.0-0.0-0.0}
$
 &$-$
 &$\hspace{0.18cm}
 2.9^{+0.0+0.1+0.2+0.1+0.1+0.1}_{-0.0-0.0-0.1-0.0-0.0-0.0}
  $
 &$-$
 \\
 $\phi_{\perp}$(rad)& $\arg\frac{{\cal A}_{\perp}}{{\cal A}_L}$
 &$\hspace{0.18cm}
 3.2^{+1.1+1.2+0.2+0.1+0.0+0.1}_{-0.0-0.2-0.4-0.2-0.1-0.2}
  $
 &$-$
 &$\hspace{0.18cm}
 3.0^{+0.0+0.0+0.2+0.0+0.0+0.0}_{-0.0-0.2-0.3-0.1-0.1-0.1}
  $
 &$-$
  \\
  \hline \hline
 $\acp^{\rm dir}(\%)$& $\frac{\overline{\Gamma}-\Gamma}{\overline{\Gamma}+\Gamma}$
 &$-52.9^{+4.2+3.0+12.7+1.9+7.5+1.0}_{-2.7-2.2-13.2-1.5-4.4-0.9}
  $
 &$-$
 &$-5.9^{+2.3+5.0+11.5+2.2+4.2+0.1}_{-2.6-5.9-10.8-4.0-3.8-0.2}
  $
 &$-$
 \\
 $\acp^{\rm dir}(L)(\%)$& $\frac{\bar{f}_L-f_L}{\bar{f}_L+f_L}$
 &$17.7^{+7.0+11.3+18.7+6.9+26.0+0.7}_{-6.0-9.9-23.0-6.1-20.9-0.8}
  $
 & $-$
 &$-71.1^{+1.3+11.2+23.8+6.6+2.4+2.3}_{-0.5-10.0-26.2-7.0-3.2-2.1}
  $
 & $-$
 \\
 $\acp^{\rm dir}(||)(\%)$& $\frac{\bar{f}_{||}-f_{||}}{\bar{f}_{||}+f_{||}}$
 &$-99.0^{+2.4+1.1+3.4+0.6+1.0+0.6}_{-0.0-0.5-1.3-0.2-0.4-0.5}
  $
 &$-$
 &$61.7^{+3.1+5.1+6.9+3.3+4.2+1.8}_{-3.9-7.7-7.6-4.1-5.4-1.8}
  $
 &$-$
 \\
 $\acp^{\rm dir}(\perp)(\%)$& $\frac{\bar{f}_\perp-f_\perp}{\bar{f}_\perp+f_\perp}$
 &$-97.8^{+3.6+1.9+3.7+1.1+1.8+1.0}_{-1.2-1.3-3.1-0.9-1.4-0.7}
$
 &$-$
 &$62.5^{+2.5+4.2+6.0+2.8+4.2+1.7}_{-3.4-7.3-7.4-3.9-5.5-1.8}
  $
 &$-$
 \\ \hline \hline
\end{tabular}}
\end{center}
\end{table}

\begin{table}[t]
\caption{ Same as Table~\ref{tab:f1rhop} but for $B_s^0 \to f_1 \omega$ decays.}
\label{tab:f1omg-s}
 \begin{center}\vspace{-0.3cm}{
\begin{tabular}[t]{c|c||c|c||c|c}
\hline  \hline
   \multicolumn{2}{c||}{Decay Modes}   &  \multicolumn{2}{c||}{$B_s^0 \to f_1(1285) \omega$} &  \multicolumn{2}{c}{$B_s^0 \to f_1(1420) \omega$}  \\
   \hline
 Parameter  & Definition & This work &   QCDF
 & This work &   QCDF   \\
\hline \hline
  BR($10^{-7}$)        & $\Gamma/ \Gamma_{\rm total}$
  &$1.9^{+0.5+0.6+0.7+0.2+0.7+0.1}_{-0.3-0.4-0.3-0.1-0.4-0.0}
  $
  &$-$
  &$\hspace{0.18cm}
 3.5^{+1.5+0.2+3.2+0.1+1.1+0.0}_{-1.1-0.3-2.2-0.2-0.8-0.1}
  $
  &$-$
 \\
 \hline \hline
 $f_L(\%)$      & $|{\cal A}_L|^2$
 &$81.8^{+1.1+4.0+10.0+2.6+0.1+0.2}_{-1.4-4.8-9.9-3.0-0.5-0.3}
  $
 &$-$
 &$50.9^{+4.0+0.6+3.4+0.4+0.3+0.6}_{-3.9-0.4-2.8-0.3-1.4-0.7}
  $
 &$-$
 \\
 $f_{||}(\%)$   & $|{\cal A}_{||}|^2$
 &$\hspace{0.18cm}
9.9^{+0.7+2.5+5.3+1.6+0.3+0.2}_{-0.6-2.1-5.5-1.4-0.1-0.2}
  $
 &$-$
 &$26.5^{+2.0+0.3+1.7+0.2+0.8+0.4}_{-2.1-0.3-2.2-0.2-0.1-0.3}
  $
 &$-$
  \\
 $f_{\perp}(\%)$& $|{\cal A}_\perp|^2$
 &$\hspace{0.18cm}
8.3^{+0.6+2.2+4.5+1.4+0.3+0.2}_{-0.5-1.8-4.6-1.2-0.0-0.1}
$
 &$-$
 &$22.6^{+1.8+0.1+1.1+0.1+0.6+0.3}_{-1.9-0.3-1.3-0.2-0.1-0.3}
  $
 &$-$
 \\
 \hline \hline
 $\phi_{||}$(rad)& $\arg\frac{{\cal A}_{||}}{{\cal A}_L}$
 &$\hspace{0.18cm}
3.9^{+0.0+0.0+0.4+0.0+0.0+0.0}_{-0.1-0.1-0.3-0.1-0.1-0.0}
$
 &$-$
 &$\hspace{0.18cm}
2.7^{+0.0+0.1+0.3+0.0+0.0+0.0}_{-0.1-0.1-0.2-0.1-0.0-0.0}
  $
 &$-$
 \\
 $\phi_{\perp}$(rad)& $\arg\frac{{\cal A}_{\perp}}{{\cal A}_L}$
 &$\hspace{0.18cm}
3.9^{+0.0+0.0+0.4+0.0+0.0+0.0}_{-0.1-0.1-0.3-0.1-0.1-0.0}
  $
 &$-$
 &$\hspace{0.18cm}
 2.7^{+0.1+0.1+0.3+0.1+0.0+0.0}_{-0.1-0.1-0.2-0.0-0.0-0.0}
  $
 &$-$
  \\
  \hline \hline
 $\acp^{\rm dir}(\%)$& $\frac{\overline{\Gamma}-\Gamma}{\overline{\Gamma}+\Gamma}$
 &$10.9^{+1.1+0.9+2.0+0.5+0.3+0.4}_{-1.1-0.5-4.6-0.3-0.9-0.4}
  $
 &$-$
 &$29.5^{+2.0+0.8+13.9+0.6+3.9+1.1}_{-2.2-0.7-7.6-0.4-4.6-1.0}
  $
 &$-$
 \\
 $\acp^{\rm dir}(L)(\%)$& $\frac{\bar{f}_L-f_L}{\bar{f}_L+f_L}$
 &$\hspace{0.18cm}
7.7^{+1.1+0.2+2.2+0.1+1.5+0.2}_{-1.1-0.1-3.8-0.0-2.2-0.3}
  $
 & $-$
 &$34.3^{+5.3+1.4+20.4+1.0+2.8+1.2}_{-4.7-1.5-11.1-0.9-3.4-1.2}
  $
 & $-$
 \\
 $\acp^{\rm dir}(||)(\%)$& $\frac{\bar{f}_{||}-f_{||}}{\bar{f}_{||}+f_{||}}$
 &$23.5^{+0.1+0.1+5.3+0.1+4.7+1.1}_{-0.1-0.0-3.7-0.0-5.2-0.9}
  $
 &$-$
 &$23.9^{+0.1+0.0+6.5+0.0+4.7+1.0}_{-0.2-0.0-4.1-0.0-5.4-1.1}
  $
 &$-$
 \\
 $\acp^{\rm dir}(\perp)(\%)$& $\frac{\bar{f}_\perp-f_\perp}{\bar{f}_\perp+f_\perp}$
 &$27.6^{+0.0+0.3+7.6+0.2+5.3+1.1}_{-0.3-0.4-5.0-0.3-6.2-1.2}
$
 &$-$
 &$25.4^{+0.0+0.1+5.2+0.1+5.2+1.1}_{-0.1-0.1-4.2-0.0-5.9-1.1}
  $
 &$-$
 \\ \hline \hline
\end{tabular}}
\end{center}
\end{table}

\item[(6)]
The {\it CP}-averaged branching ratios for penguin-dominated $B^0 \to f_1 \rho^0$, color-suppressed tree-dominated $B^0 \to f_1 \omega$, and pure penguin $B^0 \to f_1 \phi$ decays with the CKM suppressed $\bar b \to \bar d$ transition in the pQCD approach have been given in Tables~\ref{tab:f1rho0-d}, \ref{tab:f1omg-d}, and \ref{tab:f1phi-d}, in which only $B^0 \to f_1(1285) \omega$ has a large and measurable decay rate, $1.0^{+0.6}_{-0.4} \times 10^{-6}$, and the other five decays have such small branching ratios in the range of
$10^{-9} - 10^{-7}$ that it is hard to detect them precisely in a short period. Note that the ideal mixing has been assumed for
$\omega$ and $\phi$ mesons, i.e., $\omega \equiv (u\bar u+ d\bar d)/\sqrt{2}$ and $\phi \equiv s\bar s$. By employing the same distribution amplitudes but with slightly different decay constants for $\rho$ and $\omega$, the corresponding $(u\bar u - d\bar d)/\sqrt{2}$ and $(u\bar u + d\bar d)/\sqrt{2}$ components have 
dramatically different effects, i.e., being  destructive(constructive)
to $B^0 \to f_1 \rho^0(\omega)$ decays. Together with
interferences at different levels between $f_{1q} (\rho^0, \omega)$ and $f_{1s} (\rho^0, \omega)$, we finally obtain $Br(B^0 \to f_1(1285) \rho^0)_{\rm pQCD} \gtrsim Br(B^0 \to f_1(1420) \rho^0)_{\rm pQCD}$ and $Br(B^0 \to f_1(1285) \omega)_{\rm pQCD} > Br(B^0 \to f_1(1420) \omega)_{\rm pQCD}$ within uncertainties, but with a very consistent decay rate and decay pattern as given in the QCDF approach. Careful analysis shows that $B^0 \to f_1 \rho^0$ decays only include negligible color-suppressed tree contributions. For the $B^0 \to f_1 \phi$ mode, the {\it CP}-averaged branching ratios predicted in the pQCD approach are $8.9^{+5.5}_{-3.8} \times 10^{-9}$ and $3.7^{+2.8}_{-2.4} \times 10^{-9}$, respectively, which are basically consistent with but slightly larger than those
obtained in the QCDF approach.

\item[(7)]
As shown in Tables~\ref{tab:f1rho0-s}-\ref{tab:f1phi-s}, the $B_s^0 \to f_1 V$ decays are studied for the first time in the literature. The {\it CP}-averaged branching ratios of $B_s^0 \to f_1 (\rho^0, \omega, \bar K^{*0})$ predicted in the pQCD approach are of the order of
$10^{-7}$ within large theoretical errors, apart from $B_s^0
\to f_1 \phi$ modes with large decay rates around ${\cal O}(10^{-5})$. In light of the measurements of $B_d^0 \to K^+ K^-$ with decay rate $1.3 \pm 0.5 \times 10^{-7}$ and
$B_s^0 \to \pi^+ \pi^-$ with branching ratio
$7.6 \pm 1.9 \times 10^{-7}$~\cite{Agashe:2014kda,Aaltonen:2011jv,Amhis:2014hma},
it is therefore expected that the above-mentioned
$B_s^0 \to f_1 V$ decay modes can be
generally accessed at the running of LHCb and the forthcoming
Belle-II
experiments with a large number of $B_s^0 \bar B_s^0$ events
in the near future.
The interferences between $B_s^0 \to f_{1q} V$ and $B_s^0 \to f_{1s} V$ channels lead to the following relations in $B_s^0 \to f_1 V$ decays with errors:
\beq
Br(B_s^0 \to f_1(1285) (\rho^0, \omega))_{\rm pQCD} &<&
Br(B_s^0 \to f_1(1420) (\rho^0, \omega))_{\rm pQCD}\;, \non
Br(B_s^0 \to f_1(1285) (\bar K^{*0}, \phi))_{\rm pQCD} &\sim&
Br(B_s^0 \to f_1(1420) (\bar K^{*0}, \phi))_{\rm pQCD}\;.
\eeq
Note that, unlike $B^0 \to f_1 (\rho^0, \omega)$ decays, $B_s^0 \to f_1 (\rho^0, \omega)$ ones are all governed by the penguin-dominated
amplitudes with very small, color-suppressed tree contributions. Because of dominant factorizable emission contributions with a $B_s^0 \to f_{1s}$ transition and no $B_s^0 \to (\rho^0, \omega)$ transition, then $Br(B_s^0 \to f_1(1285) (\rho^0, \omega))$ is
smaller than $Br(B_s^0 \to f_1(1420) (\rho^0, \omega))$
as a naive expectation.
Relative to CKM-favored $B \to f_1 K^*$ decays, the $B_s^0 \to f_1 \bar K^{*0}$ ones have significantly smaller branching ratios because they involve a suppressed factor 0.22 in the decay amplitudes.
The penguin-dominated $B_s^0 \to f_1 \phi$ decays with negligibly
small
color-suppressed tree amplitudes have the branching ratios
as $14.7^{+8.7}_{-6.4} \times 10^{-6}$ and $16.2^{+9.9}_{-7.6} \times 10^{-6}$, respectively. When the tree contaminations are turned off, the decay rates become $14.9 \times 10^{-6}$ and $16.1 \times 10^{-6}$ correspondingly, as far as the central values are concerned. As shown in Table~\ref{tab:DAs-f1phi-s}, one can easily observe that the overall constructive(destructive) interferences in three polarizations between $B_s^0 \to f_{1q} \phi$ and $B_s^0 \to f_{1s} \phi$ modes result in the approximately equivalent {\it CP}-averaged branching ratios as mentioned previously. Furthermore, the dominance of the $B_s^0 \to f_{1s} \phi$ channel leads to a decay rate of $B_s^0 \to f_1(1420) \phi$ similar to that of $B_s^0 \to \phi \phi$~\cite{Zou:2015iwa}, while the comparable $B_s^0 \to f_{1q} \phi$ and $B_s^0 \to f_{1s} \phi$
with constructive effects make $Br(B_s^0 \to f_1(1285) \phi)$ highly different from $Br(B_s^0 \to \omega \phi)$, with a factor around ${\cal O}(10^2)$, which will be
tested by the near future LHCb and/or Belle-II measurements.
Because of the possibilities of new discoveries, the search for
NP in the $B_s$ system will be the main focus of the
forthcoming experiments at LHCb and Belle-II. Several
charmless penguin-dominated $B_s$ decays such as
$B_s^0 \to \phi \phi$ can provide ideal places to
search for NP.
In light of the similar behavior between $f_1$ and $\phi$ and
the comparable and large decay rates between $B_s^0 \to f_1 \phi$ and $B_s^0 \to \phi \phi$, it is therefore expected that the $B_s^0 \to f_1 \phi$ decays can provide effective constraints on the $B_s^0-\bar B_s^0$ mixing phase, CKM unitary triangle, and even
NP signals complementarily.

\item[(8)]
Frankly speaking, as can easily be seen in Tables~\ref{tab:f1rhop}-\ref{tab:f1phi-s}, the theoretical predictions
calculated in the pQCD approach suffer from large
errors induced by the still less constrained uncertainties in the light-cone distribution amplitudes
involved in both initial and final states. Here, we then define some interesting ratios of the branching ratios
for the selected decay modes. As generally expected, if the selected decay modes in a ratio
have similar dependence on a specific input parameter, the error induced by the uncertainty of this input parameter
will be largely canceled in the ratio, even if one cannot make an explicit factorization for this parameter.
From the experimental side, we know that the ratios of the branching ratios generally could be measured
with a better accuracy than that for individual branching ratios.
For the sake of the possibility of the experimental measurements, we here define
the following nine ratios out of the branching ratios
of ten decay modes, i.e., $B^+ \to f_1 \rho^+$, $B^{+,0} \to
f_1 K^{*+,0}$, $B^0 \to f_1 \omega$, and $B_s \to f_1 \phi$,
with relatively large branching ratios around $10^{-6}$:
\beq
R_{f_1\rho}^u &\equiv& \frac{Br(B^+ \to f_1(1285) \rho^+)}{Br(B^+ \to f_1(1420) \rho^+)}
 = 4.81^{+0.21}
       _{-0.35}\;, \qquad
R_{f_1 K^*}^u \equiv \frac{Br(B^+ \to f_1(1285) K^{*+})}{Br(B^+ \to f_1(1420) K^{*+})}
 = 1.44^{+0.69}_{-0.56}
\;, \label{eq:r-f1rhokst-u} \\
R_{f_1 K^*}^d &\equiv& \frac{Br(B^0 \to f_1(1285) K^{*0})}{Br(B^0 \to f_1(1420) K^{*0})}
= 1.14^{+0.54}_{-0.47}
\;, \qquad
R_{f_1 \omega}^d \equiv \frac{Br(B^0 \to f_1(1285) \omega)}{Br(B^0 \to f_1(1420) \omega)}
= 5.29^{+0.58}_{-0.71}
\;,\label{eq:r-f1kstomg-d}
\eeq
 \beq
R_{f_1 \phi}^s &\equiv& \frac{Br(B_s^0 \to f_1(1285) \phi)}{Br(B_s^0 \to f_1(1420) \phi)}
= 0.91^{+0.40}_{-0.30} \;; \label{eq:r-f1phi-s}
\eeq
\beq
R_{\rho/K^*}^{uu}[f_1(1285)] &\equiv&
 \frac{Br(B^+ \to f_1(1285) \rho^+)}{Br(B^+ \to f_1(1285) K^{*+})}
 = 1.72^{+0.86}_{-0.88} \;,
 \label{eq:r-uu-12} \\
R_{\rho/K^*}^{uu}[f_1(1420)] &\equiv&
 \frac{Br(B^+ \to f_1(1420) \rho^+)}{Br(B^+ \to f_1(1420) K^{*+})}
 = 0.52^{+0.36}_{-0.32} \;,
 \label{eq:r-uu-14} \\
R_{\phi/K^*}^{sd}[f_1(1285)] &\equiv&
 \frac{Br(B_s^0 \to f_1(1285) \phi)}{Br(B^0 \to f_1(1285) K^{*0})}
 = 2.97^{+1.16}_{-0.94} \;,
 \label{eq:r-sd-12}\\
R_{\phi/K^*}^{sd}[f_1(1420)] &\equiv&
 \frac{Br(B_s^0 \to f_1(1420) \phi)}{Br(B^0 \to f_1(1420) K^{*0})}
 = 3.71^{+0.86}_{-0.90} \;,
 \label{eq:r-sd-14}
\eeq
where the individual errors have been added in quadrature.
One can see from the numerical results in the above equations that the total error has been reduced to
$\sim 10\%$ for the ratio $R^u_{f_1 \rho}$, but still remains large, around $\sim 70\%$, for the ratio
$R^{uu}_{\rho/K^*}[f_1(1420)]$.
These ratios will be tested by future precise $B$ meson experiments
and could be used to explore the flavor symmetry in these modes and to further determine
the mixing angle $\phi_{f_1}$ between $f_{1q}$ and $f_{1s}$ states in the quark-flavor basis.
Note that the variations of hadronic parameters in $\rho$, $K^*$, and $\phi$ distribution
amplitudes are not considered in the last four ratios for convenience.

\end{enumerate}

\begin{table}[t]
\caption{ Same as Table~\ref{tab:f1rhop} but for $B_s^0 \to f_1 \phi$ decays.}
\label{tab:f1phi-s}
 \begin{center}\vspace{-0.3cm}{
\begin{tabular}[t]{c|c||c|c||c|c}
\hline  \hline
   \multicolumn{2}{c||}{Decay Modes}   &  \multicolumn{2}{c||}{$B_s^0 \to f_1(1285) \phi$} &  \multicolumn{2}{c}{$B_s^0 \to f_1(1420) \phi$}  \\
   \hline
 Parameter  & Definition & This work &   QCDF
 & This work &   QCDF   \\
\hline \hline
  BR($10^{-6}$)        & $\Gamma/ \Gamma_{\rm total}$
  &$14.7^{+6.1+3.3+3.0+1.7+3.9+0.1}_{-4.1-2.7-2.6-1.4-2.8-0.0}
  $
  &$-$
  &$\hspace{0.18cm}
16.2^{+5.9+2.0+7.4+1.3+1.8+0.0}_{-4.1-1.9-5.7-1.6-1.6-0.0}
  $
  &$-$
 \\
 \hline \hline
 $f_L(\%)$      & $|{\cal A}_L|^2$
 &$56.7^{+0.6+2.4+3.2+1.5+0.6+0.1}_{-0.4-2.3-3.7-1.5-1.0-0.1}
  $
 &$-$
 &$82.1^{+1.8+2.0+3.2+1.1+2.4+0.1}_{-1.9-1.8-3.1-0.9-3.6-0.0}
  $
 &$-$
 \\
 $f_{||}(\%)$   & $|{\cal A}_{||}|^2$
 &$23.7^{+0.2+1.2+1.9+0.7+0.5+0.0}_{-0.3-1.3-1.9-0.8-0.4-0.1}
  $
 &$-$
 &$10.5^{+1.1+1.0+1.8+0.5+2.1+0.0}_{-1.0-1.1-1.7-0.6-1.4-0.0}
  $
 &$-$
  \\
 $f_{\perp}(\%)$& $|{\cal A}_\perp|^2$
 &$19.6^{+0.2+1.2+1.7+0.7+0.5+0.1}_{-0.3-1.1-1.4-0.7-0.2-0.0}
$
 &$-$
 &$\hspace{0.18cm}
7.4^{+0.8+0.7+1.4+0.4+1.5+0.0}_{-0.8-0.9-1.5-0.5-1.0-0.0}
  $
 &$-$
 \\
 \hline \hline
 $\phi_{||}$(rad)& $\arg\frac{{\cal A}_{||}}{{\cal A}_L}$
 &$\hspace{0.18cm}
2.9^{+0.1+0.0+0.1+0.0+0.0+0.0}_{-0.0-0.0-0.0-0.0-0.0-0.0}
$
 &$-$
 &$\hspace{0.18cm}
2.6^{+0.0+0.0+0.2+0.0+0.0+0.0}_{-0.0-0.0-0.0-0.0-0.0-0.0}
  $
 &$-$
 \\
 $\phi_{\perp}$(rad)& $\arg\frac{{\cal A}_{\perp}}{{\cal A}_L}$
 &$\hspace{0.18cm}
2.9^{+0.1+0.1+0.1+0.0+0.0+0.0}_{-0.0-0.0-0.0-0.0-0.0-0.0}
  $
 &$-$
 &$\hspace{0.18cm}
2.6^{+0.0+0.0+0.2+0.0+0.0+0.0}_{-0.0-0.0-0.0-0.0-0.0-0.0}
  $
 &$-$
  \\
  \hline \hline
 $\acp^{\rm dir}(\%)$& $\frac{\overline{\Gamma}-\Gamma}{\overline{\Gamma}+\Gamma}$
 &$-5.3^{+0.3+0.7+0.7+0.4+0.8+0.2}_{-0.2-0.4-0.5-0.3-0.7-0.1}
  $
 &$-$
 &$\hspace{0.18cm}
2.5^{+0.1+0.7+0.4+0.5+0.2+0.1}_{-0.1-0.6-0.4-0.4-0.3-0.1}
  $
 &$-$
 \\
 $\acp^{\rm dir}(L)(\%)$& $\frac{\bar{f}_L-f_L}{\bar{f}_L+f_L}$
 &$-7.2^{+0.5+1.1+1.2+0.7+0.9+0.3}_{-0.4-1.0-1.1-0.6-1.0-0.2}
  $
 & $-$
 &$\hspace{0.18cm}
2.4^{+0.1+0.6+0.4+0.4+0.5+0.1}_{-0.1-0.5-0.4-0.3-0.2-0.1}
  $
 & $-$
 \\
 $\acp^{\rm dir}(||)(\%)$& $\frac{\bar{f}_{||}-f_{||}}{\bar{f}_{||}+f_{||}}$
 &$-2.7^{+0.1+0.3+0.4+0.1+0.4+0.1}_{-0.0-0.1-0.3-0.0-0.3-0.1}
  $
 &$-$
 &$\hspace{0.18cm}
2.6^{+0.1+1.1+0.2+0.7+0.3+0.1}_{-0.1-0.7-0.4-0.4-0.2-0.1}
  $
 &$-$
 \\
 $\acp^{\rm dir}(\perp)(\%)$& $\frac{\bar{f}_\perp-f_\perp}{\bar{f}_\perp+f_\perp}$
 &$-2.8^{+0.0+0.2+0.4+0.1+0.4+0.1}_{-0.0-0.1-0.4-0.1-0.4-0.1}
$
 &$-$
 &$\hspace{0.18cm}
3.1^{+0.2+1.1+0.2+0.9+0.4+0.1}_{-0.1-0.9-0.4-0.6-0.3-0.1}
  $
 &$-$
 \\ \hline \hline
\end{tabular}}
\end{center}
\end{table}

\begin{table}[hbt]
\caption{The decay amplitudes(in units of $10^{-3}\; \rm{GeV}^3$) of the $B^+ \to f_{1q} \rho^+$ and $B^+ \to f_{1s} \rho^+$ channels in the
$B^+ \to f_1 \rho^+$ decays with three polarizations in the pQCD approach,
where only the central values are quoted
for clarification. Note that the numerical results in the parentheses are the corresponding amplitudes without annihilation contributions.
}
\label{tab:DAs-f1rho-u}
 \begin{center}\vspace{-0.3cm}{
\begin{tabular}[t]{c|c||c|c||c|c}
\hline  \hline
   \multicolumn{2}{c||}{Decay Modes}   &  \multicolumn{2}{c||}{$B^+ \to f_1(1285) \rho^{+}$} &  \multicolumn{2}{c}{$B^+ \to f_1(1420) \rho^{+}$}  \\
   \hline
   \multicolumn{2}{c||}{Channels} & $B^+ \to \rho^{+} f_{1q}$
   & $B^+ \to \rho^{+} f_{1s}$
 & $B^+ \to \rho^{+} f_{1q}$ & $B^+ \to \rho^{+} f_{1s}$   \\
 \hline \hline
 \multicolumn{2}{c||}{$A_L$}
 &$\begin{array}{c}-2.217 - i\; 3.790 \;
 \\ (-2.359 - i\; 3.718)\end{array}$
 &$\begin{array}{c}-0.127 + i\; 0.058 \;
 \\ (-0.127 + i\; 0.058)\end{array}$
 &$\begin{array}{c}-0.987 - i\; 1.688 \;
 \\ (-1.050 - i\; 1.655)\end{array}$
 &$\begin{array}{c} 0.285 - i\; 0.131 \;
 \\ (0.285 - i\; 0.131)\end{array}$
 \\
 \hline
\multicolumn{2}{c||}{$A_N$}
 &$\begin{array}{c}-0.166 - i\; 0.424 \;
 \\ (-0.179 - i\; 0.447)\end{array}$
 &$\begin{array}{c}-0.089 + i\; 0.041 \;
 \\ (-0.089 + i\; 0.041)\end{array}$
 &$\begin{array}{c}-0.073 - i\; 0.187 \;
 \\ (-0.079 - i\; 0.197)\end{array}$
 &$\begin{array}{c} 0.201 - i\; 0.091 \;
 \\ (0.201 - i\; 0.091)\end{array}$
  \\
 \hline
\multicolumn{2}{c||}{$A_T$}
 &$\begin{array}{c}-0.224 - i\; 0.757 \;
 \\ (-0.325 - i\; 0.810)\end{array}$
 &$\begin{array}{c}-0.184 + i\; 0.080 \;
 \\ (-0.184 + i\; 0.080)\end{array}$
 &$\begin{array}{c}-0.107 - i\; 0.331 \;
 \\ (-0.152 - i\; 0.355)\end{array}$
 &$\begin{array}{c} 0.413 - i\; 0.180 \;
 \\ (0.413 - i\; 0.180)\end{array}$
 \\
 \hline \hline
\end{tabular}}
\end{center}
\end{table}

\begin{table}[hbt]
\caption{Same as Table~\ref{tab:DAs-f1rho-u} but for
$B^+ \to f_1  K^{*+}$ decays.}
\label{tab:DAs-f1kst-u}
 \begin{center}\vspace{-0.3cm}{
\begin{tabular}[t]{c|c||c|c||c|c}
\hline  \hline
   \multicolumn{2}{c||}{Decay Modes}   &  \multicolumn{2}{c||}{$B^+ \to f_1(1285) K^{*+}$} &  \multicolumn{2}{c}{$B^+ \to f_1(1420) K^{*+}$}  \\
   \hline
   \multicolumn{2}{c||}{Channels} & $B^+ \to K^{*+} f_{1q}$
   & $B^+ \to K^{*+} f_{1s}$
 & $B^+ \to K^{*+} f_{1q}$ & $B^+ \to K^{*+} f_{1s}$   \\
 \hline \hline
 \multicolumn{2}{c||}{$A_L$}
 &$\begin{array}{c}\hspace{0.18cm}
 0.284 - i\; 1.423 \;
 \\ (0.292 - i\; 0.832)\end{array}$
 &$\begin{array}{c}-0.679 - i\; 0.791 \;
 \\ (-0.672 - i\; 0.224)\end{array}$
 &$\begin{array}{c}\hspace{0.18cm}
 0.127 - i\; 0.634 \;
 \\ (0.130 - i\; 0.370)\end{array}$
 &$\begin{array}{c} 1.524 + i\; 1.776 \;
 \\ (1.510 + i\; 0.502)\end{array}$
 \\
 \hline
\multicolumn{2}{c||}{$A_N$}
 &$\begin{array}{c}-1.078 + i\; 0.436 \;
 \\ (-0.747 - i\; 0.123)\end{array}$
 &$\begin{array}{c}-0.089 + i\; 0.446 \;
 \\ (0.127 - i\; 0.027)\end{array}$
 &$\begin{array}{c}-0.465 + i\; 0.188 \;
 \\ (-0.318 - i\; 0.060)\end{array}$
 &$\begin{array}{c} 0.200 - i\; 1.003 \;
 \\ (-0.285 + i\; 0.062)\end{array}$
  \\
 \hline
\multicolumn{2}{c||}{$A_T$}
 &$\begin{array}{c}-2.166 + i\; 0.866 \;
 \\ (-1.509 - i\; 0.281)\end{array}$
 &$\begin{array}{c}-0.152 + i\; 0.896 \;
 \\ (0.287 - i\; 0.043)\end{array}$
 &$\begin{array}{c}-0.965 + i\; 0.386 \;
 \\ (-0.672 - i\; 0.125)\end{array}$
 &$\begin{array}{c} 0.340 - i\; 2.013 \;
 \\ (-0.643 + i\; 0.097)\end{array}$
 \\
 \hline \hline
\end{tabular}}
\end{center}
\end{table}

\begin{table}[hbt]
\caption{Same as Table~\ref{tab:DAs-f1rho-u} but for
$B^0 \to f_1  K^{*0}$ decays.}
\label{tab:DAs-f1kst-d}
 \begin{center}\vspace{-0.3cm}{
\begin{tabular}[t]{c|c||c|c||c|c}
\hline  \hline
   \multicolumn{2}{c||}{Decay Modes}   &  \multicolumn{2}{c||}{$B^0 \to f_1(1285) K^{*0}$} &  \multicolumn{2}{c}{$B^0 \to f_1(1420) K^{*0}$}  \\
   \hline
   \multicolumn{2}{c||}{Channels} & $B^0 \to K^{*0} f_{1q}$
   & $B^0 \to K^{*0} f_{1s}$
 & $B^0 \to K^{*0} f_{1q}$ & $B^0 \to K^{*0} f_{1s}$   \\
 \hline \hline
 \multicolumn{2}{c||}{$A_L$}
 &$\begin{array}{c}\hspace{0.18cm}
 0.563 - i\; 0.380 \;
 \\ (0.602 + i\; 0.197)\end{array}$
 &$\begin{array}{c}-0.647 - i\; 0.814 \;
 \\ (-0.665 - i\; 0.219)\end{array}$
 &$\begin{array}{c}\hspace{0.18cm}
 0.251 - i\; 0.169 \;
 \\ (0.268 + i\; 0.088)\end{array}$
 &$\begin{array}{c} 1.454 + i\; 1.829 \;
 \\ (1.495 + i\; 0.491)\end{array}$
 \\
 \hline
\multicolumn{2}{c||}{$A_N$}
 &$\begin{array}{c}-0.934 + i\; 0.649 \;
 \\ (-0.588 + i\; 0.066)\end{array}$
 &$\begin{array}{c}-0.104 + i\; 0.466 \;
 \\ (0.126 - i\; 0.027)\end{array}$
 &$\begin{array}{c}-0.416 + i\; 0.289 \;
 \\ (-0.262 + i\; 0.029)\end{array}$
 &$\begin{array}{c} 0.235 - i\; 1.047 \;
 \\ (-0.284 + i\; 0.061)\end{array}$
  \\
 \hline
\multicolumn{2}{c||}{$A_T$}
 &$\begin{array}{c}-1.949 + i\; 1.296 \;
 \\ (-1.253 + i\; 0.113)\end{array}$
 &$\begin{array}{c}-0.159 + i\; 0.920 \;
 \\ (0.289 - i\; 0.044)\end{array}$
 &$\begin{array}{c}-0.868 + i\; 0.577 \;
 \\ (-0.558 + i\; 0.050)\end{array}$
 &$\begin{array}{c} 0.358 - i\; 2.067 \;
 \\ (-0.648 + i\; 0.099)\end{array}$
 \\
 \hline \hline
\end{tabular}}
\end{center}
\end{table}

\begin{table}[hbt]
\caption{Same as Table~\ref{tab:DAs-f1rho-u} but for
$B_s^0 \to f_1 \phi$ decays.}
\label{tab:DAs-f1phi-s}
 \begin{center}\vspace{-0.3cm}{
\begin{tabular}[t]{c|c||c|c||c|c}
\hline  \hline
   \multicolumn{2}{c||}{Decay Modes}   &  \multicolumn{2}{c||}{$B_s^0 \to f_1(1285) \phi$} &  \multicolumn{2}{c}{$B_s^0 \to f_1(1420) \phi$}  \\
   \hline
   \multicolumn{2}{c||}{Channels} & $B_s^0 \to \phi f_{1q}$
   & $B_s^0 \to \phi f_{1s}$
 & $B_s^0 \to \phi f_{1q}$ & $B_s^0 \to \phi f_{1s}$   \\
 \hline \hline
 \multicolumn{2}{c||}{$A_L$}
 &$\begin{array}{c}-1.624 + i\; 0.044 \;
 \\ (-1.624 + i\; 0.044)\end{array}$
 &$\begin{array}{c}-2.502 - i\; 0.542 \;
 \\ (-2.463 - i\; 0.139)\end{array}$
 &$\begin{array}{c}-0.723 + i\; 0.020 \;
 \\ (-0.723 + i\; 0.020)\end{array}$
 &$\begin{array}{c} 5.621 + i\; 1.218 \;
 \\ (5.533 + i\; 0.312)\end{array}$
 \\
 \hline
\multicolumn{2}{c||}{$A_N$}
 &$\begin{array}{c}-1.077 + i\; 0.093 \;
 \\ (-1.077 + i\; 0.093)\end{array}$
 &$\begin{array}{c}-0.763 + i\; 0.164 \;
 \\ (-0.813 + i\; 0.081)\end{array}$
 &$\begin{array}{c}-0.480 + i\; 0.041 \;
 \\ (-0.480 + i\; 0.041)\end{array}$
 &$\begin{array}{c} 1.714 - i\; 0.368 \;
 \\ (1.827 - i\; 0.181)\end{array}$
  \\
 \hline
\multicolumn{2}{c||}{$A_T$}
 &$\begin{array}{c}-2.245 + i\; 0.163 \;
 \\ (-2.245 + i\; 0.163)\end{array}$
 &$\begin{array}{c}-1.479 + i\; 0.307 \;
 \\ (-1.576 + i\; 0.169)\end{array}$
 &$\begin{array}{c}-1.000 + i\; 0.073 \;
 \\ (-1.000 + i\; 0.073)\end{array}$
 &$\begin{array}{c} 3.322 - i\; 0.690 \;
 \\ (3.539 - i\; 0.379)\end{array}$
 \\
 \hline \hline
\end{tabular}}
\end{center}
\end{table}

\subsection{{\it CP}-averaged polarization fractions and relative phases}

In this section we will analyze the {\it CP}-averaged polarization fractions and relative phases for 20 nonleptonic $B \to f_1 V$ decays in the
pQCD approach. Based on the helicity amplitudes, we can define the
transversity ones as follows:
\beq
{\cal A}_{L}&=& \xi m^{2}_{B} A_{L}, \qquad
{\cal A}_{\parallel}=\xi \sqrt{2}m^{2}_{B} A_{N}, \qquad
{\cal A}_{\perp}=\xi m_{V} m_{f_1}
\sqrt{2(r^{2}-1)} A_{T}\;,
\label{eq:ase}
\eeq
for the longitudinal, parallel, and perpendicular polarizations,
respectively, with the normalization factor
$\xi=\sqrt{G^2_{F}P_c/(16\pi m^2_{B}\Gamma)}$ and the ratio
$r=P_{2}\cdot P_{3}/(m_{V}\cdot
m_{f_1})$. These amplitudes satisfy the relation,
\begin{eqnarray}
|{\cal A}_{L}|^2+|{\cal A}_{\parallel}|^2+|{\cal A}_{\perp}|^2=1 
\end{eqnarray}
following the summation in Eq.~(\ref{dr1}).
Since the transverse-helicity contributions can manifest themselves through polarization observables, we therefore define {\it CP}-averaged fractions in three polarizations $f_{L}$, $f_\parallel$,
and $f_\perp$ as the following,
\beq
f_{L,||,\perp}&\equiv& \frac{|{\cal
A}_{L,||,\perp}|^2}{|{\cal A}_L|^2+|{\cal A}_{||}|^2+|{\cal
A}_{\perp}|^2} = |{\cal
A}_{L,||,\perp}|^2.\label{eq:pf}
\eeq
With the above transversity amplitudes shown in Eq.~(\ref{eq:ase}), the relative phases
$\phi_{\parallel}$ and $\phi_{\perp}$ can be defined as
 \beq
 \phi_{\parallel} &=& \arg\frac{{\cal A}_{\parallel}}{{\cal A}_L} \;,
   \qquad
 \phi_{\perp} = \arg\frac{{\cal A}_{\perp}}{{\cal A}_L} \;.
 \eeq

As aforementioned, by picking up higher power $r_i^2$ terms
that were previously neglected, especially in the virtual
gluon and/or quark propagators, the global agreement with
data for $B \to VV$ decays has been greatly improved
in the pQCD approach theoretically~\cite{Zou:2015iwa}.
In particular, the polarization fractions for
penguin-dominated $B \to VV$ decays contributed from
large transverse amplitudes are well understood with
this improvement. In the present work, we followed
this treatment in charmless hadronic $B \to f_1 V$ decays.
The theoretical predictions of polarization fractions
and relative phases
have been collected in Tables~\ref{tab:f1rhop}-\ref{tab:f1phi-s}
within errors. Based on these numerical results,
some remarks are given as follows:
\begin{itemize}
\item
Overall, as can straightforwardly be seen in Tables~\ref{tab:f1rhop}-\ref{tab:f1phi-s}, the decays with large longitudinal polarization contributions include $B^+ \to f_1 \rho^+$, $B^{+,0} \to f_1(1420) K^{*+,0}$, $B^0 \to f_1(1285) (\rho^0, \omega)$, $B^0 \to f_1 \phi$, $B_s^0 \to f_1 \rho^0$,
$B_s^0 \to f_1(1285) \omega$, and $B_s^0 \to f_1(1420) \phi$, while the $B^{+,0} \to f_1(1285) K^{*+,0}$, $B^0 \to f_1(1420) \rho^0$, and $B_s^0 \to f_1(1285) \bar K^{*0}$ modes are governed by large transverse contributions. The other channels, such as
$B_{(s)}^0 \to f_1(1420) \omega$, $B_s^0 \to f_1(1420) \bar K^{*0}$, and $B_s^0 \to f_1(1285) \phi$, have longitudinal polarization fractions around 50\% competing with transverse ones
within theoretical uncertainties. These predicted {\it CP}-averaged
polarization fractions will be tested
at LHCb and/or Belle-II
to further explore the decay mechanism with
helicities associated with experimental confirmations on the decay rates.

\item
Theoretically, the pQCD predictions of polarization fractions
$f_L$ and $f_T(=f_\parallel+f_\perp = 1- f_L)$ for $B^+ \to f_1 \rho^+$ modes are
 \beq
 f_L(B^+ \to f_1(1285) \rho^+) &=& 96.3^{+0.5}_{-0.4}\%\;, \qquad
 f_T(B^+ \to f_1(1285) \rho^+) = 3.7^{+0.3}_{-0.3}\%\;; \\
 f_L(B^+ \to f_1(1420) \rho^+) &=& 90.5^{+3.1}_{-5.1}\%\;, \qquad
 f_T(B^+ \to f_1(1420) \rho^+) = 9.5^{+3.5}_{-2.4}\%\;.
 \eeq
In the QCDF approach, the longitudinal polarization
fractions
for $B^+\to f_1 \rho^+$ decays have also been available as follows~\cite{Cheng:2008gxa}:
  \beq
 f_L(B^+ \to f_1(1285) \rho^+) &=& 90^{+4}_{-3}\%\;, \qquad
 f_L(B^+ \to f_1(1420) \rho^+) = 93^{+4}_{-3}\%\;;
  \eeq
It is obvious to see that the fractions predicted in both
pQCD and QCDF approaches are consistent with each other within errors, which will be further examined by combining with large
{\it CP}-averaged branching ratios through the
LHCb and/or Belle-II measurements in the near future.
As a matter of fact, the studies on color-allowed tree-dominated $B$ decays in the pQCD approach usually
agree with those in the QCDF one within theoretical uncertainties,
e.g., $B^0 \to \rho^+ \rho^-$~\cite{Zou:2015iwa,Cheng:2009cn}. But, it is not the case in penguin-dominated and weak-annihilation-dominated modes.

\item
For penguin-dominated $B^{+,0} \to f_1 K^{*+,0}$ decays with a $\bar b \to \bar s$ transition, one can find the polarization fractions from Tables~\ref{tab:f1kstp} and \ref{tab:f1kst0} predicted in the pQCD approach as follows: 
 \beq
 f_L(B^+ \to f_1(1285) K^{*+}) &=& 23.5^{+5.8}_{-4.0}\%\;, \qquad
 f_T(B^+ \to f_1(1285) K^{*+}) = 76.5^{+2.9}_{-4.1}\%\;; \\
 f_L(B^+ \to f_1(1420) K^{*+}) &=& 69.3^{+11.4}_{-12.5}\%\;, \qquad
 f_T(B^+ \to f_1(1420) K^{*+}) = 30.7^{+8.5}_{-8.0}\%\;,
 \eeq
and
 \beq
 f_L(B^0 \to f_1(1285) K^{*0}) &=& 15.8^{+6.7}_{-3.4}\%\;, \qquad
 f_T(B^0 \to f_1(1285) K^{*0}) = 84.2^{+2.5}_{-4.8}\%\;; \\
 f_L(B^0 \to f_1(1420) K^{*0}) &=& 71.0^{+12.0}_{-13.1}\%\;, \qquad
 f_T(B^0 \to f_1(1420) K^{*0}) = 29.0^{+9.4}_{-8.6}\%\;,
 \eeq
which show the pattern of polarization fractions in the pQCD approach,
\beq
f_L(B^{+,0} \to f_1(1285) K^{*+,0}) &<& f_T(B^{+,0} \to f_1(1285) K^{*+,0})\;, \non
f_L(B^{+,0} \to f_1(1420) K^{*+,0}) &>& f_T(B^{+,0} \to f_1(1420) K^{*+,0})\;;
\label{eq:c1}
\eeq
and
\beq
f_L(B^{+,0} \to f_1(1285) K^{*+,0}) &<& f_L(B^{+,0} \to f_1(1420) K^{*+,0})\;, \non
f_T(B^{+,0} \to f_1(1285) K^{*+,0}) &>& f_T(B^{+,0} \to f_1(1420) K^{*+,0})\;.
\label{eq:c2}
\eeq
The decay amplitudes with three polarizations presented in Table~\ref{tab:DAs-f1kst-u} show that, for $B^{+,0} \to f_1(1285)[f_1(1420)] K^{*+,0}$ decays, the significantly constructive(destructive) interferences in transverse polarizations between $B^{+,0} \to f_{1q} K^{*+,0}$ and $B^{+,0} \to f_{1s} K^{*+,0}$ finally result in somewhat smaller(larger) longitudinal polarization fractions, correspondingly, although the cancellations of the real(imaginary) decay amplitudes occur at different levels in the longitudinal polarization.

In Ref.~\cite{Cheng:2008gxa}, the authors predicted longitudinal polarization fractions for the $B^{+,0} \to f_1 K^{*+,0}$ modes in the QCDF approach as follows:
 \beq
 f_L(B^+ \to f_1(1285) K^{*+}) &=& 47^{+49}_{-45}\%\;, \qquad
 f_L(B^+ \to f_1(1420) K^{*+}) = 64^{+37}_{-61}\%\;;
 \eeq
and
 \beq
 f_L(B^0 \to f_1(1285) K^{*0}) &=& 45^{+55}_{-50}\%\;, \qquad
 f_L(B^0 \to f_1(1420) K^{*0}) = 64^{+38}_{-61}\%\; ,
 \eeq
which show the longitudinal polarization fractions roughly
competing with the transverse ones for $B^{+,0} \to f_1 K^{*+,0}$ and the relation $f_L(B^{+,0} \to f_1(1285) K^{*+,0}) \sim f_L(B^{+,0} \to f_1(1420) K^{*+,0})$ within large theoretical errors, though, as far as central values are concerned, the same pattern as in Eqs.~(\ref{eq:c1}) and (\ref{eq:c2}) can also be obtained in the QCDF framework.

However, with the same $\bar b \to \bar s$ transition, the almost
pure penguin $B_s^0 \to f_1 \phi$ decays are dominated by
longitudinal contributions with the polarization fractions as
 \beq
 f_L(B_s^0 \to f_1(1285) \phi) &=& 56.7^{+4.4}_{-4.7}\%\;, \qquad
 f_T(B_s^0 \to f_1(1285) \phi) = 43.3^{+3.3}_{-3.2}\%\;; \\
 f_L(B_s^0 \to f_1(1420) \phi) &=& 82.1^{+4.9}_{-5.5}\%\;, \qquad
 f_T(B_s^0 \to f_1(1420) \phi) = 17.9^{+4.0}_{-3.2}\%\;,
 \eeq
which are different from $B^{+,0} \to f_1 K^{*+,0}$ decays, apart from
the similar pattern $f_L(B_s^0 \to f_1(1285) \phi) < f_L(B_s^0 \to f_1(1420) \phi)$.
To our best knowledge, $B_s^0 \to f_1 V$ decays in this paper are indeed investigated theoretically for the first time in the literature. It is therefore expected that these polarization fractions combined with large {\it CP}-averaged branching ratios of the order of $10^{-5}$ will be tested soon at the LHCb and/or Belle-II experiments with a large
amount of events of $B_s\bar B_s$ production.

\item
For $B^0 \to f_1 (\rho^0, \omega, \phi)$ decays with $\bar b\to \bar d$ transition, the polarization fractions have also been
predicted in the QCDF and pQCD approaches. From Tables~\ref{tab:f1rho0-d}, \ref{tab:f1omg-d}, and \ref{tab:f1phi-d}, one can observe that the pQCD predictions of longitudinal polarization fractions agree roughly with those QCDF values within very large theoretical errors. However, in terms of central values, it is noted that the above-mentioned six modes are all governed by the longitudinal contributions in the QCDF approach, which is different from those given in the pQCD approach to some extent.

For $B^0 \to f_1 \omega$ decays for example, the leading-order
QCD dynamics and the interferences between $B^0 \to f_{1q}
\omega$ and $B^0 \to f_{1s} \omega$ make $f_L(B^0 \to f_1(1285)
\omega) = 60.1^{+8.9}_{-8.3}\%$, while $f_L(B^0 \to f_1(1420)
\omega) = 45.3^{+12.1}_{-11.7}\%$, where, in terms of the
central value, the latter
polarization fraction presents
a striking contrast to the value of 
$f_L(B^0 \to f_1(1420) \omega) = 86\%$
obtained in the QCDF approach. Due to the analogous
behavior between $f_{1}$ and $V$ and the dominance of $f_{1q}$ in
the $f_1(1285)$ state, it is then expected that the longitudinal polarization fraction $f_L(B^0 \to f_1(1285) \omega)$ is more like
that of $f_L(B^0 \to \omega \omega)$. The theoretical prediction
of $f_L(B^0 \to \omega \omega) \sim 66\%$ made in the pQCD approach~\cite{Zou:2015iwa} indeed confirms this similarity.
Of course, the analogy between $f_L(B^0 \to f_1(1285) \omega) \sim 86\%$ and $f_L(B^0 \to \omega\omega)\sim 94\%$ can also be manifested in the QCDF framework. Therefore, this phenomenology should be tested by the near future measurements at LHCb and/or Belle-II experiments to distinguish these two popular factorization approaches based on QCD dynamics.

As we know, the color-suppressed tree-dominated $B^0 \to \rho^0 \rho^0$ decay is governed by large transverse amplitudes, but with a too small branching ratio to be comparable to the data at leading order in the pQCD approach~\cite{Zou:2015iwa,Li:2006cva}.
After including partial next-to-leading order contributions such as vertex corrections, quark loop, and chromomagnetic penguin~\cite{Li:2006cva}, even the 
Glauber-gluon factor~\cite{Liu:2015sra}, the predicted branching ratio and longitudinal polarization fraction of $B^0 \to \rho^0 \rho^0$ decay are simultaneously in good agreement with the existing measurements~\cite{Amhis:2014hma}. Of course, it is noted that the small longitudinal polarization fraction $0.21^{+0.18}_{-0.22} \pm 0.13$~\cite{Adachi:2012cz} provided by the Belle Collaboration
cannot match
with that given by the BABAR~\cite{Aubert:2008au} and LHCb~\cite{Aaij:2015ria} collaborations, respectively. Therefore, it is important to make a refined measurement at the forthcoming Belle-II experiment to give a definitive conclusion. The stringent measurements on the $B^0 \to f_1 \omega$ decays are also sensitive to the color-suppressed tree-amplitude, which may tell us whether they have the same issue as the $B^0 \to \rho^0 \rho^0$ mode.

Moreover, for pure penguin $B^0 \to f_1 \phi$ decays, although the central values of longitudinal polarization fractions in the pQCD approach are somewhat smaller than those in the QCDF method, the predictions of polarization fractions within large
theoretical errors
are consistent with each other, and $B^0 \to f_1 \phi$ decays are dominated by the longitudinal polarization contributions in both the pQCD and QCDF approaches. However, the predictions of polarization fractions for $B^0 \to f_1 \rho^0$ decays in the pQCD approach
show that the $B^0 \to f_1(1285)[f_1(1420)] \rho^0$ channel seems
to be governed by the longitudinal(transverse) polarization amplitudes(see Table~\ref{tab:f1rho0-d} for detail), which indicates a significantly different understanding in the QCDF framework. In QCDF, the $B^0 \to f_1 \rho^0$ decays have similar and dominantly large longitudinal polarization fractions. These phenomenologies await precise measurements in the future to further explore the unknown dynamics in the axial-vector $f_1$ states, as well as in the decay channels.

\item
For $B_s^0 \to f_1 (\rho^0, \omega, \bar K^{*0})$ decays, the pQCD predictions of polarization fractions have been presented in Tables~\ref{tab:f1rho0-s}, \ref{tab:f1omg-s}, and \ref{tab:f1kst0b}, respectively. One can easily observe that
(a) the $B_s^0 \to f_1 \rho^0$ decays are dominated by the longitudinal contributions with polarization fractions $f_L(B_s^0 \to f_1(1285) \rho^0)= 79.8^{+2.1}_{-3.7}\% \sim f_L(B_s^0 \to f_1(1420) \rho^0)= 80.8^{+1.8}_{-2.8}\%$; (b) the longitudinal amplitudes dominate the $B_s^0 \to f_1(1285) \omega$ mode with $f_L(B_s^0 \to f_1(1285) \omega) = 81.8^{+11.1}_{-11.5}\%$ and contribute to the $B_s^0 \to f_1(1420) \omega$ channel, almost competing with the transverse ones with
$f_L(B_s^0 \to f_1(1420) \omega) = 50.9^{+5.3}_{-5.1}\%$, respectively;
and (c) the $B_s^0 \to f_1(1285) \bar K^{*0}$ decay is governed by the transverse amplitudes, contrary to $B_s^0 \to f_1(1285) (\rho^0, \omega)$, with longitudinal polarization fraction $39.2^{+9.2}_{-8.5}\%$. However, similar to the $B_s^0 \to f_1(1420) \omega$ mode, the $B_s^0 \to f_1(1420) \bar K^{*0}$ channel also has nearly equivalent contributions from both longitudinal and transverse polarizations. These predictions of $B_s^0 \to f_1 V$ decays in the pQCD approach could be tested by future measurements at LHCb and/or Belle-II, or even at Circular Electron Positron Collider(CEPC) factories.

\item
In this work, the relative phases(in units of rad) $\phi_{\parallel}$
and $\phi_{\perp}$ of $B \to f_1 V$ decays are also studied for
the first time in the literature and the relevant numerical results have been given in Tables~\ref{tab:f1rhop}-\ref{tab:f1phi-s}.
Up to now, no data or theoretical predictions of
these relative phases in the considered 20 nonleptonic decays of  $B \to f_1 V$ have been available. It is therefore expected that our predictions in the pQCD approach
could be confronted with future LHCb and/or Belle-II experiments, as well as the theoretical comparison within the framework of QCDF, SCET, and so forth.

\end{itemize}

Again, as stressed in the above section, no results are available yet for both theoretical and experimental aspects of 
$B \to f_1 V$ decays. Hence, we have to wait for the examinations to our pQCD analyses in the $B \to f_1 V$ decays from
(near) future experiments.

\subsection{Direct {\it CP}-violating asymmetries}

Now we come to the evaluations of direct {\it CP}-violating asymmetries
of $B \to f_1 V$ decays in the pQCD approach.
The direct {\it CP} violation
$\acp^{\rm dir}$ can be defined as
 \beq
\acp^{\rm dir} &\equiv& \frac{\overline{\Gamma} - \Gamma}{\overline{\Gamma} + \Gamma}
=  \frac{|\overline{ A}_{\rm final}|^2 - |{ A}_{\rm final}|^2}{
 |\overline{ A}_{\rm final}|^2+|{ A}_{\rm final}|^2},
\label{eq:acp1}
\eeq
where $\Gamma$ and ${ A}_{\rm final}$ stand for the decay rate and decay amplitude of $B \to f_1 V$,
while $\overline{\Gamma}$ and $\ov{ A}_{\rm final}$ denote the
charge conjugation ones, correspondingly. It should be mentioned that
here we will not distinguish charged $B^+$ mesons from neutral $B^0$
and $B_s^0$ ones in Eq.~(\ref{eq:acp1}) because we are only
considering the direct {\it CP} violation. Meanwhile, according
to Ref.~\cite{Beneke:2006hg}, the direct-induced {\it CP} asymmetries
can also be studied with the help of helicity amplitudes.
Usually, we need to combine three polarization fractions, as
shown in Eq.(\ref{eq:pf}), with those corresponding conjugation
ones of $B$ decays and then to quote the resultant six observables
to define direct {\it CP} violations of $B \to f_1 V$ decays
in the transversity basis as follows:
\beq
\acp^{\rm dir,\ell}&=&
\frac{\bar f_\ell- f_\ell}{\bar f_\ell+
f_\ell}\;,
\eeq
where $\ell=L,\parallel,\perp$ and the definition of
$\bar f$ is the same as that in~Eq.(\ref{eq:pf}) but for the
corresponding $\bar B$ decays.

Using Eq.~(\ref{eq:acp1}),
we calculate the pQCD predictions of direct {\it CP}-violating asymmetries
in the $B \to f_1 V$ decays and present the results as shown in
Tables~\ref{tab:f1rhop}-\ref{tab:f1phi-s}. Based on these numerical
values, some comments are in order:
\begin{enumerate}
\item[(1)]
Generally speaking, the $\Delta S=0$ decays including $B^0 \to f_1 (\rho^0, \omega)$ and $B_s^0 \to f_1 \bar {K}^{*0}$ and the $\Delta S =1$ decays such as $B^+ \to f_1 K^{*+}$ and $B_s^0 \to f_1 (\rho^0, \omega)$ have large direct {\it CP} violations $\acp^{\rm dir}$ within still large theoretical errors, except for
$B^+ \to f_1 \rho^+$, $B^0 \to f_1 (\phi, K^{*0})$,
and $B_s^0 \to f_1 \phi$ modes giving
{\it CP}-violating asymmetries less than 10\%, because of either 
extremely small penguin contaminations, e.g., $B^+ \to f_1 \rho^+$, or negligible tree pollution, e.g., $B^0 \to f_1 K^{*0}$.
In particular, the $B^0 \to f_1 \phi$ modes have zero direct {\it CP} asymmetries in the SM because of pure penguin contributions.
However,
if the experimental measurements of the direct {\it CP} asymmetries
of $B^0 \to f_1 \phi$ decays exhibit large nonzero values,
this will indicate the existence of new physics beyond the SM and will provide a
very promising place to search
for possible exotic effects.

\item[(2)]
As can be seen in Tables~\ref{tab:f1rhop} and \ref{tab:f1rho0-d}, the direct {\it CP} asymmetries of $B \to f_1 \rho$ decays in the pQCD approach are
\beq
\acp^{\rm dir}(B^+ \to f_1(1285) \rho^+) &=& -6.7^{+2.2}_{-3.0}\%\;, \quad
\acp^{\rm dir}(B^+ \to f_1(1420) \rho^+) = -3.7^{+2.1}_{-2.4}\%\;, \\
\acp^{\rm dir}(B^0 \to f_1(1285) \rho^0) &=& 18.0^{+42.9}_{-30.5}\% \;, \quad
\acp^{\rm dir}(B^0 \to f_1(1420) \rho^0) = 24.1^{+20.0}_{-24.3}\%\;;
\eeq
in which various errors
as specified previously have been added in quadrature.
One can find that the large branching ratio of the order of $10^{-5}$ combined with direct {\it CP} asymmetry around $-9.7 \sim -4.5$ \%
in $B^+ \to f_1(1285) \rho^+$ is expected to be detected in
the near future at the LHCb and/or Belle-II experiments. With 
a somewhat large decay rate ${\cal O}(10^{-6})$, the small direct {\it CP} violation in $B^+ \to f_1(1420) \rho^+$ may not be easily accessed. However, it is worth mentioning that large direct {\it CP}-violating asymmetries exist in both transverse polarizations, i.e., parallel and perpendicular, as follows:
\beq
\acp^{\rm dir, ||}(B^+ \to f_1(1420) \rho^+) &=& 13.8^{+11.7}_{-11.8}\% \;, \quad
\acp^{\rm dir, \perp}(B^+ \to f_1(1420) \rho^+) = 10.5^{+12.8}_{-13.2}\% \;,
\eeq
which may be detectable and helpful to explore the physics involved in $B^+ \to f_1(1420) \rho^+$ decays. Note that the $B^0 \to f_1 \rho^0$ modes cannot be measured in the near future due to their very small decay rates, although the seemingly large direct {\it CP} violations have been predicted in the pQCD approach.

\item[(3)]
It is interesting to note from Tables~\ref{tab:f1kstp}, \ref{tab:f1kst0}, and \ref{tab:f1phi-s} that the direct-induced {\it CP} asymmetries for the penguin-dominated $B^+ \to f_1 K^{*+}$, $B^0 \to f_1 K^{*0}$, and $B_s^0 \to f_1 \phi$ decays with contaminations arising from tree amplitudes at different levels are predicted in SM as follows:
\beq
\acp^{\rm dir}(B^+ \to f_1(1285) K^{*+}) &=& -16.0^{+5.2}_{-4.9}\% \;, \quad
\acp^{\rm dir}(B^+ \to f_1(1420) K^{*+}) = 13.9^{+5.3}_{-5.3}\% \;; \\
\acp^{\rm dir}(B^0 \to f_1(1285) K^{*0}) &=& -7.8^{+2.5}_{-2.3}\%\;, \quad
\acp^{\rm dir}(B^0 \to f_1(1420) K^{*0}) = 4.7^{+1.4}_{-1.5}\%\;; \\
\acp^{\rm dir}(B_s^0 \to f_1(1285) \phi) &=& -5.3^{+1.4}_{-1.0}\%\;, \quad
\acp^{\rm dir}(B_s^0 \to f_1(1420) \phi) = 2.5^{+1.0}_{-0.9}\%\;,
\eeq
which indicates that the former $B^+ \to f_1 K^{*+}$ decays suffer from somewhat stronger interferences induced by larger tree contributions than the latter two modes.

By combining three polarization fractions in the transversity basis with those of {\it CP}-conjugated $\bar B$ decays, we also computed the
direct {\it CP} violations of the above-mentioned decays with a $\bar b \to \bar s$ transition in every polarization in the pQCD approach correspondingly.
\begin{itemize}
\item[]{\underline{$B^+ \to f_1(1285) K^{*+}$}:}
\beq
\acp^{\rm dir,L}
&=& -94.5^{+24.0}_{-7.7}\%\;, \qquad
\acp^{\rm dir,||}
= 8.2^{+2.4}_{-2.4}\%\;,\qquad
\acp^{\rm dir,\perp}
= 7.9^{+2.4}_{-2.3}\%\;;
\eeq

\item[]{\underline{$B^+ \to f_1(1420) K^{*+}$}:}
\beq
\acp^{\rm dir,L}
&=& 25.4^{+6.7}_{-6.8}\%\;, \qquad
\acp^{\rm dir,||}
= -14.1^{+6.5}_{-7.1}\%\;,\qquad
\acp^{\rm dir,\perp}
= -9.7^{+5.2}_{-5.0}\%\;;
\eeq

\item[]{\underline{$B^0 \to f_1(1285) K^{*0}$}:}
\beq
\acp^{\rm dir,L}
&=& 1.7^{+7.6}_{-11.3}\%\;, \qquad
\acp^{\rm dir,||}
= -9.3^{+1.7}_{-1.6}\%\;,\qquad
\acp^{\rm dir,\perp}
= -9.9^{+1.6}_{-1.8}\%\;;
\eeq

\item[]{\underline{$B^0 \to f_1(1420) K^{*0}$}:}
\beq
\acp^{\rm dir,L}
&=& 3.4^{+1.5}_{-1.9}\%\;, \qquad
\acp^{\rm dir,||}
= 7.9^{+2.9}_{-2.7}\%\;,\qquad
\acp^{\rm dir,\perp}
= 8.0^{+2.1}_{-2.4}\%\;;
\eeq

\item[]{\underline{$B_s^0 \to f_1(1285) \phi$}:}
\beq
\acp^{\rm dir,L}
&=& -7.2^{+2.1}_{-1.0}\%\;, \qquad
\acp^{\rm dir,||}
= -2.7^{+0.7}_{-0.4}\%\;,\qquad
\acp^{\rm dir,\perp}
= -2.8^{+0.6}_{-0.6}\%\;;
\eeq

\item[]{\underline{$B_s^0 \to f_1(1420) \phi$}:}
\beq
\acp^{\rm dir,L}
&=& 2.4^{+1.0}_{-0.7}\%\;, \qquad
\acp^{\rm dir,||}
= 2.6^{+1.4}_{-0.9}\%\;,\qquad
\acp^{\rm dir,\perp}
= 3.1^{+1.5}_{-1.2}\%\;;
\eeq
\end{itemize}
where the various errors as specified previously
have also been added in quadrature.
These pQCD predictions and phenomenological analyses of the direct {\it CP} violations of $B^{+,0} \to f_1 K^{*+,0}$ and $B_s^0 \to f_1 \phi$ decays could be tested in future measurements. Furthermore, the $B^+ \to f_1 K^{*+}$ modes with large branching ratios and large direct {\it CP} asymmetries are likely to be detected much easier in the near future.

\item[(4)]
It is worth stressing that no theoretical predictions or experimental measurements of the direct {\it CP}-violating asymmetries of 20 nonleptonic $B \to f_1 V$ decays are available yet. Therefore, examinations of these leading order pQCD predictions have to be left to LHCb and/or Belle-II, or even CEPC experiments in the future.
\end{enumerate}

\subsection{Weak annihilation contributions in $B \to f_1 V$ decays}

As proposed in~\cite{Kagan:2004uw}, a strategy correlated with penguin annihilation contributions
was suggested to explore the $B \to \phi K^*$ polarization anomaly in SM. The subsequently systematic studies on $B \to VV$ decays combined with rich data further confirm the important role of annihilation
contributions played, in particular, in the penguin-dominated modes~\cite{Beneke:2006hg,Cheng:2008gxa,Cheng:2009cn,Cheng:2009xz,Zou:2015iwa}. Here, it should be
mentioned that, up to now, different treatments on annihilation contributions have been proposed in
QCDF, SCET, and pQCD. For the former two approaches
based on the collinear factorization theorem,
both QCDF and SCET cannot directly evaluate the diagrams with
annihilation topologies because of the existence of end-point
singularities. However, different from
parametrizing and then fitting the annihilation contributions
through
rich data in QCDF~\cite{Beneke:2001ev}, the SCET method calculates the annihilation diagrams with the help of a zero-bin subtraction scheme and, consequently, obtains a real and small value for the annihilation decay amplitudes~\cite{Arnesen:2006vb}. As mentioned in the Introduction, the pQCD approach based on the $k_T$ factorization theorem together with $k_T$ resummation and threshold resummation techniques, makes the calculations of annihilation types of diagrams free of end-point singularities with a large imaginary part~\cite{Chay:2007ep}.
Recently, experimental measurements and theoretical studies on $B \to PP, PV, VV$ decays, especially on the pure annihilation-type decays such as $B^0 \to K^+ K^-$, $B_s^0 \to \pi^+ \pi^-$~\cite{Xiao:2011tx,Aaltonen:2011jv}, indicate that the
pQCD approach may be a reliable method to deal with annihilation diagrams in heavy $b$ flavor meson decays.

Because of similar behavior between vector and $^3\!P_1$-axial-vector
mesons, it is reasonable to conjecture that the weak annihilation
contributions can also play an important role, as in the $B \to VV$
ones~\cite{Beneke:2006hg,Cheng:2008gxa,Cheng:2009cn,Zou:2015iwa},
in the $B \to AV(VA)$ modes, in particular
the penguin-dominated ones.
Therefore, we will explore the important contributions from weak annihilation diagrams
to $B \to f_1 V$ decays considered in this work.
For the sake of simplicity, we will present the central values of
pQCD predictions of the {\it CP}-averaged
branching ratios, the polarization fractions, and the direct {\it CP}-violating
asymmetries with mixing angle $\phi_{f_1} = 24^\circ$
by taking the factorizable emission plus the nonfactorizable emission decay amplitudes into account.
Some numerical results and phenomenological discussions are given as follows:

\begin{itemize}
\item{Branching ratios}

When the annihilation contributions are turned off, the {\it CP}-averaged branching ratios of $B \to f_1 V$ decays in the pQCD approach then become
\beq
Br(B^+ \to f_1(1285) \rho^+) &=& 11.2 \times 10^{-6}\;, \qquad
Br(B^+ \to f_1(1420) \rho^+) = 2.3 \times 10^{-6} \;; \\
Br(B^+ \to f_1(1285) K^{*+}) &=& 1.4 \times 10^{-6}\;, \qquad
Br(B^+ \to f_1(1420) K^{*+}) = 2.7 \times 10^{-6}\;;\\
Br(B^0 \to f_1(1285) \rho^0) &=& 1.5 \times 10^{-7} \;, \qquad
Br(B^0 \to f_1(1420) \rho^0) = 7.5\times 10^{-8}\;; \\
Br(B^0 \to f_1(1285) K^{*0}) &=& 4.3\times 10^{-7} \;, \qquad
Br(B^0 \to f_1(1420) K^{*0}) =  2.5\times 10^{-6} \;; \\
Br(B^0 \to f_1(1285) \omega) &=&  7.7\times 10^{-7}\;, \qquad
Br(B^0 \to f_1(1420) \omega) = 1.4\times 10^{-7}\;; \\
Br(B^0 \to f_1(1285) \phi) &=&  5.2 \times 10^{-9}\;, \qquad
Br(B^0 \to f_1(1420) \phi) = 1.0\times 10^{-9}\;; \\
Br(B_s^0 \to f_1(1285) \rho^0) &=&  5.0\times 10^{-8}\;, \qquad
Br(B_s^0 \to f_1(1420) \rho^0) = 2.5\times 10^{-7}\;; \\
Br(B_s^0 \to f_1(1285) \bar{K}^{*0}) &=& 3.5\times 10^{-7} \;, \qquad
Br(B_s^0 \to f_1(1420) \bar{K}^{*0}) = 2.2\times 10^{-7}\;; \\
Br(B_s^0 \to f_1(1285) \omega) &=&  7.1\times 10^{-8}\;, \qquad
Br(B_s^0 \to f_1(1420) \omega) = 3.5\times 10^{-7}\;; \\
Br(B_s^0 \to f_1(1285) \phi) &=& 14.7 \times 10^{-6} \;, \qquad
Br(B_s^0 \to f_1(1420) \phi) = 15.4 \times 10^{-6}\;;
\eeq

\item{Longitudinal polarization fractions}

By neglecting the weak annihilation contributions, the {\it CP}-averaged longitudinal polarization fractions of $B \to f_1 V$ decays in the pQCD approach are written as,
\beq
f_L(B^+ \to f_1(1285) \rho^+) &=& 96.1 \%  \;, \qquad
f_L(B^+ \to f_1(1420) \rho^+) =  90.6\%  \;; \\
f_L(B^+ \to f_1(1285) K^{*+}) &=& 42.9\% \;, \qquad
f_L(B^+ \to f_1(1420) K^{*+}) = 70.4\% \;; \\
f_L(B^0 \to f_1(1285) \rho^0) &=&  91.7\% \;, \qquad
f_L(B^0 \to f_1(1420) \rho^0) = 17.5\% \;; \\
f_L(B^0 \to f_1(1285) K^{*0}) &=& 2.8\% \;, \qquad
f_L(B^0 \to f_1(1420) K^{*0}) =  75.9\% \;; \\
f_L(B^0 \to f_1(1285) \omega) &=&  46.4\%\;, \qquad
f_L(B^0 \to f_1(1420) \omega) = 27.2\%\;; \\
f_L(B^0 \to f_1(1285) \phi) &=& 46.8\% \;, \qquad
f_L(B^0 \to f_1(1420) \phi) = 47.1\%\;; \\
f_L(B_s^0 \to f_1(1285) \rho^0) &=&  80.2\%\;, \qquad
f_L(B_s^0 \to f_1(1420) \rho^0) = 80.4\%\;; \\
f_L(B_s^0 \to f_1(1285) \bar{K}^{*0}) &=& 42.3\% \;, \qquad
f_L(B_s^0 \to f_1(1420) \bar{K}^{*0}) = 75.6\% \;; \\
f_L(B_s^0 \to f_1(1285) \omega) &=&  51.0\%\;, \qquad
f_L(B_s^0 \to f_1(1420) \omega) = 51.4\%\;; \\
f_L(B_s^0 \to f_1(1285) \phi) &=& 54.6\% \;, \qquad
f_L(B_s^0 \to f_1(1420) \phi) = 78.9\% \;;
\eeq

\item{Direct {\it CP}-violating asymmetries}

Without the contributions arising from annihilation types of diagrams, the direct {\it CP}-violating asymmetries of $B \to f_1 V$ decays in the pQCD approach are given as,
\beq
\acp^{\rm dir}(B^+ \to f_1(1285) \rho^+) &=& -6.7 \% \;, \qquad
\acp^{\rm dir}(B^+ \to f_1(1420) \rho^+) = -2.2 \% \;; \\
\acp^{\rm dir}(B^+ \to f_1(1285) K^{*+}) &=& -15.0\% \;, \qquad
\acp^{\rm dir}(B^+ \to f_1(1420) K^{*+}) =  12.8\% \;; \\
\acp^{\rm dir}(B^0 \to f_1(1285) \rho^0) &=&  -83.5\% \;, \qquad
\acp^{\rm dir}(B^0 \to f_1(1420) \rho^0) = 35.4\% \;; \\
\acp^{\rm dir}(B^0 \to f_1(1285) K^{*0}) &=& -2.1\% \;, \qquad
\acp^{\rm dir}(B^0 \to f_1(1420) K^{*0}) = 3.4\% \;; \\
\acp^{\rm dir}(B^0 \to f_1(1285) \omega) &=&  -50.8\%\;, \qquad
\acp^{\rm dir}(B^0 \to f_1(1420) \omega) = -2.0\%\;; \\
\acp^{\rm dir}(B_s^0 \to f_1(1285) \rho^0) &=&  15.2\% \;, \qquad
\acp^{\rm dir}(B_s^0 \to f_1(1420) \rho^0) = 15.3\%\;; \\
\acp^{\rm dir}(B_s^0 \to f_1(1285) \bar{K}^{*0}) &=& 20.5\% \;, \qquad
\acp^{\rm dir}(B_s^0 \to f_1(1420) \bar{K}^{*0}) = -53.2\% \;; \\
\acp^{\rm dir}(B_s^0 \to f_1(1285) \omega) &=& 25.1\% \;, \qquad
\acp^{\rm dir}(B_s^0 \to f_1(1420) \omega) = 25.1\% \;; \\
\acp^{\rm dir}(B_s^0 \to f_1(1285) \phi) &=& -5.1\% \;, \qquad
\acp^{\rm dir}(B_s^0 \to f_1(1420) \phi) = 2.5\% \;.
\eeq
Note that because of the inclusion of pure penguin amplitudes, the direct {\it CP}-violating asymmetries
of $B^0 \to f_1 \phi$ decays are still zero, which are not presented here, even if the penguin annihilation contributions
are turned off in the SM. However, it should be mentioned again that once the future experimental
measurements release evidently nonzero and large direct {\it CP} violations,
there might be NP
beyond the SM hidden in these two decay modes.

\end{itemize}

Generally speaking, compared with the numerical results by considering the weak annihilation contributions in the pQCD approach as shown
in Tables~\ref{tab:f1rhop}-\ref{tab:f1phi-s}, it is clear to see
that the branching ratios and longitudinal polarization fractions of $B^+ \to f_1 \rho^+$, $B^0 \to f_1(1420) \rho^0$, $B_s^0 \to f_1 \rho^0$,
$B_s^0 \to f_1(1420) \omega$, and $B_s^0 \to f_1(1285) \phi$ decays almost remain unchanged
when the annihilation contributions are neglected, while the other channels are affected by the
annihilation decay amplitudes at different levels. Particularly, the contributions induced by the weak annihilation diagrams can
make the $B^0 \to f_1(1285) K^{*0}$ decay rate(longitudinal polarization fraction) amazingly change from $4.3 \times 10^{-7}(2.8\%)$ to $5.0 \times 10^{-6}(15.8\%)$.
From the pQCD point of view,
because the annihilation amplitudes
can contribute to {\it CP} violation as a source of the large
strong phase, the direct {\it CP}-violating asymmetries of $B \to f_1 V$ decays without annihilation contributions will deviate from the predictions presented in Tables~\ref{tab:f1rhop}-\ref{tab:f1phi-s} more or less, except for the $B^0 \to f_1 \phi$ modes with still invariant zero direct {\it CP} violations. Of course, the above general expectations in the pQCD approach
will be examined by the relevant experiments in the future, which could be
helpful to understand the annihilation decay mechanism in vector-vector
and vector-axial-vector $B$ decays in depth.

In order to clearly examine the important contributions from annihilation diagrams, we present the explicit
decay amplitudes decomposed as $B \to f_{1q} V$ and $B \to f_{1s} V$ for $B^+ \to f_1 \rho^+$, $B^{+,0} \to f_1 K^{*+,0}$, and $B_s^0 \to f_1 \phi$ modes with
large branching ratios in Tables~\ref{tab:DAs-f1rho-u}-\ref{tab:DAs-f1phi-s} with and without annihilation contributions on three polarizations. One can easily find from Table~\ref{tab:DAs-f1kst-d}, for $B^0 \to f_1 K^{*0}$ for example, that the significant variations induced by weak annihilation contributions mainly arise in the imaginary part of decay amplitudes on every polarization. Furthermore, when the annihilation
decay amplitudes are not considered, then one can straightforwardly see from the numerical results shown in the parentheses that, combined with the dominant $A_T(B^0 \to f_{1q} K^{*0})$ amplitude, almost exact cancellation of the longitudinal polarization and somewhat stronger destructive
interferences on the other two transverse polarizations between
$B^0 \to f_{1q} K^{*0}$ and $B^0 \to f_{1s} K^{*0}$ modes in
the $B^0 \to f_1(1285) K^{*0}$ decay resulted in a significantly
smaller branching fraction, about ${\cal O}(10^{-7})$, and 
surprisingly large transverse polarization fraction, around $97 \%$.
Consequently,
lack of a large strong phase coming from annihilation contributions in the pQCD approach lead to
a much smaller direct {\it CP}-violating asymmetry in magnitude, around $2\%$. Contrary to $B^0 \to f_1(1285) K^{*0}$ decay, because of the dominance of  $B^0 \to f_{1s} K^{*0}$ on the longitudinal polarization in the $B^0 \to f_1(1420) K^{*0}$ channel, the constructive interferences between $B^0 \to f_{1q} K^{*0}$ and $B^0 \to f_{1s} K^{*0}$ modes on every polarization make the decay rate somewhat smaller, with a factor of around 0.6, and the longitudinal 
polarization fraction slightly larger than those
corresponding results shown in Table~\ref{tab:f1kst0}, although the similarly large annihilation contributions are also turned off, which can be easily seen from the decay amplitudes given in Table~\ref{tab:DAs-f1kst-d}. Again, these important annihilation contributions should be tested by future experiments to further deepen our knowledge of the annihilation decay mechanism in the heavy $b$ flavor sector.

\section{Conclusions and Summary} \label{sec:summary}

In this work, we studied 20 nonleptonic decays of $B \to f_1 V$ by employing the pQCD
approach based on the framework of the $k_T$ factorization theorem. The singularities that appeared
in collinear factorization
were then naturally smeared by picking up the transverse momentum $k_T$ of valence quarks
when the quark momentum fraction $x$ approaches the end-point region. Consequently, with
the pQCD formalism,
the Feynman diagrams of every topology can be calculated perturbatively without introducing
any new parameters, which is a unique point, different from the QCDF
and the SCET based on the collinear factorization theorem.
In order to explore the perturbative and nonperturbative QCD dynamics to further understand the helicity structure
of the decay mechanism in $B \to f_1 V$ decays, we calculated the {\it CP}-averaged branching ratios,
the polarization fractions, the direct {\it CP}-violating asymmetries,
and the relative phases of those considered decay modes, where the mixing angle $\phi_{f_1}
\sim 24^\circ$ between two axial-vector $f_1(1285)$ and $f_1(1420)$ states adopted from
the first measurements of $B_{d/s} \to J/\psi f_1(1285)$ decays in the heavy $b$ flavor sector.

From our numerical pQCD predictions and phenomenological analysis, we found the following points:
\begin{enumerate}
\item[(a)]
The large {\it CP}-averaged branching ratios for $B^+ \to f_1 \rho^+$, $B^{+,0} \to f_1 K^{*+,0}$, and $B_s^0 \to
f_1 \phi$ decays are predicted in the pQCD approach as follows:
\beq
Br(B^+ \to f_1(1285) \rho^+) &=& 11.1^{+8.7}_{-6.8} \times 10^{-6}\;, \qquad \;\;
Br(B^+ \to f_1(1420) \rho^+) = 2.3^{+1.9}_{-1.4} \times 10^{-6}\;;\\
Br(B^+ \to f_1(1285) K^{*+}) &=& 6.4^{+3.6}_{-2.5} \times 10^{-6}\;, \qquad \;
Br(B^+ \to f_1(1420) K^{*+}) = 4.5^{+1.7}_{-1.5} \times 10^{-6}\;;\\
Br(B^0 \to f_1(1285) K^{*0}) &=& 5.0^{+2.7}_{-2.1} \times 10^{-6}\;, \qquad \; \; \
Br(B^0 \to f_1(1420) K^{*0}) = 4.4^{+1.7}_{-1.5} \times 10^{-6}\;;\\
Br(B_s^0 \to f_1(1285) \phi) &=& 14.7^{+8.7}_{-6.4} \times 10^{-6}\;, \qquad \; \; \; \; \;
Br(B_s^0 \to f_1(1420) \phi) = 16.2^{+9.9}_{-7.6} \times 10^{-6}\;,
\eeq
which are expected to be measured at the running LHCb and the forthcoming Belle-II experiments
in the near future. It is noted that the decay rates and
decay pattern of $B^+ \to f_1 \rho^+$ predicted in the pQCD approach are very consistent with those
 given in the QCDF approach within theoretical errors.
But, it is not the same case for the $B^{+,0} \to f_1 K^{*+,0}$ decay modes.
The future 
experimental measurements with good precision for the branching ratios and the pattern
of $B^{+,0} \to f_1 K^{*+,0}$ decays will be helpful for us to examine these two different factorization approaches.

\item[(b)]
In order to decrease the effects of the large theoretical errors of the branching ratios induced by those input parameters,
we also define the ratios of the decay rates among the ten
$B^+ \to f_1 \rho^+$, $B^{+,0} \to f_1 K^{*+,0}$, $B^0 \to f_1 \omega$, and $B_s^0 \to f_1 \phi$ decay
modes as given in Eqs.~(\ref{eq:r-f1rhokst-u})-(\ref{eq:r-sd-14}),
where the large uncertainties of the predicted branching ratios are canceled to a large extent in such ratios.
The future experimental measurements of these newly defined ratios will be helpful
to further determine the mixing angle $\phi_{f_1}$ between $f_{1q}$
and $f_{1s}$ states for an axial-vector $f_1(1285)-f_1(1420)$ mixing system in the quark-flavor basis.

\item[(c)]
The predictions of polarization fractions for the 20 nonleptonic $B \to f_1 V$ decays are given
explicitly in the pQCD approach. Furthermore, associated with large branching ratios, the large
longitudinal(transverse) polarization fractions in $B^+ \to f_1 \rho^+$, $B^{+,0} \to f_1(1420)
K^{*+,0}$, $B^0 \to f_1(1285) \omega$, and $B_s^0 \to f_1 \phi$ [$B^{+,0} \to f_1(1285) K^{*+,0}$
and $B^0 \to f_1(1420) \omega$] decays are expected to be detected at LHCb and Belle-II
experiments and to provide useful information to understand the famous polarization puzzle in rare
vector-vector $B$ meson decays, which will be helpful to shed light on the helicity structure of the decay mechanism.

\item[(d)]
Some large direct {\it CP}-violating asymmetries of $B \to f_1 V$ decays are provided with the pQCD approach,
such as $\acp^{\rm dir}(B^+ \to f_1(1285) K^{*+}) = -16.0^{+5.2}_{-4.9}\%$,
$\acp^{\rm dir}(B^+ \to f_1(1420) K^{*+}) = 13.9^{+5.3}_{-5.3}\%$, $\acp^{\rm dir}(B^0 \to f_1(1285) K^{*0})
= -7.8^{+2.5}_{-2.3}\%$, and even $\acp^{\rm dir, ||}(B^+ \to f_1(1420) \rho^+) = 13.8^{+11.7}_{-11.8}\%$
and $\acp^{\rm dir, \perp}(B^+ \to f_1(1420) \rho^+) = 10.5^{+12.8}_{-13.2}\%$, and so forth, which
are believed to be detectable at the LHCb, Belle-II, and even the future CEPC experiments.
At the same time, a stringent examination of the zero direct {\it CP} asymmetries in the SM
of $B^0 \to f_1 \phi$ decays is of great interest to provide useful information for the possible signal of the
new physics beyond the SM.
Moreover, the theoretical estimations on physical observables of $B_s \to f_1 V$ decays are given
for the first time in the pQCD approach, which can also be tested in the future.

\item[(e)]
The weak annihilation contributions play an important role in many $B \to f_1 V$ decays.
The near future measurements with good precision on some decay modes affected significantly by
the annihilation amplitudes, such as $B^{+,0} \to f_1 K^{*+,0}$ with large branching ratios,
can provide evidence to verify the reliability of the pQCD approach on the
calculations of annihilation-type diagrams, and help us to understand the annihilation mechanism in the heavy flavor sector.

\end{enumerate}


\begin{acknowledgments}
This work is supported by the National Natural Science
Foundation of China under Grants No.~11205072, No.~11235005,
and No.~11447032
and by a project funded by the Priority Academic Program Development
of Jiangsu Higher Education Institutions (PAPD),
by the Research Fund of Jiangsu Normal University under Grant No.~11XLR38,
and by the Natural Science Foundation of Shandong Province
under Grant No.~ZR2014AQ013.
\end{acknowledgments}


\begin{appendix}

\section{ Mesonic distribution amplitudes}
\label{sec:app1}

As we know, mesonic distribution amplitudes in hadron wave functions are the essential nonperturbative inputs in the pQCD approach. Now, we will give a brief introduction to these items involved in the present work.

For the $B$ meson, the distribution amplitude in the impact $b$
space has been proposed as 
\beq
\phi_{B}(x,b)&=& N_Bx^2(1-x)^2
\exp\left[-\frac{1}{2}\left(\frac{xm_B}{\omega_b}\right)^2
-\frac{\omega_b^2 b^2}{2}\right] \;,
\eeq
in Ref.~\cite{Keum:2000ph} and widely
adopted, for example, in~\cite{Keum:2000ph,Ali:2007ff,Zou:2015iwa,
Liu:2014doa,Liu:2014dxa,Liu:2014jsa,Liu:2015sra,Li:2010nn},
where the normalization factor $N_{B}$
is related to the decay constant $f_{B}$ through Eq.~(\ref{eq:norm}).
The shape parameter $\omega_b$ was fixed at
$0.40$~GeV by using the rich experimental data on the $B^+$ and $B^0$ mesons, 
with $f_{B}= 0.19$~GeV, based on many calculations of form factors~\cite{Lu:2002ny} and
other well-known modes of $B^+$ and $B^0$ mesons~\cite{Keum:2000ph}
in the pQCD approach.
Here, the assumption of isospin symmetry has been made. For the $B_s^0$ meson, relative to the lightest
$u$ or $d$ quark, the heavier $s$ quark leads to a somewhat larger momentum fraction than that
of the $u$ or $d$ quark in the $B^+$ or $B^0$ mesons.
Therefore, by taking a small SU(3) symmetry-breaking effect into account,
we adopt the shape parameter
$\omega_{b} = 0.50$~GeV with $f_{B} = 0.23$~GeV
for the $B_s$ meson~\cite{Ali:2007ff}, and the
corresponding normalization
constant is $N_{B} = 63.67$.
In order to estimate the theoretical uncertainties
induced
by the inputs, we consider varying the shape parameter $\omega_{b}$
by 10\%, i.e., $\omega_b = 0.40 \pm 0.04$~GeV for $B^+$ and $B^0$ mesons and $\omega_{b} = 0.50 \pm 0.05$~GeV for the $B_s^0$ meson,
respectively.

The twist-2 light-cone distribution amplitudes $\phi_{V}$ and $\phi_{V}^T$ can be parametrized as
\beq
\phi_{V}(x)&=&\frac{3f_{V}}{\sqrt{2N_c}} x
(1-x)\left[1+3a_{1V}^{||}\, (2x-1)+ a_{2V}^{||}\, \frac{3}{2} ( 5(2x-1)^2  - 1 )\right]\;,\\
\phi_{V}^T(x)&=&\frac{3f^T_{V}}{\sqrt{2N_c}} x
(1-x)\left[1+3a_{1V}^{\perp}\, (2x-1)+ a_{2V}^{\perp}\,
\frac{3}{2} ( 5(2x-1)^2  - 1 )\right]\;,\label{phiV}
\eeq
in which, $f_{V}$ and $f_V^T$ are the decay constants of the vector meson
with longitudinal and transverse polarization, respectively, whose values are
shown in Table \ref{f-vector}.
\begin{table}[hbt]
 \caption{ Input values of the decay constants  of the light vector
mesons (in MeV)~\cite{Beringer:1900zz,Ball:2004rg}}
\label{f-vector}
 \begin{center}\vspace{-0.3cm}{
\begin{tabular}[t]{cccccccc}
\hline\hline
 $f_\rho $ & $ f_\rho^T $ & $ f_\omega $ & $ f_\omega^T $
 & $ f_{K^*} $ & $ f_{K^*}^T $ & $f_\phi $ & $
f_\phi^T $  \\
 $ 209\pm 2$& $ 165\pm 9 $&
 $ 195\pm 3$&
 $ 145\pm 10$&
$ 217\pm 5$&
 $185\pm 10$&
 $ 231\pm 4$&
 $ 200\pm 10$\\
\hline \hline
\end{tabular}}
\end{center}
\end{table}
The decay constants can be extracted from $V^0\to l^+l^-$ and $\tau \to V^- \bar\nu$~\cite{Li:2006jv,Ball:2007rt}.
The Gegenbauer moments taken from the  recent updates~\cite{Ball:2007rt}
are collected in Table~\ref{tab:Gegmom-V}.
\begin{table}[hbt]
\caption{ Gegenbauer moments in the distributions amplitudes of the lightest vector mesons taken at $\mu=1$ GeV~\cite{Ball:2007rt}}
\label{tab:Gegmom-V}
 \begin{center}\vspace{-0.3cm}{
\begin{tabular}[t]{c|c||c|c||c|c||c|c}
\hline  \hline
     \multicolumn{4}{c||}{$K^*$ meson} &
     \multicolumn{2}{c||}{$\rho$ and $\omega$ mesons} &  \multicolumn{2}{c}{$\phi$ meson} \\
   \hline
   $a^{\parallel}_1$
   & $a_2^{\parallel}$
 & $a_1^\perp$ & $a_2^\perp$   &$a_2^\parallel$
   & $a_2^\perp$
 & $a_2^\parallel$ & $a_2^\perp$   \\
 \hline \hline
  $0.03\pm0.02 $
 &$0.11\pm0.09 $
 &$0.04\pm0.03 $
 &$0.10\pm0.08 $
 &$0.15\pm0.07 $
 &$0.14\pm0.06 $
 &$0.18\pm0.08 $
 &$0.14\pm0.07 $
 \\
\hline  \hline
\end{tabular}}
\end{center}
\end{table}

The asymptotic forms of the twist-3 distribution amplitudes
$\phi^{t,s}_V$ and $\phi_V^{v,a}$ are~\cite{Li:2004mp,Li:2004ti}
\beq
\phi^t_V(x) &=& \frac{3f^T_V}{2\sqrt {2N_c}}(2x-1)^2,\;\;\;\;\;\;\;\;\;\;\;
  \hspace{0.5cm} \phi^s_V(x)=-\frac{3f_V^T}{2\sqrt {2N_c}} (2x-1)~,\\
\phi_V^v(x)&=&\frac{3f_V}{8\sqrt{2N_c}}(1+(2x-1)^2),\;\;\;\;\; \ \
 \phi_V^a(x)=-\frac{3f_V}{4\sqrt{2N_c}}(2x-1).
\eeq

For the axial-vector state $f_{1q(s)}$, its
twist-2 light-cone distribution amplitudes
can generally be expanded as the Gegenbauer polynomials~\cite{Yang:2007zt}:
\beq
 \phi_{f_{1q(s)}}(x)&=&\frac{f_{f_{1q(s)}}}{2\sqrt{2N_c}} 6 x  (1-x) \left[ 1  + a_{2}^\parallel\, \frac{3}{2} ( 5(2x-1)^2  - 1 )
\right]\;,\\
\phi_{f_{1q(s)}}^T(x)&=& \frac{f_{f_{1q(s)}}}{2\sqrt{2N_c}}6 x (1-x)
\left[  3 a_{1}^\perp\, (2x-1) \right] \;,
\eeq

For twist-3 ones, we use the following
form as in Ref.~\cite{Li:2009tx}:
\beq
\phi_{f_{1q(s)}}^s(x)&=&\frac{f_{f_{1q(s)}}}{4\sqrt{2N_c}} \frac{d}{dx}\Biggl[ 6x
(1-x) (  a_{1}^\perp (2x-1) )\Biggr]\;, \\
\phi_{f_{1q(s)}}^t(x)&=&\frac{f_{f_{1q(s)}}}{2\sqrt{2N_c}}\Biggl[
\frac{3}{2}\,a_{1}^\perp\,(2x-1) (3 (2x-1)^2-1)\Biggr],
\eeq
\beq
\phi_{f_{1q(s)}}^v(x)&=&\frac{f_{f_{1q(s)}}}{2\sqrt{2N_c}}
\Biggl[\frac{3}{4}
(1+(2x-1)^2) \Biggr]\;,\qquad
\phi_{f_{1q(s)}}^a(x)=\frac{f_{f_{1q(s)}}}{8\sqrt{2N_c}}
\frac{d}{dx}\Biggl[6 x (1-x) \Biggr]\;.
\eeq
where $f_{f_{1q(s)}}$ is the ``normalization" constant for both
longitudinally and transversely polarized mesons and
the Gegenbauer moments $a_{2(1)}^{\parallel(\perp)}$ can be found in
Table~\ref{tab:Gegmom-A}.
\begin{table}[hbt]
\caption{ Same as Table~\ref{tab:Gegmom-V} but for light axial-vector
$f_{1q}$ and $f_{1s}$ states~\cite{Yang:2007zt}.}
\label{tab:Gegmom-A}
 \begin{center}\vspace{-0.3cm}{
\begin{tabular}[t]{c|c||c|c}
\hline  \hline
     \multicolumn{2}{c||}{$f_{1q}$ state}
     &  \multicolumn{2}{c}{$f_{1s}$ state}  \\
   \hline
   $a_2^\parallel$
   & $a_1^\perp$
 & $a_2^\parallel$ & $a_1^\perp$   \\
 \hline \hline
 $-0.05 \pm 0.03 $
 &$-1.08 \pm 0.48 $
 &$0.10^{+0.15}_{-0.19} $
 &$0.30^{+0.00}_{-0.33} $
 \\
 \hline \hline
\end{tabular}}
\end{center}
\end{table}

\end{appendix}


\end{document}